\documentclass[fleqn,usenatbib]{mnras}

\usepackage{newtxtext,newtxmath}

\usepackage[T1]{fontenc}

\DeclareRobustCommand{\VAN}[3]{#2}
\let\VANthebibliography\thebibliography
\def\thebibliography{\DeclareRobustCommand{\VAN}[3]{##3}\VANthebibliography}

\usepackage{fancybox}
\usepackage{graphicx}	
\usepackage{amsmath}	

\title[Multi-stream radial structure of CDM haloes]{Multi-stream radial structure of cold dark matter haloes from particle trajectories: deep inside splashback radius}

\author[Y.Enomoto et al.]{
Yohsuke Enomoto,$^{1}$\thanks{E-mail: enomoto@tap.scphys.kyoto-u.ac.jp}
Takahiro Nishimichi,$^{2,3,4}$
Atsushi Taruya$^{3,4}$
\\
$^{1}$Department of Physics, Kyoto University, Kyoto 606-8502, Japan\\
$^{2}$Department of Astrophysics and Atmospheric Sciences, Faculty of Science,
Kyoto Sangyo University, Motoyama, Kamigamo, Kita-ku, Kyoto 603-8555, Japan\\
$^{3}$Centre for Gravitational Physics and Quantum Information, Yukawa Institute for Theoretical Physics, Kyoto University, Kyoto 606-8502, Japan\\
$^{4}$Kavli Institute for the Physics and Mathematics of the Universe (WPI),
The University of Tokyo Institutes for Advanced Study, The University of Tokyo,\\ 5-1-5 Kashiwanoha, Kashiwa, Chiba 277-8583, Japan
}

\thisfancyput(14.5cm,0.5cm){\large{YITP-23-117}}

\date{Accepted XXX. Received YYY; in original form ZZZ}

\pubyear{2023}

\begin{document}
\label{firstpage}
\pagerange{\pageref{firstpage}--\pageref{lastpage}}
\maketitle

\begin{abstract}
By tracking trajectories of dark matter (DM) particles accreting onto haloes in cosmological $N$-body simulations, we investigate the radial phase-space distribution of cold dark matter (CDM) haloes, paying attention to their inner regions deep inside the halo boundary called the splashback radius, where the particles undergo multi-stream flows. 
Improving the analysis by Sugiura et al., we classify DM particles 
by the number of apocenter passages, $p$, and count it up to $p=40$ for each halo over a wide mass range. Quantifying the radial density profile for particles having the same value of $p$, we find that it generally exhibits a double-power law feature, whose indices of inner and outer slopes are well-described by $-1$ and  $-8$, respectively. 
Its characteristic scale and density are given as a simple fitting function of $p$, with a weak halo mass dependence. Interestingly, summing up these double-power law profiles beyond $p=40$ reproduces well the total density profile of simulated haloes. The double-power law nature is persistent and generic not only in mass-selected haloes but also in haloes selected in different criteria. 
Our results are compared with self-similar solutions that describe the stationary and spherical accretion of DM.
We find that even when introducing a non-zero angular momentum, none of them explain the radial multi-stream structure. 
The analysis with particle trajectories tracing back to higher redshifts suggests that the double-power law nature has been established during an early accretion phase and remains stable.

\end{abstract}




\section{Introduction}

It is widely accepted that dark matter is an important constituent that dominates $~80$\% of the matter components of the universe. Dark matter thus plays a very crucial role in cosmic structure formation driven by the gravitational instability. In particular, the dark matter is supposed to be non-relativistic and to have a negligibly small velocity dispersion, which accounts for the early growth of cosmic density fields just after the recombination epoch, referred to as the cold dark matter (CDM). In the process of cosmic structure formation, the gravitational collapse of cold dark matter is followed by the formation of dark matter haloes, i.e., self-gravitating bound systems composed of dark matter, and a sufficient amount of baryon has been accumulated by their potential well. This explains why the dark matter haloes are considered as an ideal site of galaxy and star formation \citep[][]{1977MNRAS.179..541R,1978MNRAS.183..341W}, and are important building blocks to explain the observed large-scale structure. Since the structural properties of dark matter haloes are known to be very sensitive to their formation and merging histories, the dark matter halo offers unique testing ground for structure formation scenarios, and there have been so far numerous discussions based on the observed small-scale structures \citep[see][for comprehensive reviews]{Bullock_Boylan-Kolchin2017}.

Theoretically, CDM haloes are regarded as a collisionless bound system, and their evolved structure in general depends on the initial conditions. However, early numerical simulations have shown that the radial density profile of each halo, $\rho(r)$, has a similar shape, and is described in a universal fashion by the so-called Navarro-Frenk-White (NFW) profile \citep{1997ApJ...490..493N,2014MNRAS.441..378L}. One striking feature of this profile is that haloes commonly have a shallow cusp with the density slope of $-1$, i.e., $\rho(r)\propto r^{-1}$. Although it has been later suggested that the density profile proposed by \citet{einasto1965construction}  provides a more accurate description \citep{2004MNRAS.349.1039N,2008MNRAS.387..536G,2014MNRAS.441.3359D,2020Natur.585...39W}, a cuspy structure of CDM haloes still persists in cosmological $N$-body simulations, and its physical origin remains unresolved.

For more physical insight supported by various observations, CDM is considered to have initially a negligible velocity dispersion. This implies that in six-dimensional phase space (i.e., velocity and position in three-dimensional space), the initial distribution of DM is described by the three-dimensional sheet. Due to the collisionless nature, the topology of such a structure is preserved during the formation and evolution process of haloes. One thus anticipates that the phase-space distribution of haloes is still described by the three-dimensional sheet, but it exhibits a complex folded sheet structure having a multi-valued velocity flow for a given position \citep{CausticsWhite2009,CausticVogelsberger2009,Vogelsberger2011,Colombi2021}. This multi-stream structure would be a unique and distinctive feature of CDM, and provide a clue to discriminate from other DM candidates. A natural expectation may be that the universal density profiles seen in the configuration space are a direct consequence of the phase-space structures having some universal features. In this respect, it would be very interesting and useful to quantitatively investigate the phase-space structure of CDM haloes.  In fact, the multi-stream nature of haloes has attracted recent attention, highlighted with a renewed interest as the outer boundary of the multistream region, referred to as the splashback radius \citep{2014ApJ...789....1D,2014JCAP...11..019A}. There have been numerous works to investigate the splashback radius \citep{2016ApJ...825...39M,2017ApJ...841...18B,2019MNRAS.485..408C,2020MNRAS.499.3534X,Shin2021}.

Motivated by these, \cite{2020MNRAS.493.2765S} (hereafter S20) developed a method to reveal the multi-stream nature of haloes in radial phase space. Extending the SPARTA algorithm in \citet{2017ApJS..231....5D} and using the cosmological $N$-body simulations, they succeeded to clarify outer part of the multi-stream flows of CDM haloes. Further, in comparison with the self-similar solution by \citet{1984ApJ...281....1F}, their radial phase-space structures are quantified, finding that $\sim30$\% of the simulated haloes are well-described by the self-similar solutions with a wide range of mass accretion rate. As will be discussed in more detail, their method relies on many simulation snapthots at different redshifts in order to track back the trajectory of each DM particle in haloes identified at the present time. They then count, the number of apocentre passages for each DM particle orbiting around the halo centre. Denoting it by $p$, a family of particles having the same value of $p$ characterises a specific stream of multi-stream DM flow in radial phase space. In this way, S20 investigated the multi-stream properties using the DM particles up to $p=5$.

Note that the radial streams of $p\leq5$  are still away from the halo centre, and hence one expects that their structures are sensitive to the outer environment, which often exhibits irregular and extended structures in the presence of the merging haloes/subhaloes. This may partly explain why only the $\sim30$\% of haloes is described by the self-similar solutions. In other words, if one succeeds in revealing inner multi-stream structures, each stream may exhibit a universal feature, giving a clue to clarify the origin of cuspy structure in radial density profiles. In this respect, characterizing the inner multi-stream structure would provide a more fundamental characteristic useful to describe the physical properties of CDM haloes.

Along the line of this, the goal of this paper is, therefore, to characterise the inner multi-stream structure based on the method developed by S20 with a substantial improvement in both the simulation data set and numerical analysis. The present paper is regarded as a follow-up paper of \citet{2023ApJ...950L..13E} (hereafter E23), which highlights our major finding that the radial density profile of each stream characterised by the number of apocentre passages $p$ is commonly described by a simple double-power law function irrespective of $p$. On top of this, the paper further includes extensive discussions on the robustness of these findings together with a comprehensive study of the radial phase-space distributions in comparison with self-similar solutions. 
We also carefully describe the method of counting the number of apocenter passages of $N$-body particles.

This paper is organized as follows. 
In Section~\ref{sec:datas}, we introduce our simulations and the halo catalog. 
Then, we introduce the methods counting $p$ of simulation particles in Section~\ref{subsec:halo_centre} and \ref{subsec:counting}, and stacking and fitting procedures in Section~\ref{subsubsec:density_profiles}.
We show the results for individual halos in Section~\ref{subsec:individual_profiles} and stacked profiles in Section~\ref{subsec:dep_on_mass} with introducing the double-power law density profile found in E23.
In Section~\ref{subsec:alpha_beta_free}, we show the optimized $\chi^2$ of fitting for stacked profiles and investigate the best-fit values of the inner and outer slope of the double-power law. 
Then we investigate the dependence of the two free parameters on $p$ and the halo mass in Section~\ref{subsec:universal_stream_profiles}.
We explore the dependency on the concentration and on the recent accretion rate in Section~\ref{subsec:dep_on_MAR}, discuss the physical meaning of the value of inner slopes in Section~\ref{subsec:SScompar}, and the evolution of $\rho(r;p)$ to explore why the universal feature appears in Section~\ref{subsec:dp_emerge}.
Finally, Section~\ref{sec:conclusions} provides conclusions and prospects for future study.

\section{Data}\label{sec:datas}

\subsection{$N$-body simulations}\label{subsec:n-body}
Our analysis is based on two cosmological $N$-body simulations (LR and HR) performed in a periodic comoving box with a side length of $41\,h^{-1}\mathrm{Mpc}$, loaded with different numbers of particles as listed in Table~\ref{tab:simulation_parameters}.
We assume a flat-geometry $\Lambda$CDM universe consistent with the recent result from the Planck satellite:
$\Omega_m = 0.3089,\Omega_\Lambda=0.6911,h=0.6774,n_s=0.9667,\sigma_8=0.8259$  \citep{2016A&A...594A..13P}. 
The initial conditions are generated by adding displacements to particles arranged in a regular lattice based on second-order Lagrangian perturbation theory \citep{Scoccimarro1998,Crocce2006}, sourced by a Gaussian random field drawn from the linear matter power spectrum computed using the CLASS Boltzmann solver \citep{Blas2011}. 
The LR and HR simulations share the same random realization to allow for straightforward comparison.
We then evolve the positions and velocities of the particles using a TreePM code \textsc{GINKAKU} (Nishimichi, Tanaka \& Yoshikawa in prep.). 
This code is developed based on a public library, the Framework for Developing Particle Simulators \citep[\textsc{FDPS};][]{2016PASJ...68...54I,2018PASJ...70...70N}, which is aimed at efficient particle simulations in modern supercomputers. Working in a hybrid MPI-openmp parallel mode, the library allows an efficient domain decomposition as well as communication between processors. The short-range tree force is accelerated by the use of SIMD instructions implemented in the \textsc{Phantom-GRAPE} library \citep{Nitadori2006-ek,2012NewA...17...82T,2013NewA...19...74T}, while the details of the long-range PM force can be found in \citet{Yoshikawa_2005} (see also \citealt{2009PASJ...61.1319I,2012arXiv1211.4406I}).

To track the trajectories of particles, we utilized 1001 snapshots of LR, which were uniformly sampled in redshift from $z=5$ to $z=0$. Since LR comprises approximately five times the number of snapshots used in S20, we anticipate that we can analyze particle orbits in greater detail, particularly those that undergo apocentre passages more than five times.
In order to assess the convergence of the density profile considering the limited softening length, we rely on the $z=0$ snapshot of HR. 
Note that we have not retained particle snapshots at higher redshifts for the HR simulation due to constraints in disk storage capacity. 
Therefore, we investigate the evolution of the phase-space structure based on LR data, and then verify the accuracy of the density profile at $z=0$ through HR data.

\subsection{Halo catalogue}\label{subsec:halo catalogue}
\begin{table}
	\centering
	\caption{The parameters of $N$-body simulations. $N^3$ denotes the number of simulation particles, $m_p$ the mass of simulation paticles, $\epsilon$ the softening length, $N_\mathrm{snaps}$ the number of snapshots.}
	\label{tab:simulation_parameters}
	\begin{tabular}{lcccc} 
		\hline
		Name & $N$ & $m_p$ & $\epsilon$ & $N_\mathrm{snaps}$\\
		 & & ($h^{-1} M_\odot$)& ($h^{-1}\mathrm{kpc}$) & \\
		\hline
		LR & $500^3$ &  $4.72716\times 10^{7}$ & 4.10 & $1001$\\
		HR & $2000^3$ & $7.38619\times 10^{5}$ & 1.025 & $1$ ($z=0$)\\
		\hline
	\end{tabular}
\end{table}

We identified haloes at $z=0$ using a 6D phase-space temporal friends-of-friends halo finder \textsc{Rockstar} \citep[][]{2013ApJ...762..109B}. 
As explained in more detail in Section~\ref{subsec:halo_centre}, we tracked both the centre and the surrounding particles of the haloes identified at $z=0$ backward in time to establish the most massive main progenitor branch.

While \textsc{Rockstar} can identify haloes undergoing major mergers, the positions, velocities, and particle memberships of such haloes are often highly uncertain. Additionally, as we will discuss detail in Section~\ref{subsec:halo_centre}, it is typically challenging to rigorously track the main progenitor branch of merger trees for these haloes, especially when they are in close proximity to other massive haloes. 
To circumvent these challenges, we concentrate on relaxed haloes that are less influenced by recent mergers, using two parameters: the spin parameter $\lambda$ and the offset parameter $X_{\mathrm{off}}$, both calculated by the \textsc{Rockstar} halo finder. 
Here, the parameter $X_\mathrm{off}$ represents the distance between the density peak location and the centre of mass of particles, normalized by the virial radius $R_\mathrm{vir}$. 
On the other hand, $\lambda$ denotes the amplitude of angular momentum divided by $\sqrt{2}R_\mathrm{vir} V_\mathrm{vir} M_\mathrm{vir}$, and is referred to as the Bullock spin parameter in \citet[][]{2013ApJ...762..109B}.
Following \citet{2016MNRAS.457.4340K}, we consider haloes that meet either of the following conditions as undergoing major mergers and remove them from our halo catalogue:
\begin{equation}\label{eq:equilibrium1}
    \lambda>0.07,\:\:X_{\mathrm{off}}>0.07.
\end{equation}
\begin{figure}
    \centering
    \includegraphics[width=\columnwidth]{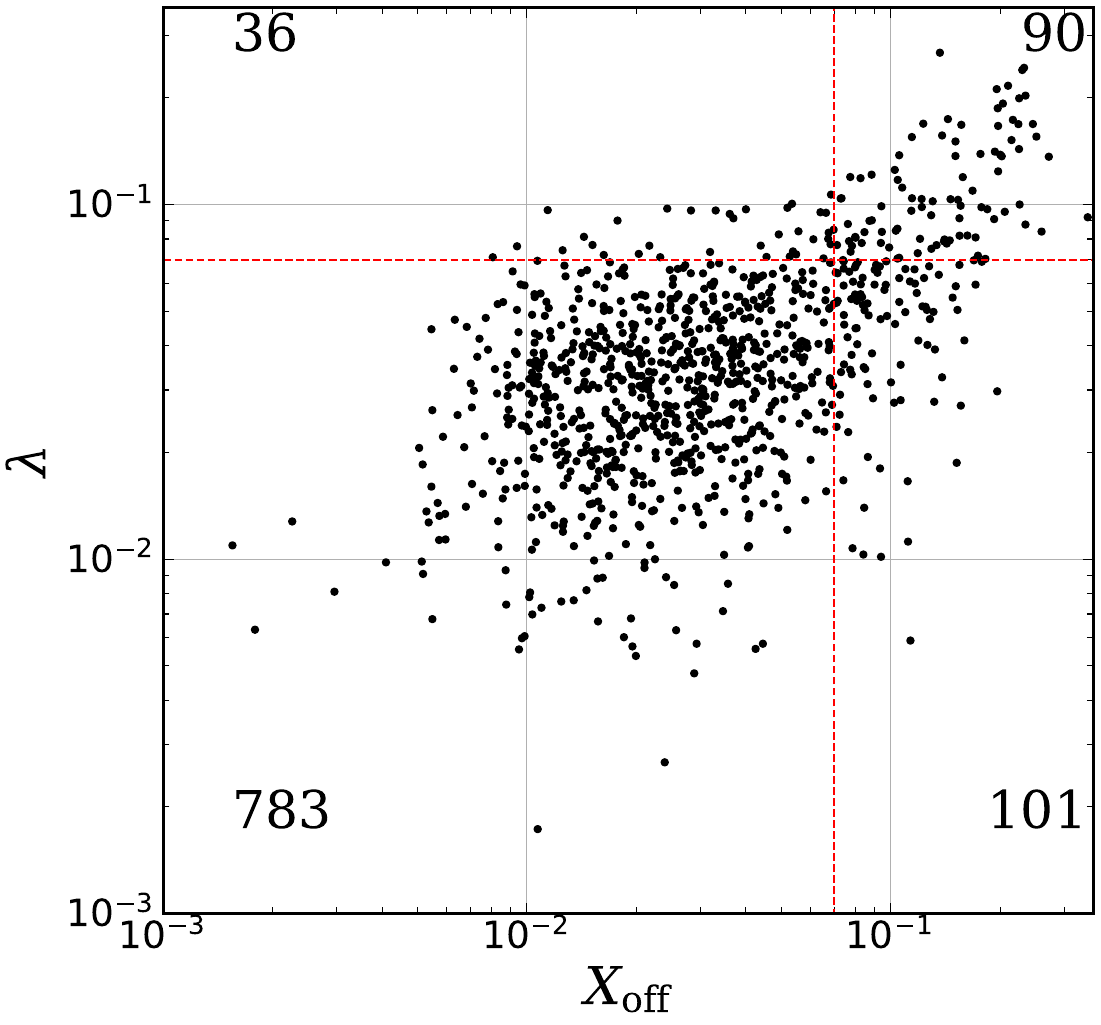}
    \caption{Distribution of the offset parameter $X_\mathrm{off}$ and spin parameter $\lambda$ for 1010 halos (black dots). 
    The red dotted lines represent the criteria from equation~\ref{eq:equilibrium1}. 
    The numbers near the four corners indicate the number of haloes contained in each of the areas demarcated by the red dashed lines.}
    \label{fig:xoffspinb}
\end{figure}
Fig.~\ref{fig:xoffspinb} illustrates the distribution of $X_\mathrm{off}$ and $\lambda$, and the outcomes of applying the criteria in equation~\eqref{eq:equilibrium1}. The distribution closely resembles that shown in Fig.~4 of \citet{2016MNRAS.457.4340K}.

\begin{figure}
    \centering
    \includegraphics[width=\columnwidth]{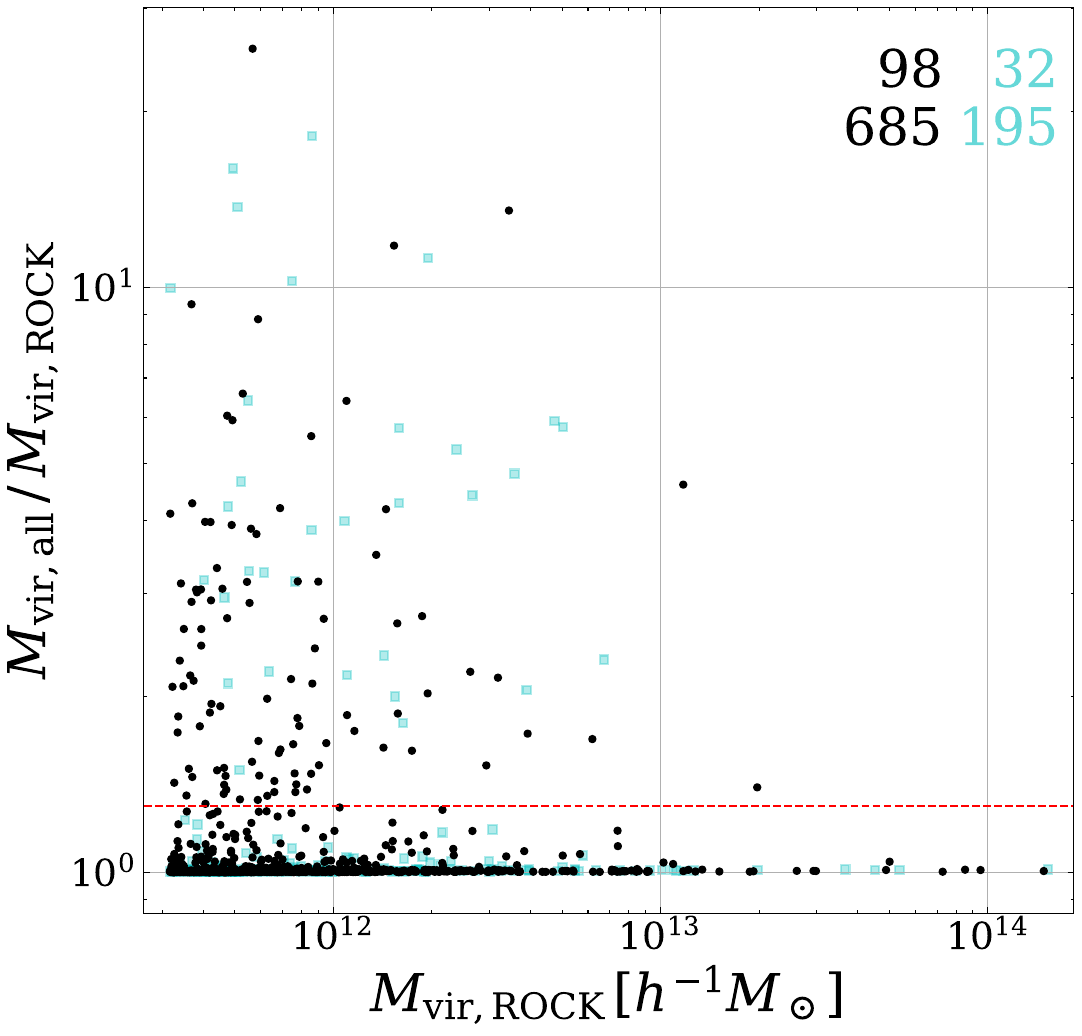}
    \caption{Distribution of $M_\mathrm{vir,all}/M_\mathrm{vir,ROCK}$ for the 1010 haloes in the original \textsc{Rockstar} catalogue. 
    The black dots denote 783 haloes after removing those experiencing major mergers (equation~\ref{eq:equilibrium1}), while the cyan boxes denote those removed. The subhalo condition (equation~\ref{eq:subhaloes}) is shown by the horizontal red dashed line. We consider only those below this line.
    The four numbers in the figure legend represent the number of haloes: for the criterion equation~\eqref{eq:equilibrium1} (left: retained, right: removed), and for equation~\eqref{eq:subhaloes} (upper: removed, lower: retained).}
    \label{fig:malltomrock}
\end{figure}

In addition to the haloes undergoing major mergers, we also exclude subhaloes. 
Subhaloes are typically surrounded by particles that belong to a distinct nearby host halo. 
This complicates the verification of whether the DM particles are bound by subhaloes or not, and it makes it challenging to accurately determine their density profiles. 
Thus, we exclude subhaloes from our halo catalog through the following procedure.
First, we calculate the mass, denoted as $M_{\mathrm{vir,all}}$, which represents the total mass of particles within the virial radius $R_{\mathrm{vir}}$ as calculated by the \textsc{Rockstar} halo finder ($R_{\mathrm{vir,ROCK}}$). The \textsc{Rockstar} halo finder computes $R_{\mathrm{vir,ROCK}}$ and the virial mass $M_{\mathrm{vir,ROCK}}$ after removing unbound particles, ensuring that $M_{\mathrm{vir,all}}$ is always greater than $M_{\mathrm{vir,ROCK}}$. 
For host haloes, most of the particles within $R_{\mathrm{vir,ROCK}}$ are bound to the host halo, so $M_{\mathrm{vir,ROCK}} \approx M_{{\mathrm{vir,all}}}$. In contrast, for subhaloes, many particles around them are bound to the host haloes and not to the subhaloes themselves, resulting in $M_{{\mathrm{vir,all}}} \gg M_{\mathrm{vir,ROCK}}$.
Here, there is no theoretical value that strictly distinguishes between host haloes and subhaloes.
To address this, we identify haloes that satisfies the condition:
\begin{equation}\label{eq:subhaloes}
    M_{\mathrm{vir,all}} > 1.3 M_{\mathrm{vir,ROCK}}
\end{equation}
as subhaloes, and remove them from our halo catalogue. Fig.~\ref{fig:malltomrock} displays the distribution of $M_{\mathrm{vir,all}}/M_{\mathrm{vir,ROCK}}$ and the outcomes of applying equation~\eqref{eq:subhaloes}. 
We retain those plotted by the black dots below the horizontal dashed line after the two conditions (685 haloes).
We can see that the two conditions are almost independent: the ratio of the haloes removed by the second condition does not depend on the first condition. 

In addition to the criteria discussed above, we introduce two additional criteria to exclude two more haloes that are considered to be undergoing major mergers, as discussed in Section~\ref{subsec:halo_centre}. Table~\ref{tab:halo_catalog} provides a summary of the resulting halo catalog.
We also present the mass ranges from S to XL that we will use for stacking analysis after Section~\ref{sec:results}.

In summary, even after excluding subhaloes and the haloes that are undergoing major mergers, nearly $70\%$ of all haloes remain in our catalog, with equations~\eqref{eq:equilibrium1} and \eqref{eq:subhaloes} serving as the primary criteria for constructing our halo catalog.

\begin{table*}
\centering
\caption{The number of haloes meeting selection criteria in each mass range. 
The upper three rows describe the characteristics of our halo samples divided into four mass bins.
The first and second rows respectively show the ranges of halo mass and radius, which are calculated by the \textsc{Rockstar} halo finder. 
On the other hand, the third row shows the splashback radius of the stacked density profiles, defined by the radius at which the density slope takes a minimum value. The estimated values shown here are those normalized by the mean virial radius for each mass range. The rest of the lower rows summarize the number of total haloes in each mass range, denoted by $N_{\rm all}$, as well as the number of samples after setting the criteria shown in the left column. The number of haloes shown at each row is the result when further adding the criterion to the upper rows. 
}
	\label{tab:halo_catalog}
	\begin{tabular}{lccccc}
		Mass range & S & M &L &XL & Total\\
  \hline
        $M_\mathrm{vir}[10^{11}h^{-1}M_\odot]$& $[3.16,5.71]$ & $[5.71,24.2]$ & $[24.2,134]$ & $[134,1530]$ & $[3.16,1530]$ \\
        $R_\mathrm{vir}[h^{-1} \mathrm{Mpc}]$& $[0.14,0.17]$ & $[0.17,0.27]$ & $[0.27,0.48]$ & $[0.48,1.08]$ & $[0.14,1.08]$ \\
        $R_\mathrm{sp}[R_\mathrm{vir}]$ & 0.64
        & 0.70 & 0.65&0.96 & \\
        
  \hline \hline
		$N_\mathrm{all}$ & 445 & 433 & 113 & 19 & 1010\\
		\hline
		$+\mathrm{Eq.~\eqref{eq:equilibrium1}}$ & 350 & 342 & 77 & 14& 783\\
		$+\mathrm{Eq.~\eqref{eq:subhaloes}}$ & 301 & 301 & 70 & 13 & 685\\
		$+\mathrm{Eq.~\eqref{eq:residual_mass}}$ & 300 & 301 & 70 & 13& 684\\
		$+\mathrm{Eq.~\eqref{eq:fract_cendif}}$ & 300 & 300 & 70 & 13& 683\\
		
	\end{tabular}
\end{table*}

\section{Tracking haloes and particles}\label{sec:methods}
To analyse each stream within the multi-stream region, we place our focus on the number of apocentre passages, denoted as $p$, for particles and categorize them according to their respective $p$ values.
In principle, the apocentre of a particle orbiting around the halo centre is defined as the point at which its radial velocity changes from positive to negative along its trajectory.
Therefore, once we have established the position and bulk motion of the halo, represented as $\boldsymbol{x}_\mathrm{h}$ and $\boldsymbol{v}_\mathrm{h}$, at each redshift, and determine the starting time for counting $p$, denoted as $t_s$, the process of counting $p$ for particles within the halo becomes straightforward: (1) calculate the radial velocity $v_r$ of each particle relative to the centre of the halo at every snapshot after $t_s$, (2) sum up the instances where the radial velocity changes from positive to negative until $z=0$. 

In the following, we introduce the method for calculating $\boldsymbol{x}_\mathrm{h}$, $\boldsymbol{v}_\mathrm{h}$ and $t_s$ of haloes in Section~\ref{subsec:halo_centre}. The procedure outlined in steps (1) and (2) above is presented in Section~\ref{subsec:counting}.

\subsection{Halo centre and merger trees}\label{subsec:halo_centre}
For a particle with position $\boldsymbol{r}$ and peculiar velocity $\boldsymbol{v}$, the radial velocity $v_r$ is defined as: 
\begin{equation}
    v_r = \frac{(\boldsymbol{x}-\boldsymbol{x}_\mathrm{h})\cdot (\boldsymbol{v}-\boldsymbol{v}_\mathrm{h})}{|\boldsymbol{x}-\boldsymbol{x}_\mathrm{h}|}.
\end{equation}
We monitor the sign of this quantity over different snapshots to determine $p$.
Therefore, to calculate $p$, we must determine the values of $\boldsymbol{x}_\mathrm{h}$ and $\boldsymbol{v}_\mathrm{h}$ for each halo in our catalog at each snapshot. 
While these values can be found by applying the halo finder to the snapshots at higher redshifts, they change discontinuously over time because different simulation particles are used to determine them in each snapshot. This leads to discontinuous variations in $v_r$ between snapshots, making it challenging to accurately count $p$. 

To overcome this challenge,  we calculate $\boldsymbol{x}_\mathrm{h}$ and $\boldsymbol{v}_\mathrm{h}$ as the average position and velocity of a fixed list of particles over time containing 1000 particles that can be considered the oldest progenitors. These progenitors are identified by tracking the massive branch of merger trees for each halo. 
By averaging the positions and velocities of these 1000 particles, we can calculate continuously varying $\boldsymbol{x}_\mathrm{h}$ and $\boldsymbol{v}_\mathrm{h}$ between snapshots while reducing the impact of individual particle noise.

The particles in the oldest progenitor are gravitationally well-bound within the halo at $z=0$ and move in tandem with it. Thus, these particles are suitable for representing the position and bulk motion of the halo centre. 
Here, the specific number 1000 is chosen to ensure numerical noise is minimised and can be adjusted as long as it provides a sufficient particle count.
In Appendix~\ref{app:1} we vary this number and confirm that the results for $p\leq 40$ remain consistent quantitatively.
Therefore, for the subsequent analysis, we focus solely on the particles with $p\leq 40$.

To construct merger trees, we initiate the process by determining the particles that compose each halo at $z=0$. This is achieved through a series of steps. First, we identify the centre of the halo using the shrinking sphere method. We then expand this sphere around the fixed centre until the overdensity within it decreases to the virial overdensity $\Delta_\mathrm{vir}$, as defined in \cite{1998ApJ...495...80B}. The particles found within this last sphere are designated as members of the halo.  
In the shrinking-sphere method, we initially set the centre of the halo, as calculated by the \textsc{Rockstar} finder, as the starting point for the sphere. We then systematically reduce the radius of the sphere in a logarithmically equally manner through 120 bins, ranging from $R_\mathrm{vir,ROCK}$ down to the radius closest to the first centre (typically corresponds to $\mathcal{O}(10^{-3}) \times  R_\mathrm{vir,ROCK}$). 
The shrinking process concludes when the number of particles contained within the sphere falls below 100. We denote the centre-of-mass position of these 100 particles as $\boldsymbol{x}_\mathrm{h,ss}$. Note that the shrinking sphere method converges towards the primary peak of the halo. Consequently, this approach may not be suitable for haloes classified as subhaloes or undergoing major mergers, which exhibit another peak within $R_\mathrm{vir}$. This is another reason why we focus on relaxed and host haloes. 

\begin{figure}
	\includegraphics[width=\columnwidth]{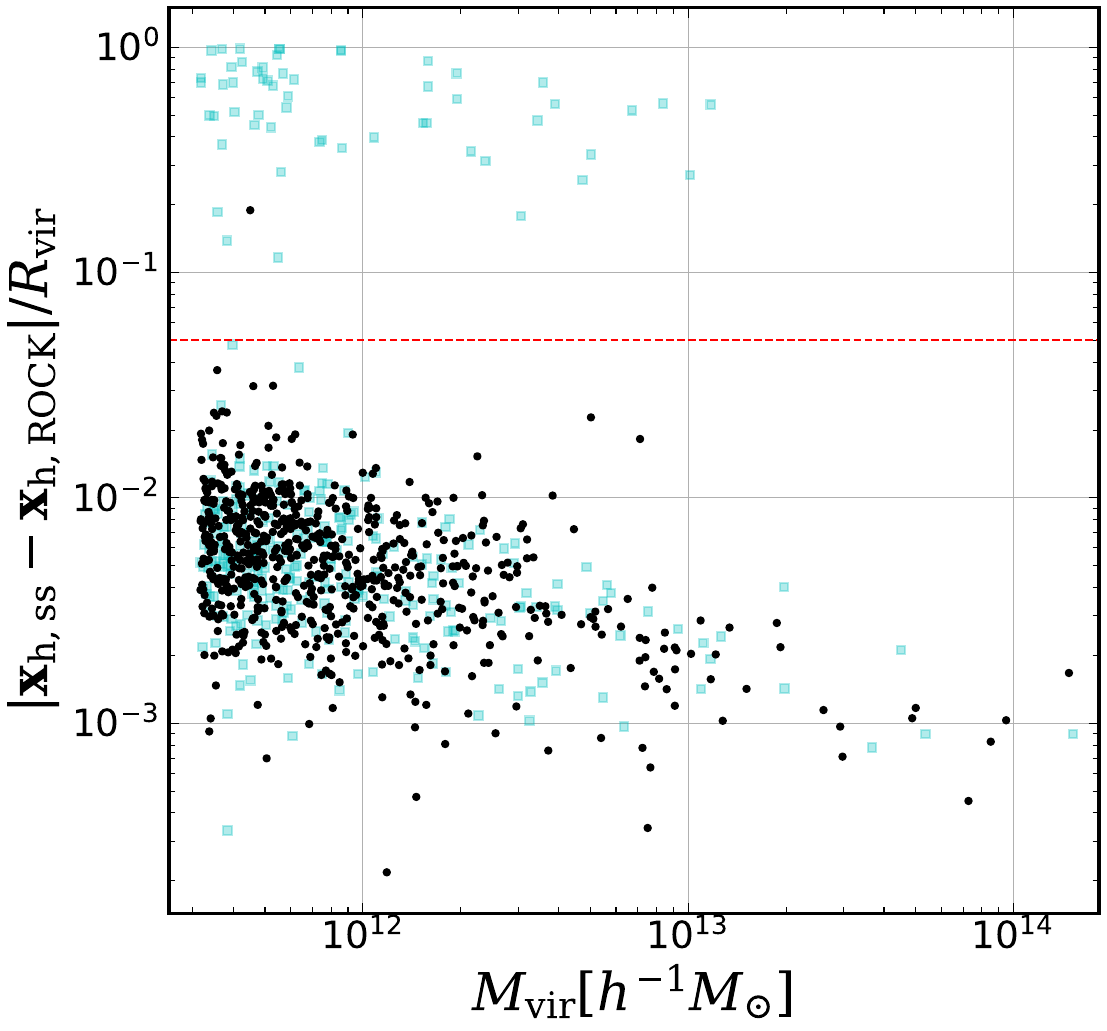}
    \caption{Difference between the centres of haloes at $z=0$, as calculated by \textsc{Rockstar} ($\boldsymbol{x}_\mathrm{h,ROCK}$) and using the shrinking-sphere method ($\boldsymbol{x}_\mathrm{h,ss}$). In this figure, $R_\mathrm{vir}$ in the denominator of the vertical axis and $M_\mathrm{vir}$ in the horizontal axis are determined by \textsc{Rockstar} ($R_\mathrm{vir,ROCK}$ and $M_\mathrm{vir,ROCK}$, respectively). Cyan boxes represent haloes excluded by the criteria in equations~\eqref{eq:equilibrium1} or \eqref{eq:subhaloes}, while black dots denote the haloes that are retained.
    We can observe that the haloes show a bimodal distribution in this plane: those appearing in the upper part of the figure are mostly discarded already by the previous two criteria. We decide to exclude one more halo with a large mismatch in the two definitions of the centre appearing in the upper part (see text for detail). The horizontal red dashed line shows a mismatch of $0.05\,R_\mathrm{vir,ROCK}$  (equation~\ref{eq:residual_mass}) introduced to remove this halo. 
    }
    \label{fig:residual_mass}
\end{figure}

The results are shown in Fig.~\ref{fig:residual_mass}. 
Most of the haloes that passed the criteria of equations~\eqref{eq:equilibrium1} and \eqref{eq:subhaloes} exhibit close agreement between $\boldsymbol{x}_\mathrm{h,ss}$ and $\boldsymbol{x}_\mathrm{h,ROCK}$ to within $5\%$ of $R_\mathrm{vir,ROCK}$. 
However, there is one halo, despite satisfying the two previous criteria, with a mass of $\sim 5\times 10^{11} M_\odot$ and a relatively large positional difference ($\sim 20 \%$ of $R_\mathrm{vir}$). 
In Appendix~\ref{excep1}, we carefully investigate this halo and find that it possesses a secondary peak, with both the peaks meeting the previously defined criteria. In general, haloes exhibiting significant deviations in $|\boldsymbol{x}_\mathrm{h,ss}-\boldsymbol{x}_\mathrm{h,ROCK}|$ can be considered as subhaloes or haloes undergoing major mergers. 
Therefore, we introduce an additional criterion based on Fig.~\ref{fig:residual_mass} to exclude haloes that meet the following condition:
\begin{equation}\label{eq:residual_mass}
    |\boldsymbol{x}_\mathrm{h,ss}-\boldsymbol{x}_\mathrm{h,ROCK}|/R_\mathrm{vir,ROCK} >0.05.
\end{equation}

Next, we select the member particles of haloes in a snapshot one step before $z=0$ from those retained at $z=0$, employing the same method. This involves determining the centre of the halo using the shrinking-sphere method and expanding the sphere until the overdensity within it decreases to $\Delta_\mathrm{vir}$. These steps are repeated, tracking back to $z=5$, until the number of particles constituting the halo is below 1000. When the number goes below 1000, we slightly expand the sphere to ensure that it encompasses exactly 1000 particles, and we define the starting time of the counting of $p$, $t_s$, by the cosmic time of this snapshot. Note that when we move to the next snapshot (i.e., the one preceding it in cosmic time) in this procedure, we exclusively take into account the particles that are retained in the previous snapshot for the shrinking-sphere process. This guarantees that the list of member particles can only diminish as we move towards higher redshifts. Consequently, we define the particles within the oldest progenitor as consistently located near the halo centre from $t_s$ until the present. 

As a result of this procedure, we can define the virial mass $M_\mathrm{vir}(z)$ and the virial radius $R_\mathrm{vir}(z)$ at each redshift as the mass contained within the sphere with an overdensity of $\Delta_\mathrm{vir}$ and its corresponding radius.
These values are utilised in the calculation of the accretion rate in Section~\ref{subsec:dep_on_MAR}.
For clarity, we distinguish these values from those computed by \textsc{Rockstar} at $z=0$ by denoting the latter simply as $R_\mathrm{vir}$ and $M_\mathrm{vir}$ (i.e., without the argument $z$) in what follows.
\begin{figure}\label{fig:lambda_and_xoff}
    \centering
    \includegraphics[width=\columnwidth]{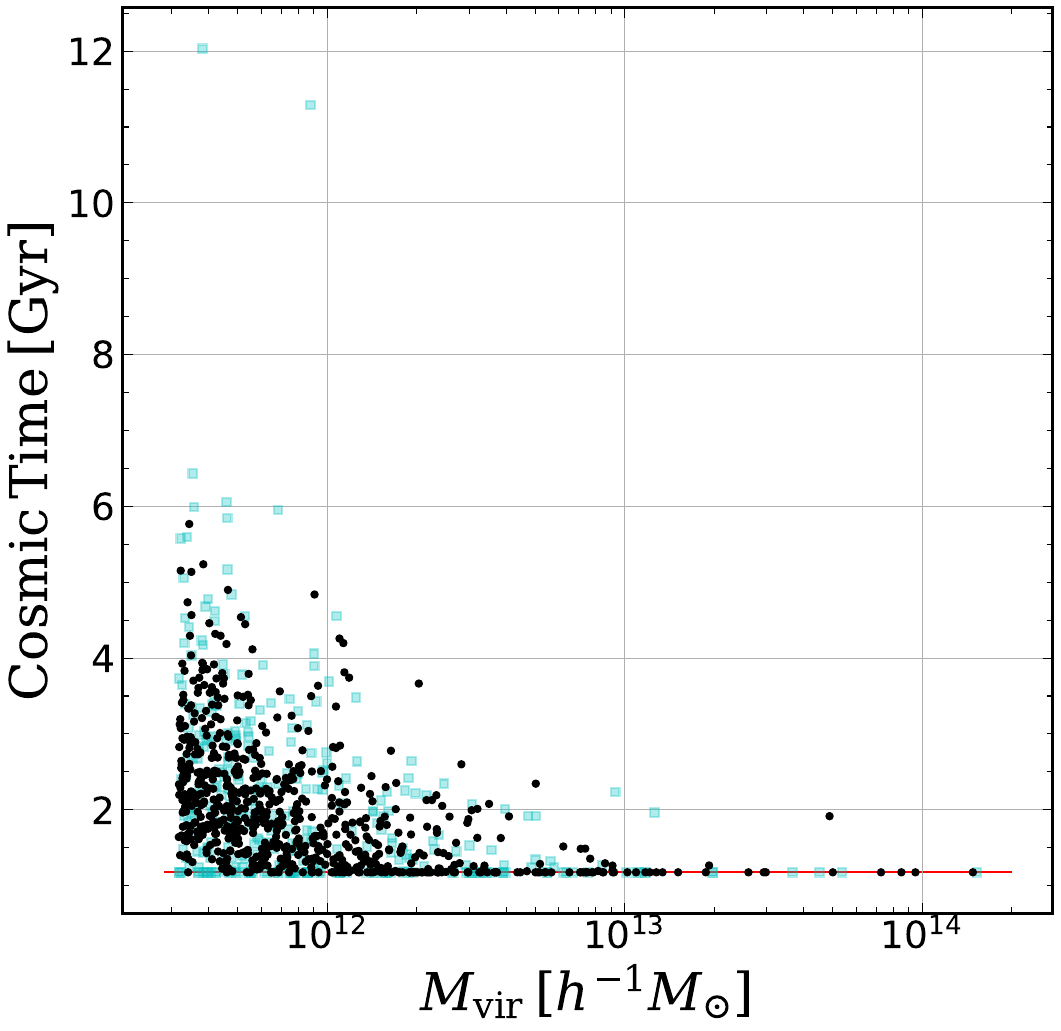}
    \caption{The starting time of the counting of $p$, denoted as $t_s$, of each halo. Cyan and black symbols represent the same halos as in Fig.~\ref{fig:residual_mass}. The horizontal red solid line shows $z=5$. 
    Out of 684 haloes depicted as black dots, 665 have $t_s$ values below $4\mathrm{Gyr}$. 
    There are 97 black dots on the red line, indicating that they are tracked until $z=5$. Given that massive haloes tend to have correspondingly massive progenitors, most of their progenitors still have virial masses larger than the sum of 1000 particles $(\sim 4.7\times 10^{10} M_\odot)$ even at $z=5$.}
    \label{fig:mah_breaktime}
\end{figure}

Fig.~\ref{fig:mah_breaktime} illustrates the distribution of $t_s$, representing the cosmic time when the particle count in the progenitor of each halo falls below 1000. The majority of haloes exhibit $t_s$ values below $4\mathrm{Gyr}$, allowing us to trace the particles for over $\sim 10\mathrm{Gyr}$. However, note that in the case of 97 out of the 684 haloes in our catalogue, the tracking process extends all the way back to $z=5$, which is our first snapshot, without dropping below 1000 particles. While we can still define the centres of these 97 haloes using all the particles from their progenitor at $z=5$, the varying number of particles used to determine their centres may introduce systematic effects. 

To address this issue, we choose to select 1000 particles from the progenitors at $z=5$, with a particular focus on their phase space. Initially, we calculate the position and velocity dispersions, denoted as $\sigma_x$ and $\sigma_v$, respectively, for the particles in the progenitor at $z=5$. Subsequently, we define the phase space distance, $d_\mathrm{ps}$, for each particle with respect to the average position $\boldsymbol{x}_\mathrm{ave}$ and velocity $\boldsymbol{v}_\mathrm{ave}$ for each particle using the following equation:
\begin{equation}\label{eq:dps}
    d_\mathrm{ps}(\boldsymbol{x},\boldsymbol{v}) = \left(\frac{|\boldsymbol{x}-\boldsymbol{x}_\mathrm{ave}|^2}{\sigma_x^2} +  \frac{|\boldsymbol{v}-\boldsymbol{v}_\mathrm{ave}|^2}{\sigma_v^2} \right)^{\frac{1}{2}},
\end{equation}
where $\boldsymbol{x}$ and $\boldsymbol{v}$ represent the positions and velocities of the particles, respectively. Finally, we select 1000 particles with the smallest values of $d_\mathrm{ps}$. This selection process ensures that we are focusing on particles that are well bound within the halo progenitor.

\begin{figure}
    \centering
    \includegraphics[width=\columnwidth]{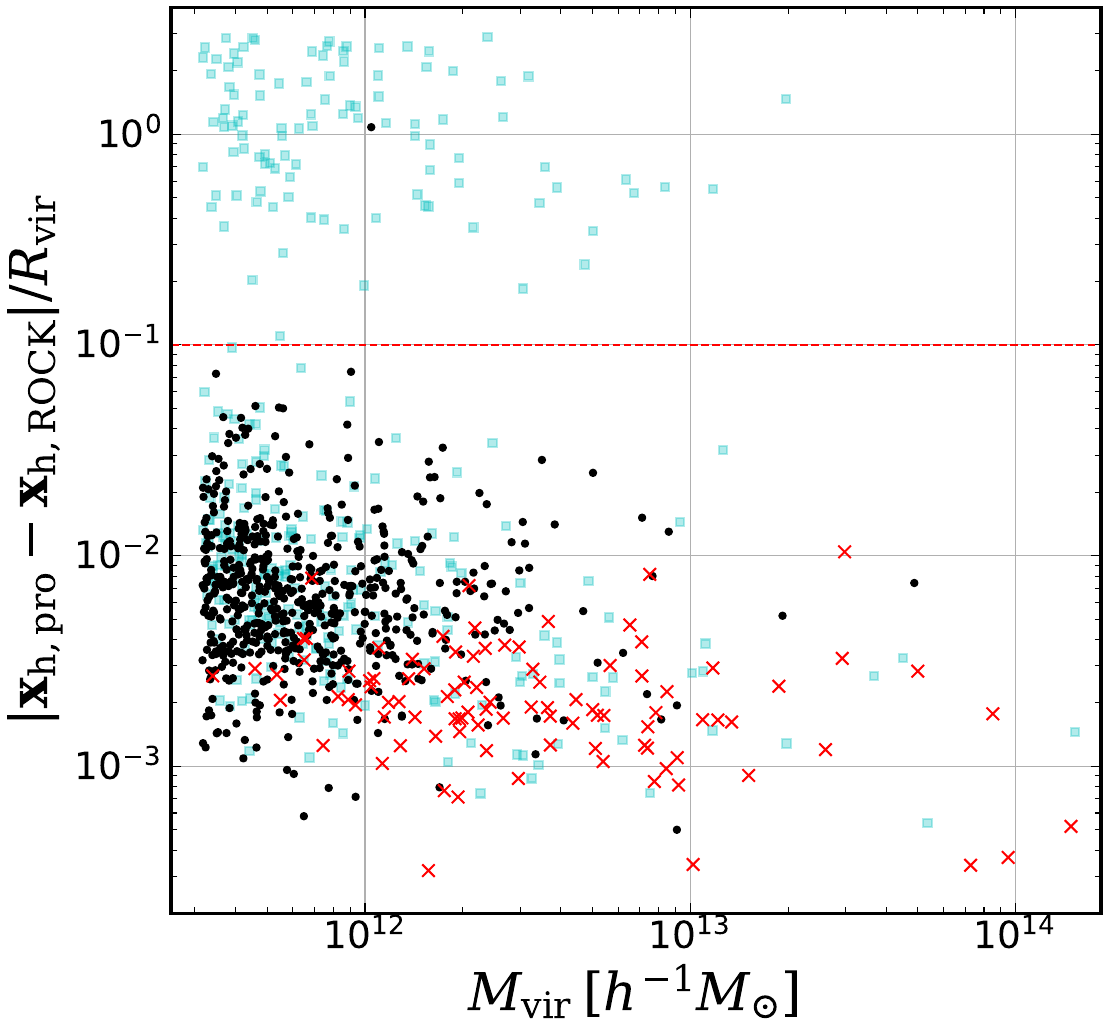}
    \caption{Difference between the centre of the halo at $z=0$ calculated by \textsc{Rockstar} ($\boldsymbol{x}_\mathrm{h,ROCK}$) and the average position over 1000 particles in the oldest progenitor ($\boldsymbol{x}_\mathrm{h,pro}$). 
    The cyan boxes and black dots are for the same haloes as in Fig.~\ref{fig:mah_breaktime}. 
    The red crosses represent 97 haloes located on the red line in Fig.~\ref{fig:mah_breaktime}. Their $\boldsymbol{x}_\mathrm{h,pro}$ is calculated by averaging over 1000 particles with the smallest phase-space distance, $d_\mathrm{ps}$ (defined in equation~\eqref{eq:dps}) at $z=5$. 
    The red dashed line indicates the criterion from equation~\eqref{eq:fract_cendif}.}
    \label{fig:fract_cendif}
\end{figure}

Fig.~\ref{fig:fract_cendif} illustrates the difference between the positions of halo centres at $z=0$, calculated by averaging over the positions of 1000 progenitor particles ($\boldsymbol{x}_\mathrm{h,pro}$), and those determined by the \textsc{Rockstar} halo finder ($\boldsymbol{x}_\mathrm{h,ROCK}$). 
Apart from one halo with a mass of $\sim 10^{12} M_\odot$, which exhibits a substantial difference ($\sim R_\mathrm{vir}$), we observe that all 683 $\boldsymbol{x}_\mathrm{h,pro}$ align with $\boldsymbol{x}_\mathrm{h,ROCK}$ within an accuracy of less than $10 \%$ of $R_\mathrm{vir}$, with 564 $\boldsymbol{x}_\mathrm{h,pro}$ falling below $1 \%$. This demonstrates that we can faithfully track the motion of the progenitor centres from $t_s$ to the present using the fixed list of 1000 particles. These particles are the oldest population that occupies the central region of a halo throughout its evolution. We utilise these particles to define the position and bulk motion of the centres, denoted as $\boldsymbol{x}_\mathrm{h,pro}$ and $\boldsymbol{v}_\mathrm{h,pro}$, respectively, at intermediate redshifts.

In Appendix~\ref{excep2}, we show that the exceptional halo we mentioned above possesses a significant secondery density peak that hinders us from accurately tracing the primary peak. This halo can be considered to be currently undergoing a major merger. Therefore, we introduce a final criterion below and remove haloes that satisfy it:
\begin{equation}\label{eq:fract_cendif}
    |\boldsymbol{x}_\mathrm{h,pro}-\boldsymbol{x}_\mathrm{h,ROCK}|/R_\mathrm{vir,ROCK} >0.1.
\end{equation}

In summary, we constructed merger trees for each halo and selected 1000 particles in their progenitor; then we defined $\boldsymbol{x}_\mathrm{h}$ and $\boldsymbol{v}_\mathrm{h}$ as the average position and velocity of these 1000 particles at each snapshot. 
As demonstrated in Fig.~\ref{fig:fract_cendif}, $\boldsymbol{x}_\mathrm{h}$ matches those calculated by \textsc{Rockstar} at $z=0$.

\subsection{Counting particles' apocentre passages}\label{subsec:counting}
We can now calculate the radial velocity, $v_r$, of each particle at each of the 1001 snapshots using $\boldsymbol{x}_\mathrm{h}$, $\boldsymbol{v}_\mathrm{h}$ defined above, and count the number of apocentre passages from the change in sign of $v_r$. 
However, before proceeding with the actual counting of $p$, we need to account for particles residing in subhaloes. Our treatment of this issue is based on the method used in S20.

Generally, particles consisting a subhalo follow orbits relative to the its centre, and the sign of their radial velocities relative to the host halo can sometimes transition from negative to positive (and vice versa) due to this orbital motion within the subhalo. While we define $p$ as the number of apocentre passages relative to the host halo, this effect inadvertently increases the count of $p$. To mitigate this effect, 
we take into consideration the direction of each particle, denoted as $\hat{\boldsymbol{r}}\left(t\right)$, from the centre of the host halo as follows. 

Suppose that a particle undergoes an apocentre passage at time $t_1$ and experiences a change in the sign of its radial velocity from positive to negative. Later, at time $t_2$, the positive-to-negative sign transition occurs for the first time after $t_1$. If $\hat{\boldsymbol{r}}\left(t_1\right) \cdot \hat{\boldsymbol{r}}\left(t\right) >0$ always holds for $t_1<t\leq t_2$, it implies that the particle remains on the same side of the halo, not completing a full orbit before the next apocentre passage. In such cases, the sign change at $t=t_2$ is likely due to orbital motion within the subhalo. We can show that a particle within a static spherical potential undergoes an apocentre passage after moving at least $180^{\circ}$ around the centre of that potential since the previous apocentre passage (see Sec.3.1 of \citet{2008gady.book.....B}). Therefore, we count the number of apocentre passages only when $\hat{\boldsymbol{r}}\left(t_1\right) \cdot \hat{\boldsymbol{r}}\left(t\right) <0$ holds at least once for $t_1<t\leq t_2$. Note that this condition ensures that the particle has orbited at least $90^{\circ}$ from the previous apocentre passage. Nevertheless, we confirmed that this works in practice to remove fake apocentre passages due to the internal motion within a subhalo.
 
\begin{figure}
	\includegraphics[width=\columnwidth]{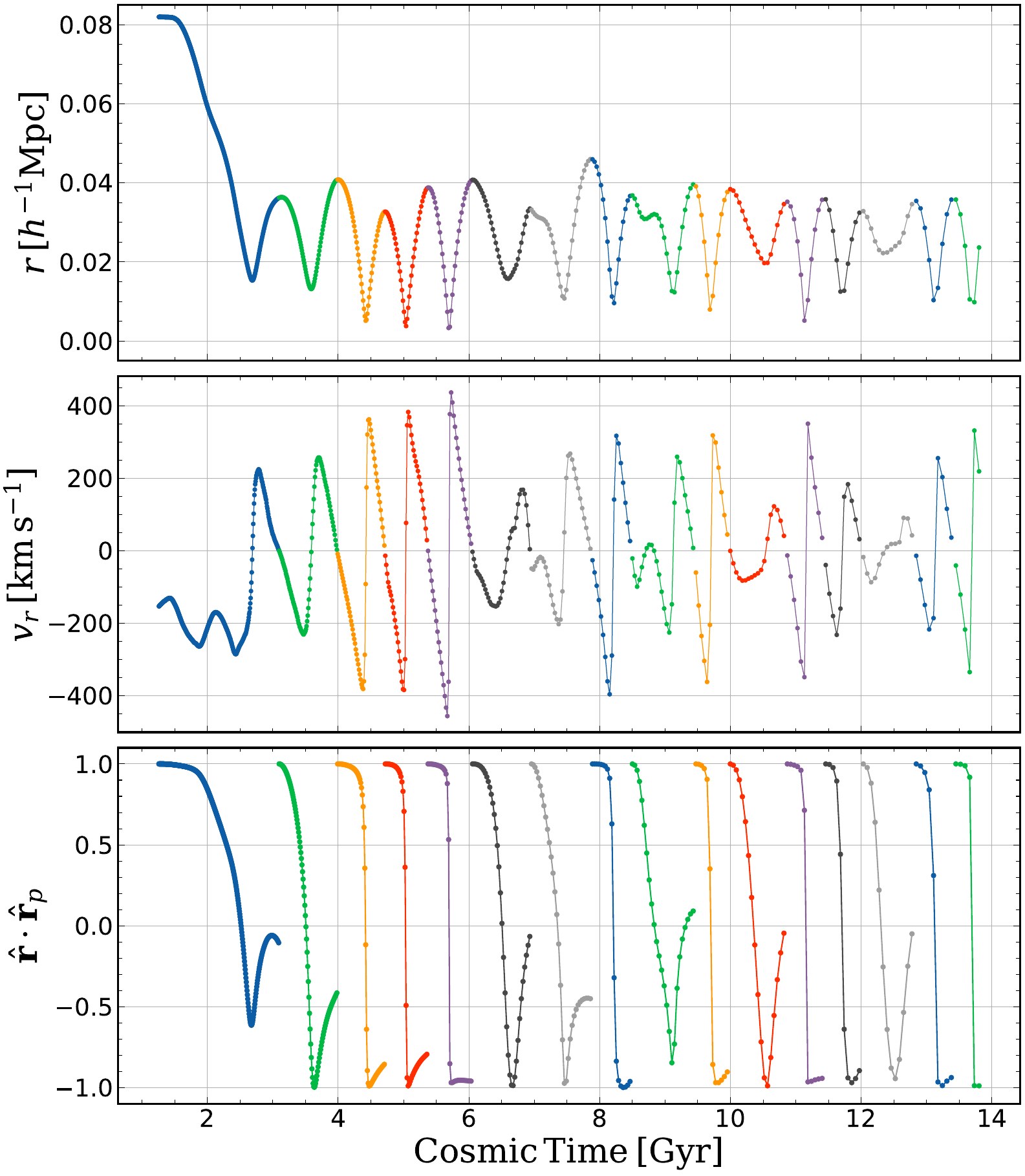}
    \caption{
    Example orbit of a particle around the centre of a progenitor halo.  
    The rows display, from top to bottom, the radial distance $r$, radial velocity $v_r$ and inner product, $\hat{\boldsymbol{r}}\cdot \hat{\boldsymbol{r}}_p$, where $\hat{\boldsymbol{r}}$ represents the particle's direction from the centre at a given time, and
    $\hat{\boldsymbol{r}}_p$ is the direction evaluated at the previous apocentre passage. Points are plotted in different colours corresponding to the value of $p$. We ensure that we only increment $p$ after the sign of $\hat{\boldsymbol{r}} \cdot \hat{\boldsymbol{r}}_p$ turns negative since the previous apocentre passage. Note that this particle has $p=15$ at the present time.}
    \label{fig:example1_figure}
\end{figure}

As an example, Fig.~\ref{fig:example1_figure} shows the counting of $p$ for an orbiting particle using the method described above. In this figure, the colour of the points changes as the particle passes through its orbital apocentre, demonstraining the effectiveness of the method. Around $t=9\,\mathrm{Gyr}$, we can see that the sign of $v_r$ changes from positive to negative. However, the value of $p$ is not incremented because the particle has not traveled much from the previous apocentre passage (i.e., $\hat{\boldsymbol{r}}\left(t_1\right) \cdot \hat{\boldsymbol{r}}\left(t\right)$ remains positive since $t_1$). Indeed, the time since the previous apocentre passage is short compared to the typical intervals between other apocentre passages, suggesting that the sign change is due to internal motion with the subhalo.
We can also see that the interval between the data points in the figure is sufficiently dense to count the apocentre passages. If the number of snapshots is insufficient, the algorithm cannot detect the apocentre passages, potentially leading to miscounts of $p$. To ensure accuracy, we conduct a verification in Appendix~\ref{app:1}, confirming that the $1001$ snapshots we used offer sufficient time resolution to avoid miscounting $p$ for $p\leq 40$.
Furthermore, in Appendix~\ref{app:2}, we examine a  convergence study and check if our simulation setup has enough mass resolution to fairly trace the multi-stream flow for a large number of apocenter passages, again confirming that the present mass resolution provides a reliable estimate of $p$ for $p\leq40$. 

\subsection{Density profiles and fitting procedure}\label{subsubsec:density_profiles}
Before presenting our main results, we provide a summary of some technical aspects related to calculating density profiles, stacking procedure, and fitting methodology. 
In Section~\ref{subsec:individual_profiles}, we showcase density profiles in physical coordinates. 
To construct these density profiles, we employ logarithmically spaced radial bins ranging from $2$ times the softening length to $2.5\,R_\mathrm{vir}$.
Then, in Section~\ref{subsec:dep_on_mass}, we utilise stacked density profiles, which involves averaging over individual profiles within specific subgroups. 
We define subgroups according to, for example, the number of particles with a certain value of $p$ and the mass.
In our stacking procedure, we first normalise the radial distances of particles within individual haloes by their respective values of $R_\mathrm{vir}$, and create individual density profiles. 
Subsequently, we construct stacked density profiles by averaging over these individual profiles in the subgroup. 
For the stacked profiles for each mass bin, we define 80 logarithmically spaced radial bins spanning from $0.0025\,R_\mathrm{vir}$ to $2.5\,R_\mathrm{vir}$. However, we exclude bins that fall below $1.2$ times the maximum value of the softening length, scaled by $R_\mathrm{vir}$, for halos within the subgroup to avoid numerical artifact due to the limited force resolution.
By comparing the LR and HR simulations, we confirm that the softening effect does not significantly impact the stacked profiles beyond this radius.
These specific radii will be explicitly referenced in Section~\ref{subsec:universal_stream_profiles}.

In Section~\ref{sec:results}, we will also present fitted curves for the density profiles.
Our fitting procedure rely on the standard minimisation method for $\chi^2$, which is defined as follows:
\begin{align}\label{eq:chi2}
    \chi^2 \equiv \sum_\mathrm{i} \left(\frac{\ln{\rho_\mathrm{i,fit}}-\ln{\rho_\mathrm{i,data}}}{\sigma_\mathrm{i}/\rho_\mathrm{i,data}} \right)^2.
\end{align}
Here, $\rho_\mathrm{i,fit}$ represents the fitted density in the $i$-th radial bin, while $\rho_\mathrm{i,data}$ denotes the individual or stacked density profile. 
$\sigma_\mathrm{i}$ is the uncertainty of $\rho_\mathrm{i,data}$, determined by the Poisson scatter for individual profiles and the root mean square deviation of the profiles stacked within subgroups.
For our analysis, we consider all radial bins above the aforementioned lowest radial bins for each case.

\section{Results}\label{sec:results}

In this section, we present the results of the halo density profiles for particles having different numbers of apocentre passages, $p$. We first pick up representative halos and show their individual profiles in Section~\ref{subsec:individual_profiles}. We then consider the stacked halo profiles and characterise them in detail with a double-power law profile in Section~\ref{subsec:dep_on_mass}. To elucidate their statistical properties, we in particular discuss their dependence on the halo mass.

\begin{figure*}
    \centering
    \includegraphics[width=2\columnwidth]{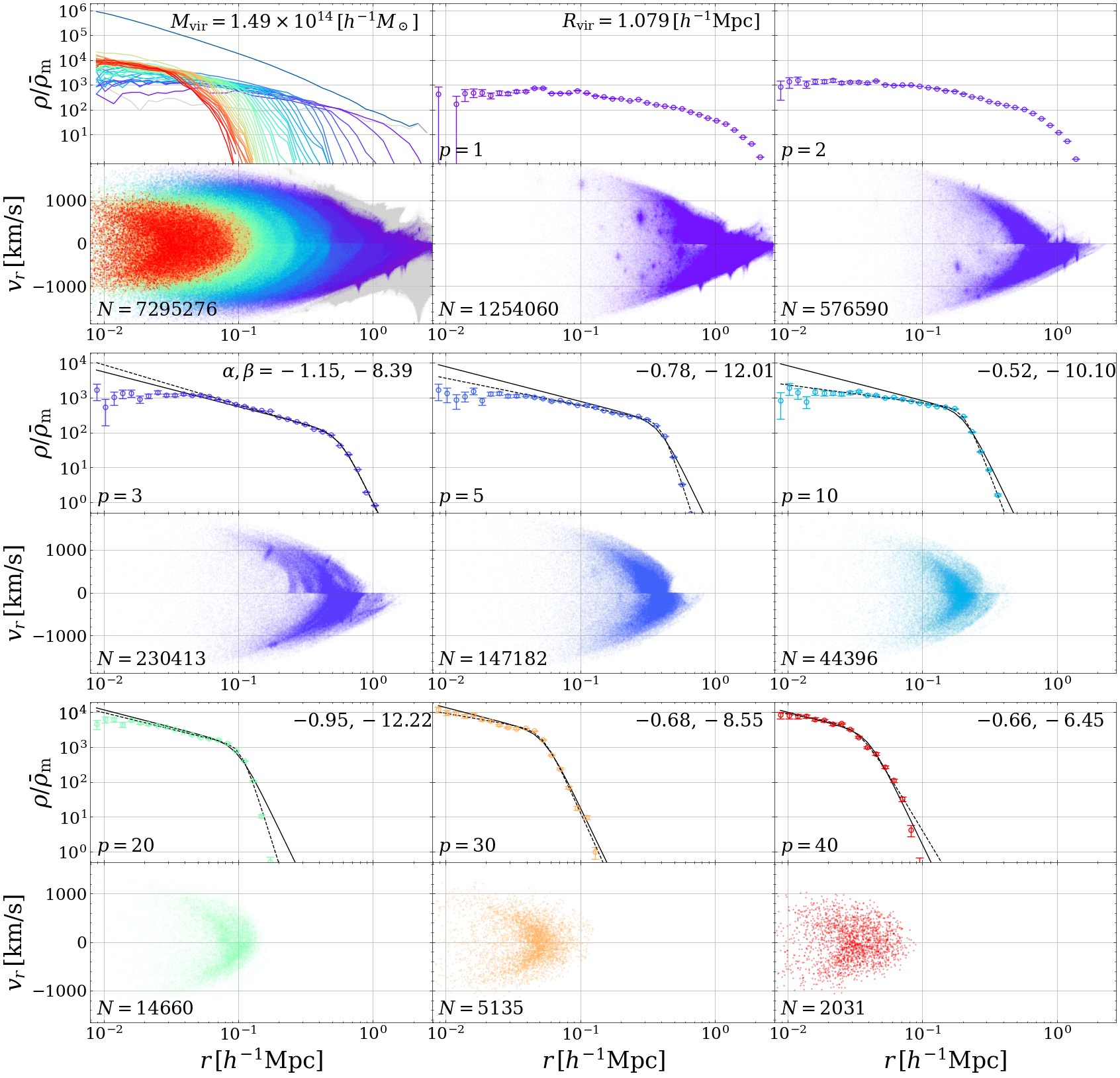}
    \caption{Radial density profiles and radial phase space distributions of particles for $p=1,2,3,5,10,20,30,40$.
    $N$ is the number of particles plotted in each phase space. 
    Black solid (dashed) lines show the fits to the density profiles with equation.~\eqref{eq:stream}  (equation.~\eqref{eq:double_power}), and the inner and outer slopes $\alpha,\beta$ for the dashed lines are also shown.
    In the left top panel, the density profiles and the phase-space distribution for total and $1\leq p\leq40$ (as the same colour coding of other panels), $p=0$ (light grey).
    As we mentioned in Section~\ref{subsubsec:density_profiles}, here we set the minimum radial bin to $2$ times the softening length for the density profiles.}
    \label{fig:radialprop1}
\end{figure*}
\begin{figure*}
    \centering
    \includegraphics[width=2\columnwidth]{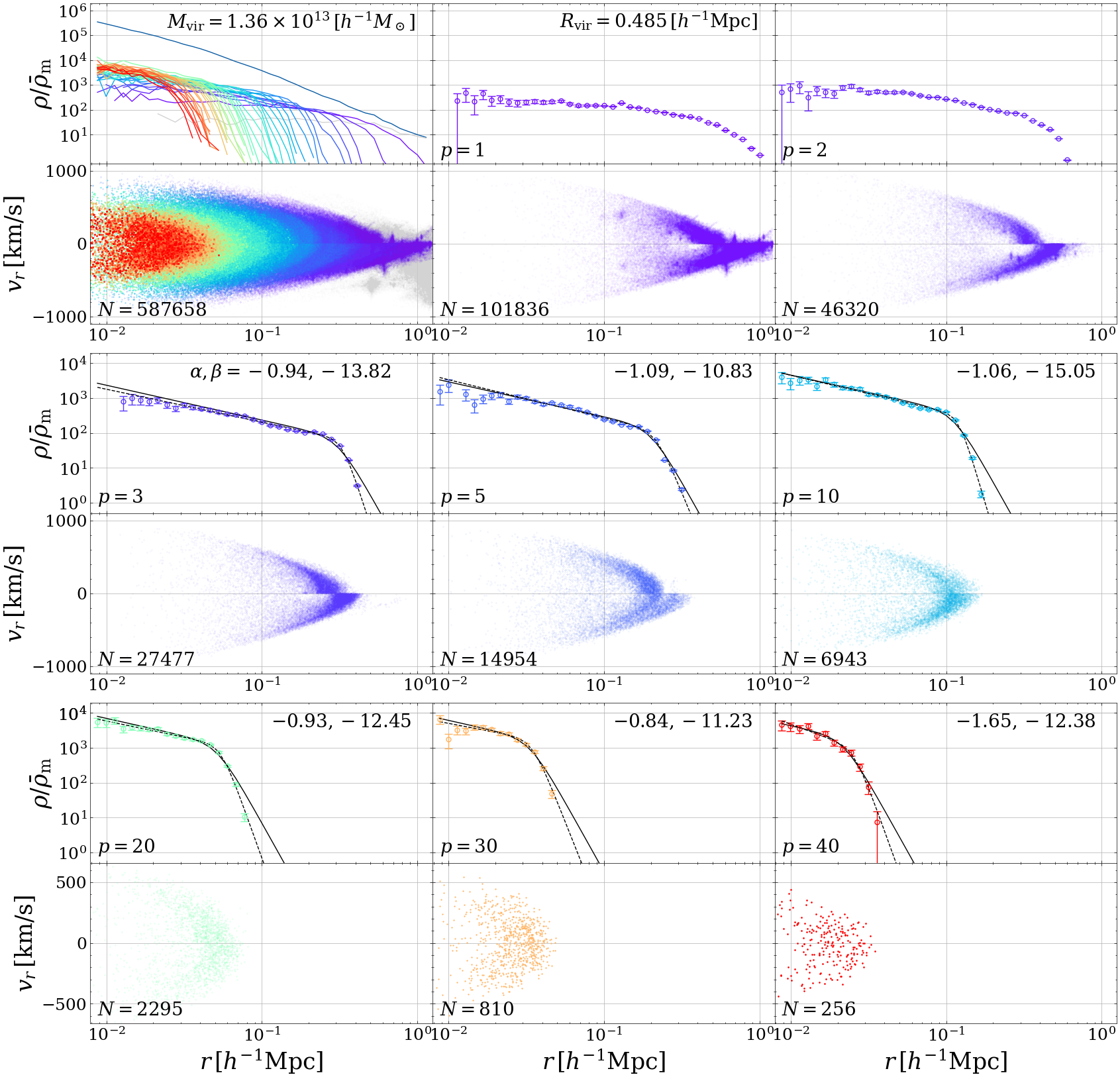}
    \caption{Same as Fig.~\ref{fig:radialprop1} but halo mass is $1.36\times10^{13}[h^{-1}M_\odot]$.}
    \label{fig:radialprop18}
\end{figure*}
\begin{figure*}
    \centering
    \includegraphics[width=2\columnwidth]{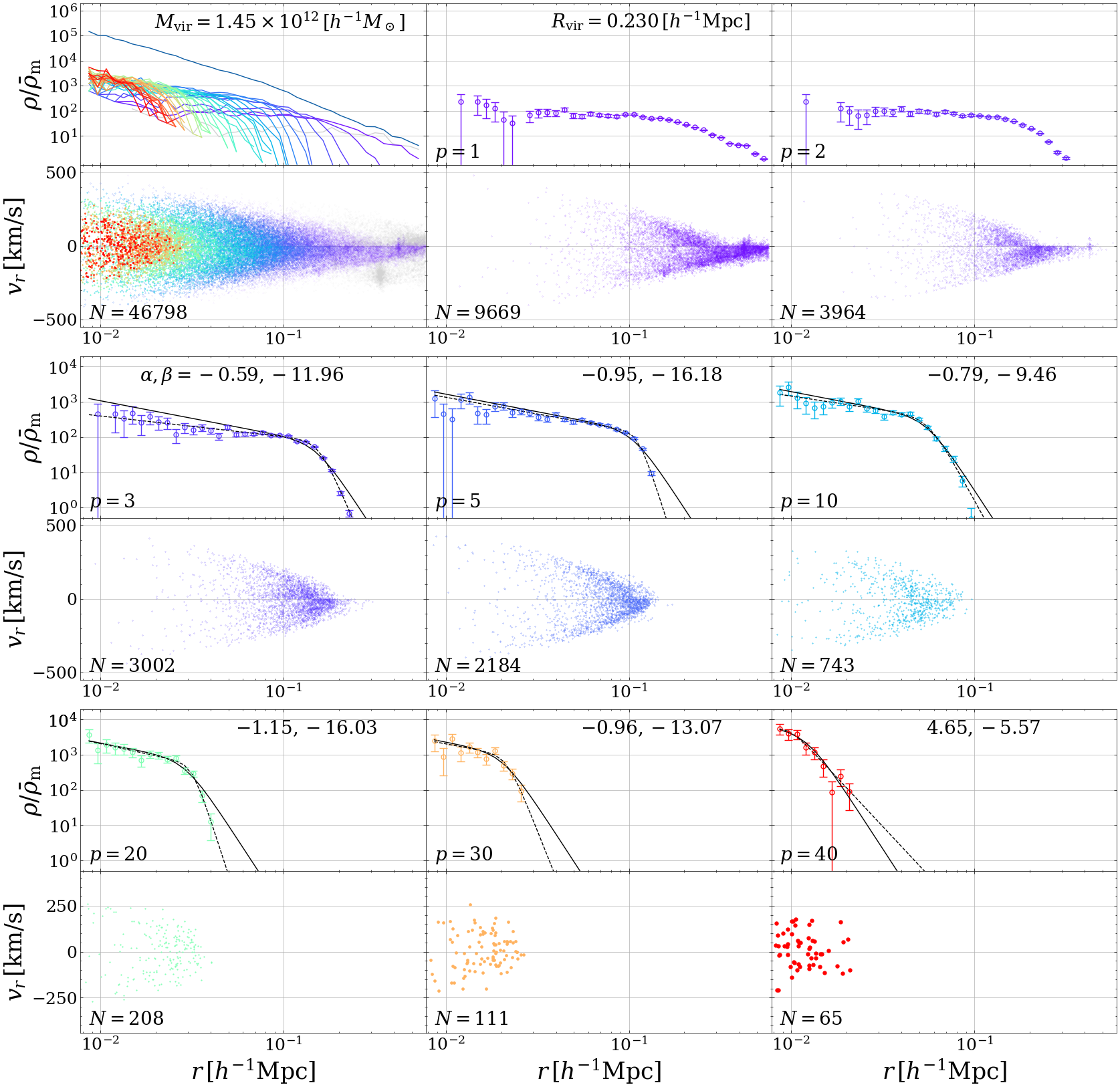}
    \caption{Same as Fig.~\ref{fig:radialprop1} but halo mass is $1.45\times10^{12}[h^{-1}M_\odot]$.}
    \label{fig:radialprop235}
\end{figure*}
\subsection{Individual halo profiles}\label{subsec:individual_profiles}

Applying the method and algorithm in Section~\ref{sec:methods} to 683 halos in our simulations, summarised in Table~\ref{tab:halo_catalog}, we are now able to characterise the radial phase-space structure by looking at the particle distributions classified with different numbers of apocentre passages. To do so, we first pick up three representative halos whose masses are $1.49\times10^{14}$, $1.36\times10^{13}$, and $1.45\times10^{12}$ $h^{-1}M_\odot$, and plot their radial phase-space and density profiles in  Figs.~\ref{fig:radialprop1}-\ref{fig:radialprop235}.
In these figures, we show density profiles (top row) and phase-space distributions (bottom row) for particles with $p=1$ to $40$ (top left panel) and for $p=1,2,3,5,10,20,30,40$ (other panels).

Despite a wide range of mass scales, overall trends in the density profiles and phase-space structures are mostly similar for the three cases. Looking in particular at the outer part of the radial phase-space distribution, we see a substantial number of clumps with various spatial and velocity extents at $p=1$ and $2$. 
Although unrelaxed haloes are removed in our samples, this implies that the outer parts of haloes are generally not truly relaxed, having a large fluctuation in phase-space density due to the remnant of subhaloes that have recently accreted or undergone mergers. Note that this scatter has been reported in \citet{Diemer2022} as the halo-to-halo scatter in characterizing the infalling profiles. 


On the other hand, focusing on the inner structures (i.e., large values of $p$), the phase-space distribution becomes rather smooth, and we see fewer substructures.
Furthermore, the density profile for a large value of $p$ generically exhibits double-power law features consisting of a shallow inner slope and a steep outer slope. 
At the transition scale where the slope suddenly changes, we see that particles in phase space are accumulated near zero velocity, associated with an apocentre passage for particles having the same $p$. In other words, this transition scale roughly corresponds to the radial caustic of each multi-stream flow. S20 reported that about $\sim30$\% of haloes have structures quantitatively similar to those predicted by the self-similar solution of \citet[]{1984ApJ...281....1F}. 
Although their self-similar solutions generally predict a spiky density structure at the caustics, such a spike is smeared in reality in the presence of non-zero tangential velocities and non-sphericity, leading to what is seen in Figs.~\ref{fig:radialprop1} to \ref{fig:radialprop235}, i.e., a smooth transition in slope. 

Here, we investigate in detail the structure of the density profile for each $p$, and try to describe it with the following functional form:
\begin{align}\label{eq:double_power}
    \rho(r;p) = \frac{A(p)}{ \bigl\{r/S(p)\bigr\}^{-\alpha(p)}\,\Bigl[ 1+\bigl\{r/S(p)\bigr\}^{\alpha(p)-\beta(p)}\Bigr] },
\end{align}
which behaves like $\rho\propto r^{\alpha(p)}$ at $r\ll S(p)$ and $\rho\propto r^{\beta(p)}$ at $r\gg S(p)$, under the assumption of $\alpha(p)<0$ and $\beta(p)<0$.  
Here, the four parameters, $A(p)$, $S(p)$, $\alpha(p)$ and $\beta(p)$, are determined by fitting equation~\eqref{eq:double_power} to the measured density profile for each $p$ following the method introduced in Section~\ref{subsubsec:density_profiles}.

The results are shown by black dashed lines in the upper panels of Figs.~\ref{fig:radialprop1}--\ref{fig:radialprop235}. Here, the fitting to equation~\eqref{eq:double_power} was performed for profiles with $p\geq3$, since the profiles for $p=1$ and $2$ clearly show non-double-power law features due partly to the contributions of substructure, as seen in the radial phase-space distribution. Figs.~\ref{fig:radialprop1}--\ref{fig:radialprop235} show that the fitting form of equation~\eqref{eq:double_power}
describes the measured profile fairly well, especially around the transition scales. 
In particular, a better fitting result is obtained as increasing $p$.
For a smaller value of $p$, a deviation from the double-power law function is found at the inner part of massive halos (see Figs.~\ref{fig:radialprop1} and \ref{fig:radialprop18}), and they tend to have a flat core. However, statistical variations seem large due to the lack of particles, and we cannot judge whether this deviation is significant or not on the basis of individual haloes. We will thus consider stacked halo profiles, and quantify statistically the goodness of fit to the double-power law function. 

\subsection{Stacked profiles for mass selected samples}\label{subsec:dep_on_mass}

In this subsection, using the selected $683$ halos, we analyse stacked density profiles
for four halo mass bins S -- XL summarised in Table \ref{tab:halo_catalog}.
Fitting them to the double-power law function, statistical properties of the radial profile for different $p$ are elucidated, particularly focusing on their mass dependence.

In Section \ref{subsec:alpha_beta_free}, fitting to the double-power law profile is examined for stacked profiles of each $p$, and we show that the fitting function with $\alpha=-1$ and $\beta=-8$ reproduces well the measured profiles for all mass bins. 
With these fixed slopes, we quantitatively investigate the dependence of the fitting parameters, i.e., characteristic density $A(p)$ and radius $S(p)$, on the number of apocentre passages $p$ and halo mass in Section~\ref{subsec:universal_stream_profiles}.

\begin{figure}
\centering\includegraphics[width=\columnwidth]{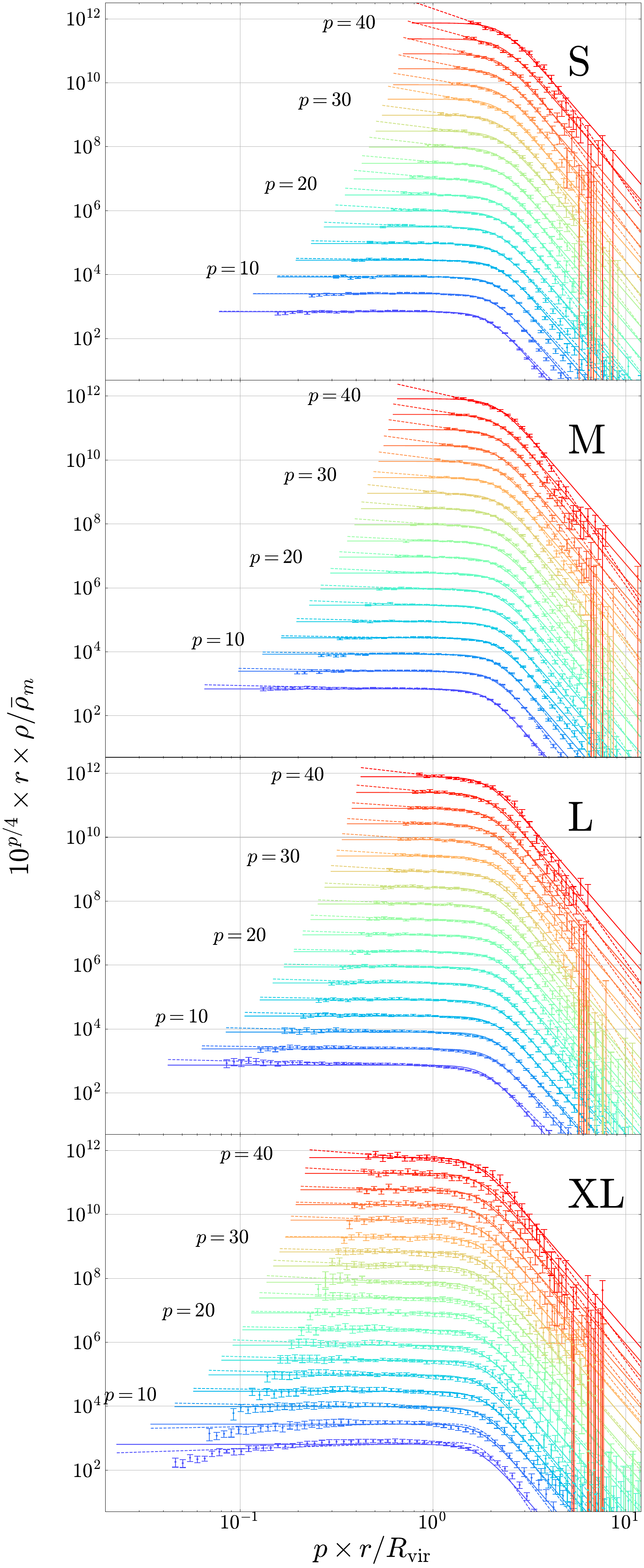}
    \caption{Stacked density profiles for $p=4$ to $40$ for every $2$ (markers). 
    The error bars indicate the root-mean-square deviation of profiles divided by the square root of the number of stacked halos.
    Dashed lines show the best-fit models when all the four parameters varied in equation~\eqref{eq:double_power}, while solid lines are for fixed slope parameters (i.e., $\alpha=-1$ and $\beta=-8$; equation~\ref{eq:stream}).
    Note that the vertical axis denotes $r\times \rho$ and both axes are rescaled according to $p$ for clarity.
    The best-fit parameters for the two cases are shown in Figs.~\ref{fig:stacked_dpfitparams} and \ref{fig:stacked_dpfitparams_A10B80}, respectively.}
    \label{fig:rdpfit_rhor_all}
\end{figure}

\begin{figure}
\centering\includegraphics[width=\columnwidth]{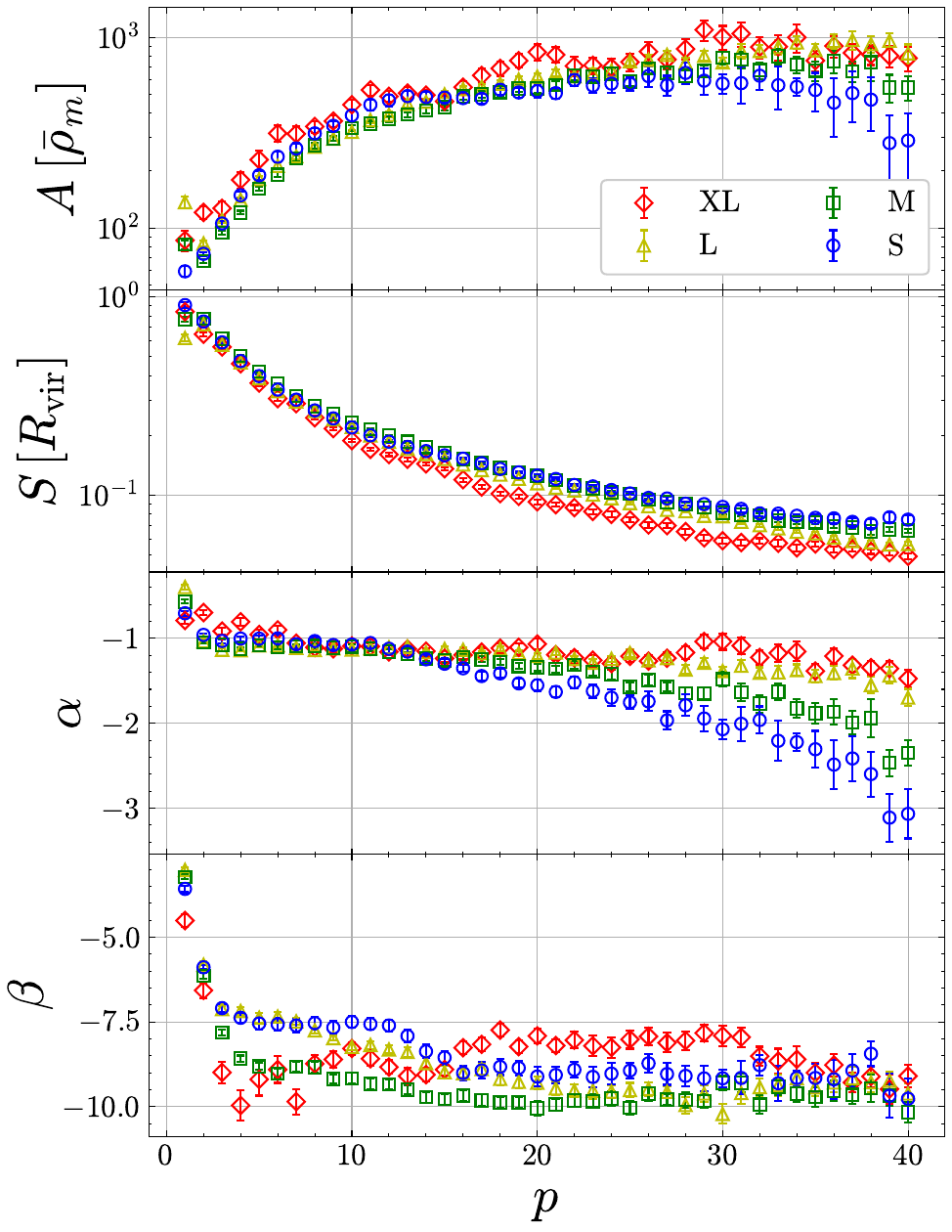}
    \caption{Best-fit parameters for the model~\eqref{eq:double_power} shown in  Fig.~\ref{fig:rdpfit_rhor_all} (markers).
    The errorbars denote the  square root of the diagonal part of the covariance matrix of the optimized parameters.}
    \label{fig:stacked_dpfitparams}
\end{figure}

\subsubsection{Fitting to a general form of double-power law profile}
\label{subsec:alpha_beta_free}

\begin{figure}
\centering\includegraphics[width=\columnwidth]{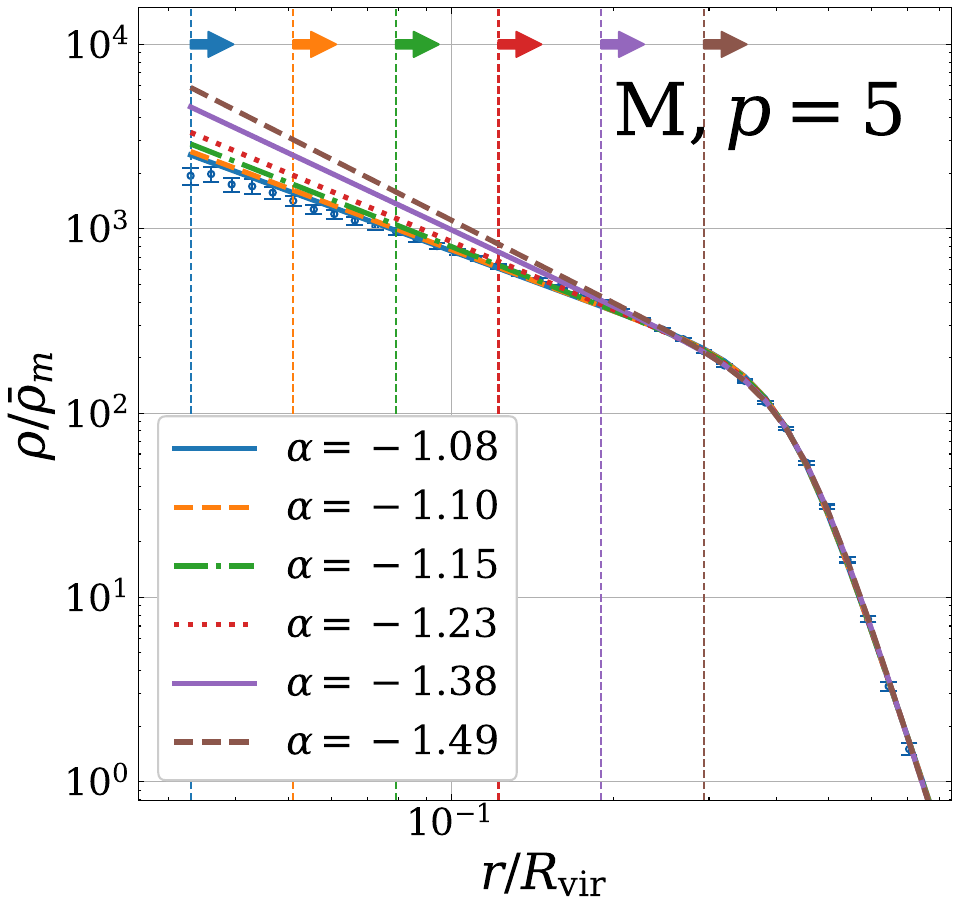}
    \caption{Dependence of the fit on the lower limit used in the fitting.
    Vertical dotted lines indicate the lower limit. The best-fit curves are indicated by the same colour as the lower limits. The best-fit inner slope is shown in the legend.}
    \label{fig:depon_NOB}
\end{figure}

First, we show the stacked density profiles and their fitting results to the double-power law form in equation \eqref{eq:double_power}. In Fig.~\ref{fig:rdpfit_rhor_all}, we scale the radius and density by factors of $p$ and $10^{p/4}$, respectively. We plot various stream profiles ranging from $p=4$ to $40$ for each mass bin. Dashed lines represent the best-fit curves, which extend down to half the resolution limit of the LR run. Overall, the results accurately reproduce the measured profiles. In Fig.~\ref{fig:stacked_dpfitparams}, the best-fit values of the parameters characterizing the double-power law profiles, i.e., the characteristic density $A$ (upper left), scale $S$ (lower left), and inner and outer slopes, $\alpha$ (upper right) and $\beta$ (lower right), are plotted as a function of $p$, with errorbars estimated from the curvature of the $\chi^2$ function in equation \eqref{eq:chi2}, where $\sigma_i$ is set to the RMSD of the stacked profile over each mass bin. There are clear trends for the dependence on $p$. While the characteristic density and scale respectively increase and decrease in a monotonic manner, the outer slope $\beta$ is almost constant except smaller $p$ and fluctuates around $\beta\sim-8\sim-10$. Also, the inner slope $\alpha$ remains almost constant over $p$ for massive halo samples, L and XL, but a notable steepening with $p$ is found for lighter samples, increasing also errorbars. Looking at Fig.~\ref{fig:rdpfit_rhor_all}, the measured profiles for a large value of $p$ tend to have less sampling points at inner radii. In particular, for light mass bins (S and M),  it seems insufficient to resolve the converged inner slope. This explains why the fitted values of $\alpha$ for S and M systematically decreases with $p$. In order to quantitatively access the impact of the lack of sampling points on the determination of $\alpha$, in Fig.~\ref{fig:depon_NOB}, we focus on the stacked halos for $p=5$ in halo sample M. We then reduce the number of inner sampling points by hand, and examine the same fitting procedure as we adopted in Fig.~\ref{fig:rdpfit_rhor_all}.
As we see clearly, removing the inner sampling points systematically changes the best-fit values indicated in the figure legend, apparently steepening the inner slope $\alpha$.

In order to ameliorate the biased estimation for slopes, instead of treating $\alpha$ and $\beta$ as free parameters, we set both (or either) of them to some fixed values and quantify the goodness of fit for parameter estimations. Fig.~\ref{fig:rdpfit_chi2} shows the reduced $\chi^2$ for various fitting results. From top to bottom panels, results for the halo samples from S to XL are respectively shown for even numbers of $p$, ranging from $p=4$ to $40$.  The left panel examines the cases fixing $\alpha$, taking only $\beta$ to be free, and the goodness of fit is shown as a function of the fixed value of $\alpha$. For profiles with a small value of $p\lesssim10-15$, we see some trends that the reduced $\chi^2$ takes a minimum value around $\alpha\sim1$. On the other hand, for profiles with a larger value of $p$, the reduced $\chi^2$ has no minimum, but it slightly increases with $\alpha$, thus preferring a smaller $\alpha$. The latter trends are indeed explained by a lack of enough support for fitting range at inner radius, and consistent with what we saw in Fig.~\ref{fig:depon_NOB}. Apart from this point, we may set $\alpha$ to $1$ as an optimal value to describe the inner region of multi-stream profiles. Then, we next examine the cases if both $\alpha$ and $\beta$ are fixed in the fitting analysis, and the goodness of fit is similarly evaluated for various value of $\beta$, fixing $\alpha$ to $1$. The results are shown in right panel of Fig.~\ref{fig:rdpfit_chi2}, plotted as function of $\beta$. We see the trend that the reduced $\chi^2$ becomes minimum around $\beta\sim8$ for profiles with most of $p$, irrespective of halo mass bins.

\subsubsection{Universal nature of multi-stream flows}
\label{subsec:universal_stream_profiles}

\begin{figure*}
\centering\includegraphics[width=2\columnwidth]{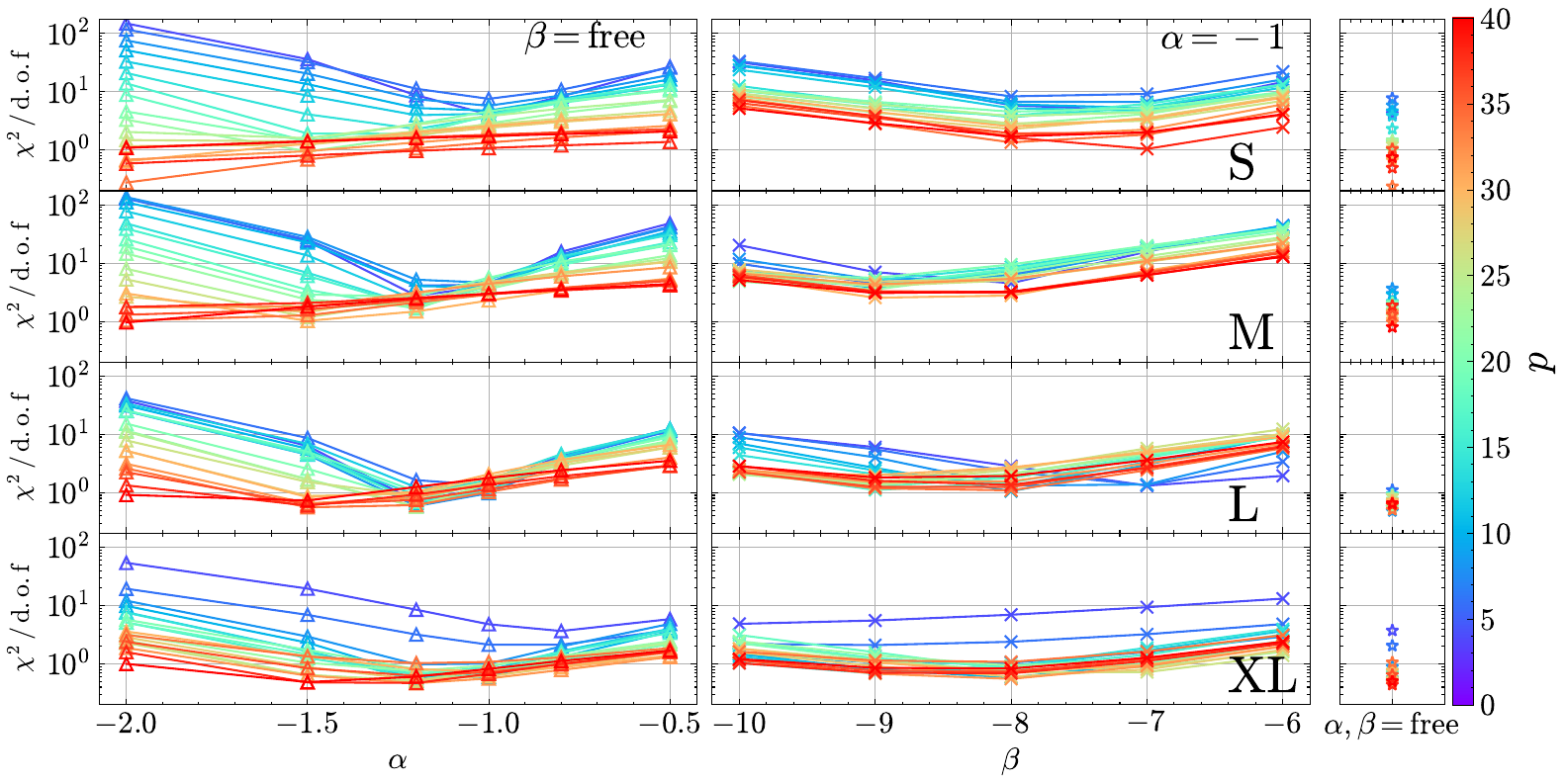}
    \caption{
    Dependence of $\chi^2$ on the slope parameters for each density profile (every $2$ between $p=4$ -- $40$ as the colour bar indicates).
    In the left column, we fix the inner slope $\alpha$ to the value indicated by the horizontal axis and optimise the other three parameters in equation~\eqref{eq:double_power}. 
    In the middle column, we fix the inner slope $\alpha$ to $-1$ and see the dependence of $\chi^2$ on the outer slope $\beta$, optimising the two remaining parameters.
    In the right column, we show the minimum $\chi^2$ when all the four parameters are varied.
}
    \label{fig:rdpfit_chi2}
\end{figure*}

Based on the results in Fig.~\ref{fig:rdpfit_chi2}, we propose to use the following function to characterise the universal behavior of the radial density profile of each stream: 
\begin{align}\label{eq:stream}
    \rho(r;p) = \frac{A(p)}{\bigl\{r/S(p)\bigr\}\,\Bigl[ 1+\bigl\{ r/S(p)\bigr\}^{7}\Bigr]}.
\end{align}

Adopting equation \eqref{eq:stream},  fitting results for the individual and stacked profiles  are respectively shown in Figs.~\ref{fig:radialprop1}--\ref{fig:radialprop235} and Fig.~\ref{fig:rdpfit_rhor_all}, depicted in all cases as solid curves. 
Overall, equation \eqref{eq:stream} reproduces mostly the measured profiles in $N$-body simulations for a large value of $p$. 
In particular, the stacked profiles in Fig.~\ref{fig:rdpfit_rhor_all} show good agreements with the model with $\alpha=-1$ and $\beta=-8$ as expected.
On the other hand, we see a small discrepancy in the outer slope, $\beta$, for individual haloes, which show steeper profiles with $\beta \lesssim -10$. Although we scale the radius before the stacking analysis, the slope after stacking tends to be shallower due to the remaining individuality of haloes, such as the characteristic scale, $S(p)$. However, as we will see shortly, the total profile reconstructed by adding the profiles with individual $p$ values is insensitive to the precise value of $\beta$, as long as its absolute value is large.

Returning to the stacked profile, another point to be noted is flatter inner profiles found at $p\lesssim10$ in the halo sample XL. This could be explained by the sensitivity of these low-$p$ orbits to recent mass accretion or merger history \cite{2020MNRAS.493.2765S}. Indeed, the trend tends to be erased after several orbits, reaching a universal slope of $-1$ for $p\gtrsim10$. There is thus no clear evidence for the flaw in the fitting form in equation~\eqref{eq:stream}, and we conclude that the measured profile for each stream is quantitatively described in a universal manner by the double-power law form in equation~\eqref{eq:stream}. 

\begin{figure}
\centering\includegraphics[width=\columnwidth]{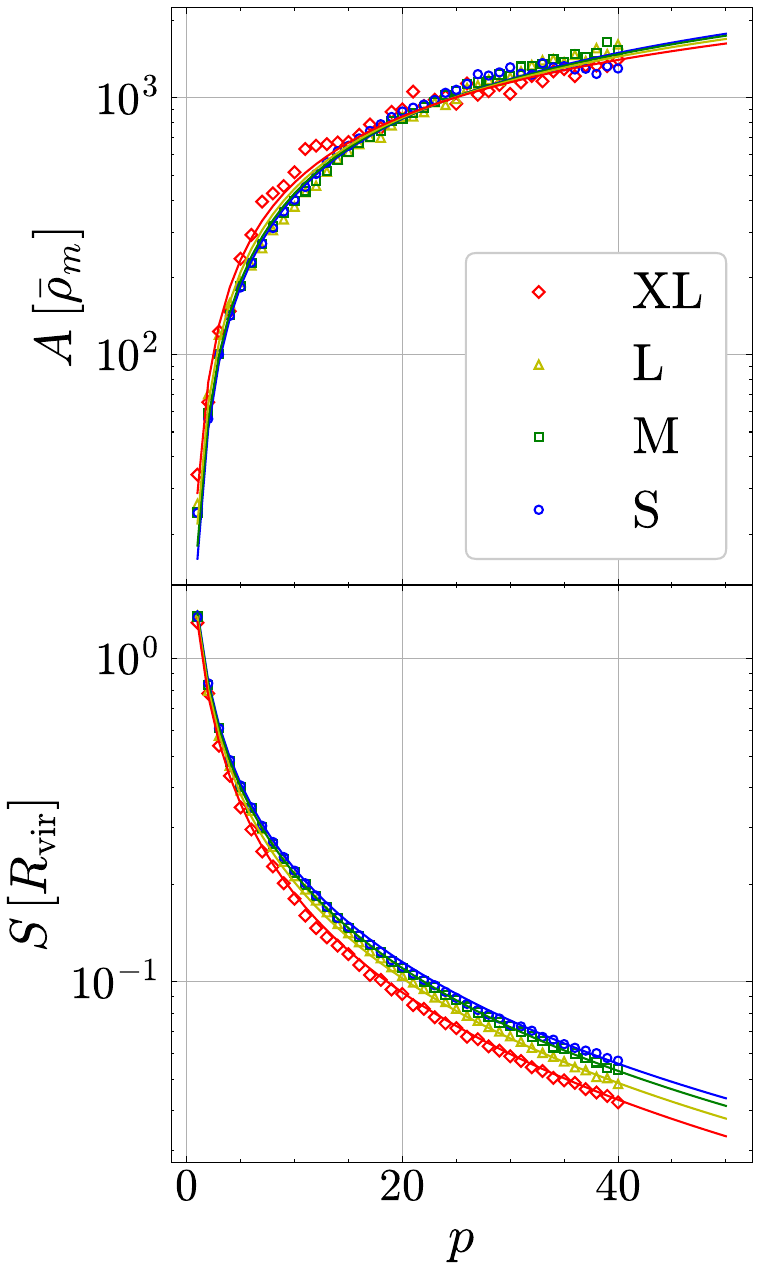}
    \caption{Best-fit values of the characteristic density $A$ and scale radius $S$ of the double-power law model with the fixed slopes, given at equation~\eqref{eq:stream}.
    The solid lines are the fitting formulas summarized in equations~\eqref{eq:fitA} and \eqref{eq:fitS}, with the halo mass $M_\mathrm{vir}$ estimated from the averaged mass in each halo sample.
    For ease of viewing, the error bars are omitted, but they are generally similar to those in Fig.~\ref{fig:stacked_dpfitparams}
    }\label{fig:stacked_dpfitparams_A10B80}
\end{figure}

In equation~\eqref{eq:stream}, the free parameters, the characteristic density $A$ and scale $S$, are determined by the fitting analysis for the stacked halo profiles. We find the following fitting formulas accurately describe the best-fit values of $A$ and $S$ for mass selected halos \citep{2023ApJ...950L..13E}: 
\begin{align}\label{eq:fitA}
\begin{split}
    \log_{10}{\Bigl\{A_\mathrm{fit}(p)/\overline{\rho}_{\rm m}\Bigr\}}  = 4.89-0.119\log_{10}{\left(M_\mathrm{vir,10}\right)}  \\ +\Bigl\{-3.89+0.243\log_{10}{\left(M_\mathrm{vir,10}\right)}\Bigr\}\,p^{-9/40},
\end{split}
\end{align}
\begin{align}\label{eq:fitS}
\begin{split}
    \log_{10}{\Bigl\{S_\mathrm{fit}(p)/R_{\rm vir}\Bigr\}} = 
    2.46-0.0474\log_{10}{\left(M_\mathrm{vir,10}\right)}  \\ +
    \Bigl\{-2.29-0.0639\log_{10}{\left(M_\mathrm{vir,10}\right)}\Bigr\}\,p^{1/8}.
\end{split}
\end{align}
Here, we define $M_{\rm vir,10}=M_{\rm vir}/\{10^{10}\,h^{-1}\,{\rm M_\odot}\}$. Fig.~\ref{fig:stacked_dpfitparams_A10B80} shows the best-fit values of characteristic density and scale (symbols), which are compared with the above fitting formulas (solid curves). With an explicit but weak halo mass dependence given in equations~\eqref{eq:fitA} and \eqref{eq:fitS}, the formulas agree quite well with the fitting results.  

\begin{figure*}
\centering\includegraphics[width=2\columnwidth]{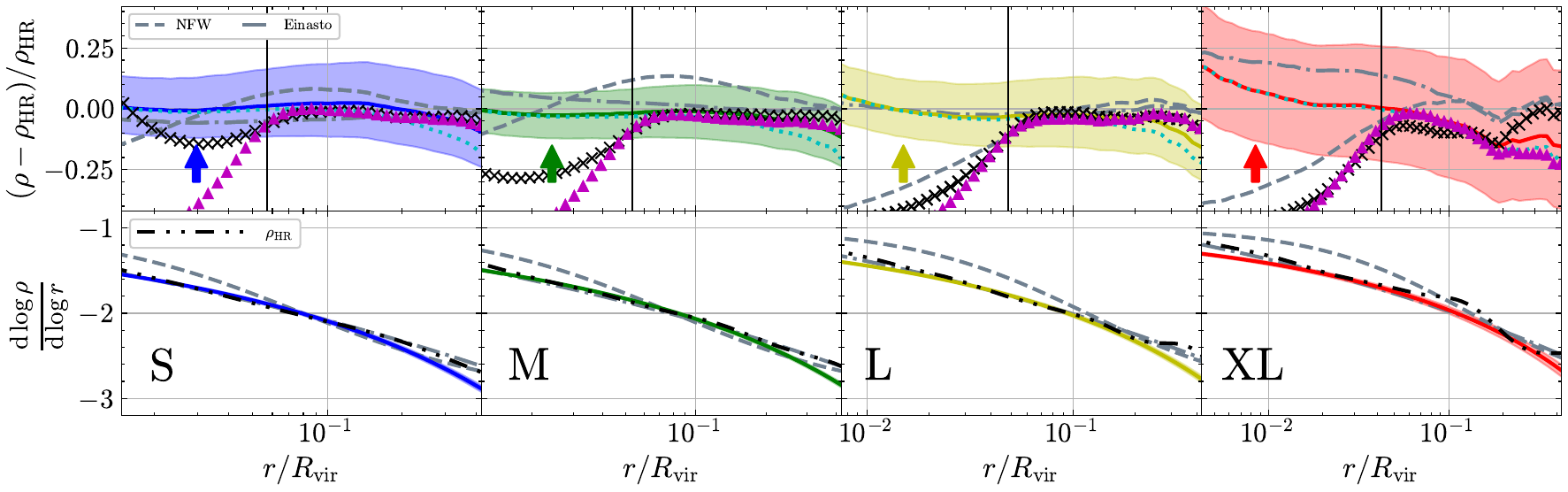}
    \caption{Comparison between the sum of fitting functions based on the LR run and the total profile in the HR run.
    In the upper panels, fractional differences between the two are shown. Solid lines depict the results when we use the model~\eqref{eq:fitA} and \eqref{eq:fitS} for the $p$ dependence of the model parameters $A(p)$ and $S(p)$ with fixed slopes ($\alpha=-1$ and $\beta=-8$). Cyan dotted lines correspond to the same model but with $\beta=-30$. Magenta triangles show the same model as the solid lines but the summation is truncated at $p=40$. The double-power law model with four free parameters fitted to the measured profiles is shown by cross symbols. Note that we also truncate the summation at $p=40$ in this case, as we cannot rigorously fit the data at higher $p$ values. Also shown are the NFW profile (dashed) and the Einasto profile (dot-dashed). The lower panel shows the logarithmic slope of the total profile (dot-dot dashed lines for the HR run). The solid, dashed, and dot-dashed lines are the same as in the upper panels.  
    }
    \label{fig:dpfitsum}
\end{figure*}

To quantitatively assess the double-power law nature of each stream by equation~\eqref{eq:stream}, we measure the total density proﬁle obtained from the HR run for halos that have been matched with the LR run. We then compare their stacked profiles over haloes in each mass bin with the predictions obtained by summing the double-power law proﬁles (equation~\ref{eq:stream}) over $p\geq1$ described by
equations~\eqref{eq:stream}--\eqref{eq:fitS}. The results are shown in Fig.~\ref{fig:dpfitsum}, where the upper and lower panels for each mass bin respectively represent the fractional difference of the total density profile with respect to the HR run, $(\rho-\rho_{\rm HR})/\rho_{\rm HR}$, and the logarithmic slope of the total profile, $d\log\rho/d\log r$. In each panel, the solid curve shows the prediction based on the double-power law profiles, 
with the shaded region indicating the uncertainties arising from those in determining the numerical coefficients of equations \eqref{eq:fitA} and \eqref{eq:fitS}. Note that the summation over the number of apocentre passages is conservatively taken up to $p = 3,000$. The change in density is less than 0.2\% over the plotted range when we instead stop at $p = 300$. For reference, we also plot the NFW profile \citep{1997ApJ...490..493N}
\begin{equation}\label{eq:NFW}
    \rho_\mathrm{NFW}(r) = \frac{\rho_s}{r/R_s\left(1+r/R_s\right)^{2}},
\end{equation}
and the Einasto profile \citep{einasto1965construction}
\begin{equation}\label{eq:Einasto}
    \rho_\mathrm{Einasto}(r) = \rho_s \exp{\left[-\frac{2}{\alpha}\left\{ \left(\frac{r}{R_s}\right)^{\alpha} -1\right\}\right]},
\end{equation}
which are obtained by fitting the profiles in the HR run over the range $2\,{\rm Max}(\epsilon_{\rm HR}/R_{\rm vir})\leq r/R_{\rm vir} \leq 0.9$, with ${\rm Max}(\epsilon_{\rm HR}/R_{\rm vir})$ being the maximum value of the ratio of softening length $\epsilon_{\rm HR}$ to $R_{\rm vir}$ estimated for individual halos in each mass bin. 
Note that we fixed $\alpha = 0.16$ in the Einasto profile following \citet{2020Natur.585...39W}.
Our model is in good agreement with the HR run for all four mass bins. Notably, we can recover the proﬁle below the scale of $1.2 {\rm Max}(\epsilon_{\rm LR}/R_{\rm vir})$ indicated by vertical arrows, corresponding to the convergence radius above which the measured profiles from the two runs agree well with each other at $\sim3 \%$ precision. This suggests that the inner slope of $\alpha=-1$ is a reasonable choice, applicable below the softening scale of the LR run.

In Fig.~\ref{fig:dpfitsum}, 
we also consider three variants: (i) summation of equation~\eqref{eq:double_power} fixing $\alpha$ and $\beta$ respectively to $-1$ and $-30$, using the characteristic density and scale given by equations \eqref{eq:fitA} and \eqref{eq:fitS} (dotted), (ii) the same as our main model depicted by the solid lines, but truncated the summation at $p=40$ (triangles) and (iii) summation of equation~\eqref{eq:double_power} with the four parameters determined to fit the individual profiles in Fig.~\ref{fig:stacked_dpfitparams} (crosses). Note that we also truncate the summation at $p=40$ for case (iii), where the best-fit values of the model parameters are not available beyond $p=40$. 

The upper panels of Fig.~\ref{fig:dpfitsum} show that the total profiles of the case (i) are hardly distinguishable with the solid curves except for the outer radii of $r/R_{\rm vir}\gtrsim0.2$ despite the rather small value of $\beta=-30$. This implies that the inner part of the total profile is insensitive to the precise value of $\beta$ as long as the profile decays quickly towards large radii, and hence the inner slope of $-1$ in equation \eqref{eq:stream} is the main clue to clarify the cuspy structure of the central halo profiles. On the other hand, due to the summation over a restricted number of $p$, the total profile predicted by the case (ii) starts to fall off as the radius decreases, although the outer profiles above the vertical solid lines mostly match the predictions based on equations \eqref{eq:stream}--\eqref{eq:fitS}. This suggests that the extrapolation beyond $p=40$ by the fitting forms~\eqref{eq:fitA} and \eqref{eq:fitS} play a significant role to fill the gap on small radii and reproduce the results of the HR run. Finally, the case (iii) with the four free parameters agrees well with case (ii) reinforcing our claim that fixing inner and outer slopes to $-1$ and $-8$ is an optimal choice, as examined in Fig.~\ref{fig:rdpfit_chi2}.  Thus, with the fitting formulas given by equations \eqref{eq:fitA} and \eqref{eq:fitS},  the double-power law form of the stream profiles at equation~\eqref{eq:stream} describes accurately spherically averaged halo structures, and provides the most optimal description among other variants of the model.   

As a result, the logarithmic slope plotted in the lower panels for the HR run (dot-dot-dashed) is accurately reproduced by our model (solid). It is interesting to observe that the Einasto profile (dot dashed) is equally in good agreement with the simulation results. We expect that the two models should depart asymptotically at the small scale limit (i.e., $-1$ vs $0$), which cannot be examined from the current simulation due to the limited dynamical range. We postpone to address the slope on even smaller scales as a future work.

\section{Discussion}\label{sec:discussion}

\subsection{Dependence on halo samples}\label{subsec:dep_on_MAR}

So far, we have studied individual stream profiles for the mass-selected halo samples and characterised their properties based on the double-power law profile. Here, to assess the robustness of our findings, we analyse a subset of $460$ halos within a specific mass range $[4.10\times10^{11},\,2.39\times10^{12}]\,h^{-1}M_\odot$, which are divided into two sub-samples based on two different criteria. One is the concentration parameter $c_{\rm vir}$, defined by the ratio $R_{\rm vir}/R_{\rm s}$. Here, the radius $R_{\rm s}$ is the scale radius of the NFW profile (equation~\eqref{eq:NFW}), and we estimate it from \textsc{Rockstar} based on the maximum circular velocity \citep{Klypin_etal2011}. Another quantity used for sample selection is the mass accretion rate defined by 
\begin{align}
    \label{eq:gammadyn1}
    \Gamma_\mathrm{dyn}(t) \equiv \frac{\log{\Bigl[M_\mathrm{vir}(t)-M_\mathrm{vir}\bigl\{t-t_\mathrm{dyn}(z)\bigr\}\Bigr]}}{\log{\Bigl[a(t)-a\bigl\{t-t_\mathrm{dyn}(z)\bigr\}\Bigr]}}, 
\end{align}
where the quantity $t_{\rm dyn}$ represents the dynamical time estimated from halo masses \citep{2017ApJS..231....5D}\footnote{
We use the virial mass, $M_\mathrm{vir}$, to measure $\Gamma_\mathrm{dyn}$, whereas \cite{2017ApJS..231....5D} uses $M_\mathrm{200m}$.}:
\begin{align}
    \label{eq:gammadyn}
    t_\mathrm{dyn}(z) \equiv \frac{2R_\mathrm{vir}}{V_\mathrm{vir}} = H(z)^{-1} \Bigl\{\frac{8}{\Delta_\mathrm{vir}(z)\Omega_\mathrm{m}(z)}\Bigr\}^{\frac{1}{2}},
\end{align}
which gives $4.04$\,Gyr at $z=0$ in our case.

\begin{figure}
\centering\includegraphics[width=\columnwidth]{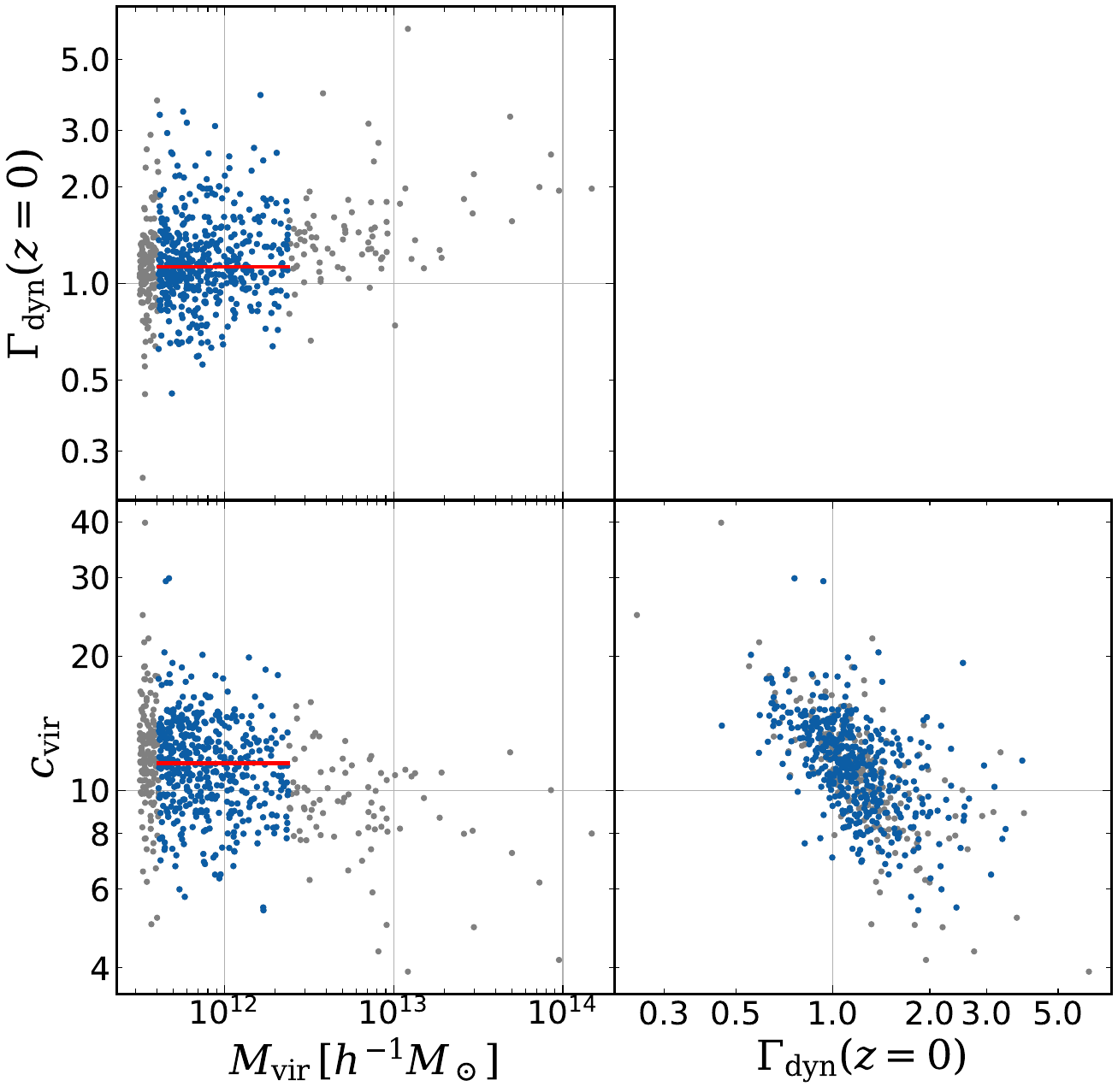}
    \caption{Distribution of concentration $c_\mathrm{vir}$, halo mass $M_\mathrm{vir}$ and mass accretion rate $\Gamma_\mathrm{dyn}$ at $z=0$.
    Thick markers indicate $460$ halos in the mass range $[4.10\time10^{11},\,2.30\times10^{12}]\,h^{-1}M_\odot$
    , which are the halos analysed in Section~\ref{subsec:dep_on_MAR}.
    The thin markers indicate the other halos in our halo catalog.
    The horizontal red lines indicate the median $\Gamma_\mathrm{dyn}$ and $c_\mathrm{vir}$ of the thick markers, which are used as boundaries of 460 samples.
    Here $c_\mathrm{vir}$ is calculated from $R_s$ which is estimated by \textsc{Rockstar}.}
    \label{fig:cmg_distribution}
\end{figure}

In both cases, we divide the halos into two halves, one with high values of these indicators and the other with low values. Fig.~\ref{fig:cmg_distribution} shows the distribution of the parameters, $c_{\rm vir}$ and $\Gamma_{\rm dyn}$, and $M_{\rm vir}$. Darker points are the halos used in the present analysis, and the boundary of the samples is shown in red lines in the upper and lower left panels. Clearly, the distribution in the measured parameters $c_{\rm vir}$ and $\Gamma_{\rm dyn}$ exhibits a tight correlation, and these parameters are anti-correlated. That is, halos with a higher concentration parameter tend to have a smaller $\Gamma_{\rm dyn}$. Therefore, we expect these two cuts to affect the results in a similar manner.

\begin{figure}
\centering\includegraphics[width=\columnwidth]{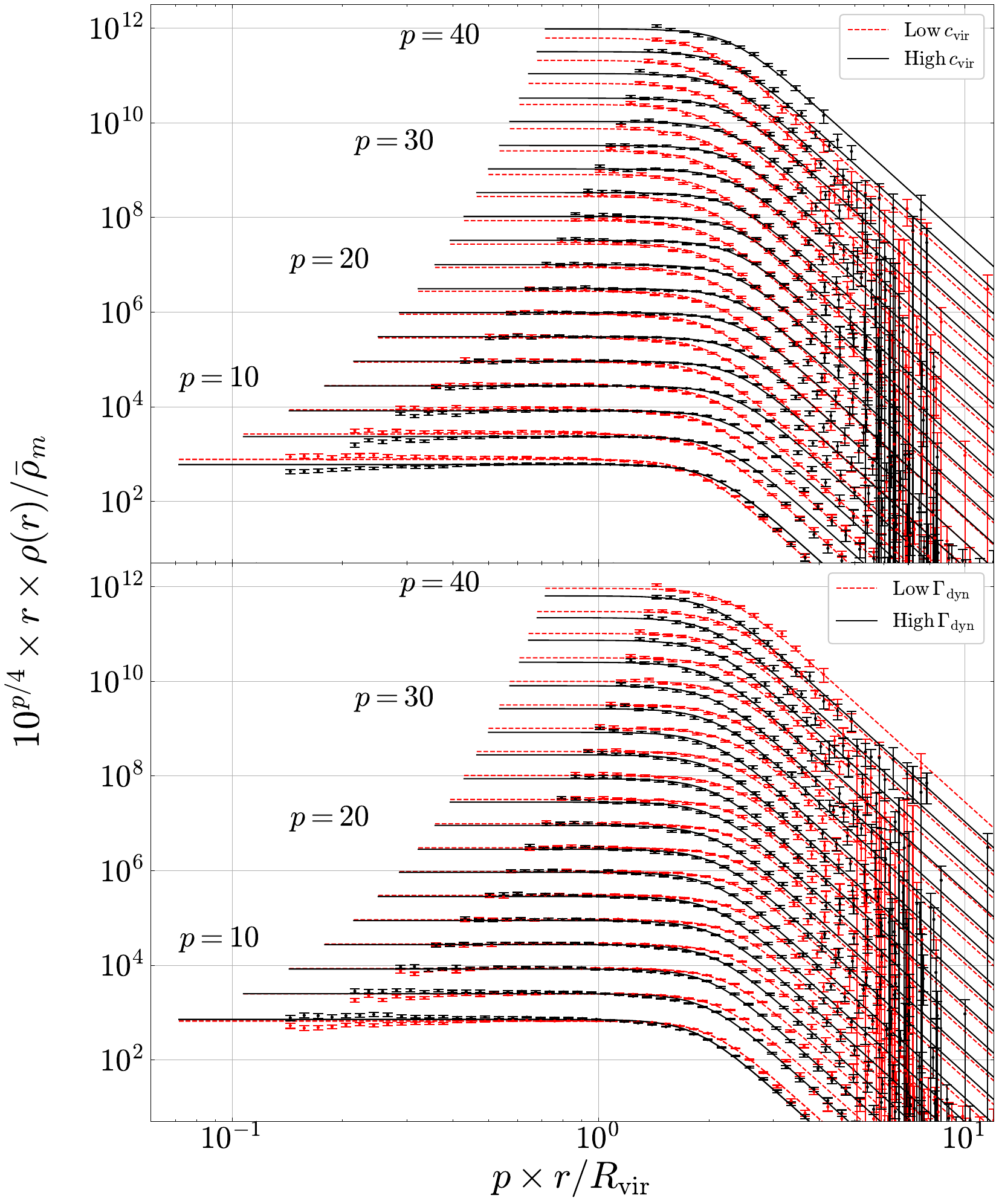}
    \caption{Stacked density profiles for $p=4$ to $40$ for every $2$ (markers) of subgroups Low $c_\mathrm{vir}$ (dotted in upper panel), High $c_\mathrm{vir}$ (solid in upper panel), Low $\Gamma_\mathrm{dyn}$ (dotted in lower panel) and High $\Gamma_\mathrm{dyn}$ (solid in lower panel).
    Same as Fig.~\ref{fig:rdpfit_rhor_all}, the vertical axis denotes $r\times \rho$ and both axes are rescaled according to $p$ for clarity.    }
\label{fig:rdpfit_rhor_subgroups}
\end{figure}

Fig.~\ref{fig:rdpfit_rhor_subgroups} shows the results of density profiles for the subsamples defined by $c_{\rm vir}$ (upper) and $\Gamma_{\rm dyn}$ (lower). In each panel, results for low and high values of $c_{\rm vir}$ or $\Gamma_{\rm dyn}$ are depicted as red and black colours, respectively. The stacked profiles for each $p$ are fitted to the double-power law form in equation (\ref{eq:stream}), and the best-fit results are plotted in dotted and solid curves for the two subsamples, respectively. We again see a good agreement between the double-power law function and measured profiles over a wide range of $p$ and radius.

A close look at each stream profile in Fig.~\ref{fig:rdpfit_rhor_subgroups} reveals that halos with high concentration or low accretion rate tend to have a large amplitude $A(p)$ and a large characteristic scale $S(p)$, in particular for $p\gtrsim14$. The opposite trends in the samples divided with the parameters $c_{\rm vir}$ and $\Gamma_{\rm dyn}$ simply come from their anti-correlation behavior seen in the lower right panel of Fig.~\ref{fig:cmg_distribution}. On the other hand, the result that high-concentration halos have large $S(p)$ seems somewhat counter-intuitive in the sense that a large value of $c_{\rm vir}$ implies, by definition, a small scale radius $R_{\rm s}$. Nevertheless, the mass inside the radii $r\leq R_{\rm s}$ actually increases as increasing $c_{\rm vir}$. Since the inner mass of the halo is mainly determined by the sum of stream profiles for large values of $p$, 
and the mass of each stream profile is proportional to $A(p)S(p)^3$, a high mass concentration leads to a large value of $A(p)S(p)^3$, consistent with what is seen in Fig.~\ref{fig:rdpfit_rhor_subgroups}\footnote{Assuming the double-power law form of equation (\ref{eq:stream}), the mass of each stream profile, $M(p)$, is analytically expressed as follows:  
\begin{align}
    M(p) &= \int_{r=0}^{r=+\infty}   
    \frac{4 \pi r^{2} A(p)}{\bigl\{r/S(p)\bigr\}\,\Bigl[ 1+\bigl\{ r/S(p)\bigr\}^{7}\Bigr]}
    dr \notag \\
    &= \frac{4 \pi}{49}\left(2\sin{\frac{\pi}{7} -\cos{\frac{\pi}{14}} + 3\cos{\frac{3\pi}{14}} } \right) A(p)S(p)^{3}, \notag 
\end{align}
which gives $M(p)\approx 0.574A(p)S(p)^3$.
}. 

The results shown in Figs.~\ref{fig:rdpfit_rhor_subgroups} suggest that the double-power law feature in the radial stream profiles generically appears, irrespective of the halo sample selection. Although the characteristic density and scale, $A(p)$ and $S(p)$, depend generally on the selection criteria, their $p$ dependence can be translated from one to another once the relationship between different samples is established. In this respect, it might be useful to re-derive the fitting formulas of $A(p)$ and $S(p)$ expressed in terms of $c_{\rm vir}$ rather than $M_{\rm vir}$, since the concentration parameter is tightly correlated with the inner structure of haloes. 

To do this, we further divide the haloes according to the value of $c_{\rm vir}$. We consider four subgroups: from $10$ to $30$, $30$ to $50$, $50$ to $70$ and $70$ to $90$ percentiles of $c_{\rm vir}$, each containing the same number of halos. We then look for a fitting function for $A(p)$ and $S(p)$, now with dependence on $c_{\rm vir}$. We find the following functions give reasonable fit to the data:
\begin{align}\label{eq:fitA_c}
    \log_{10}{\Bigl\{A_\mathrm{fit}(p)/\overline{\rho}_{\rm m}\Bigr\}}  &= 4.09 +0.133 c_\mathrm{vir} 
    -\Bigl\{1.99+0.202 c_\mathrm{vir} \Bigr\}\,p^{-4/25},
\end{align}
\begin{align}\label{eq:fitS_c}
    \log_{10}{\Bigl\{S_\mathrm{fit}(p)/R_\mathrm{vir}\Bigr\}} &= 
    1.69+0.0132 c_\mathrm{vir} -
    1.74 \,p^{4/25}.
\end{align}

\begin{figure*}
\centering\includegraphics[width=2\columnwidth]{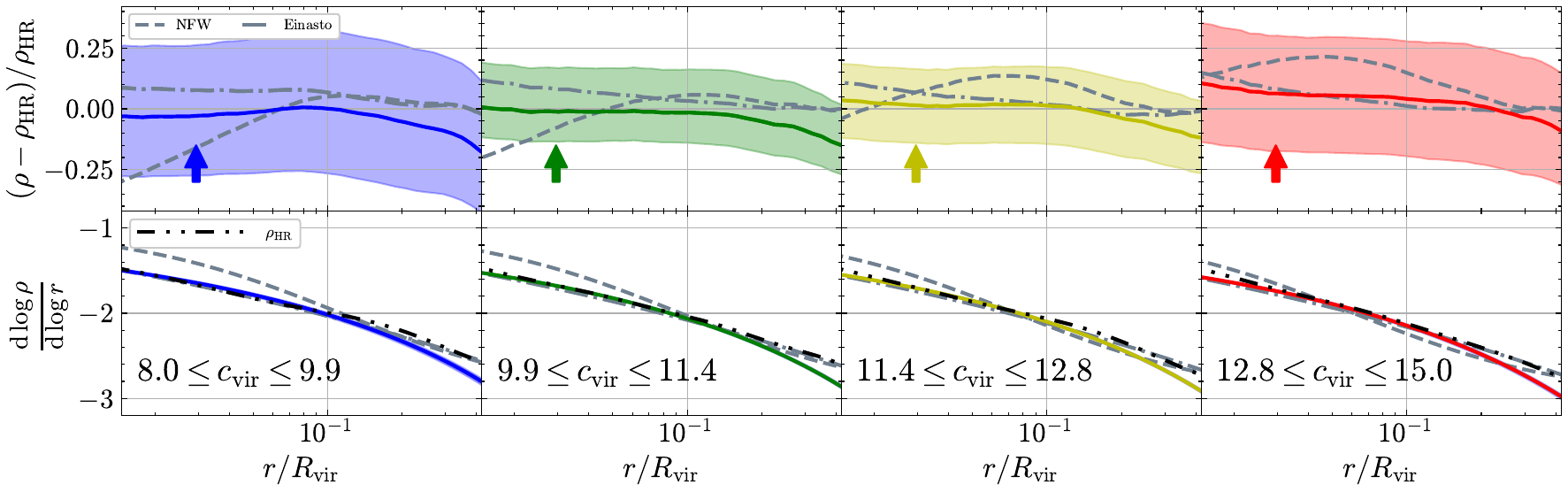}
\caption{
Same as in Fig.~\ref{fig:dpfitsum}, but we use equations~\eqref{eq:fitA_c} and \eqref{eq:fitS_c} for $A(p)$ and $S(p)$. 
In the upper panels, fractional differences between the total profiles in HR simulation and the sum of equation~\eqref{eq:stream} using equations~\eqref{eq:fitA_c} and \eqref{eq:fitS_c} for $A(p)$ and $S(p)$ (solid line), and the best fit NFW (dashed line) and Einasto (dot-dashed line) profiles for the HR total profile are shown.
The vertical arrows indicate the resolution limit of the LR simulation.
The shaded regions indicate the estimated uncertainties of the solid line, which are propagated from the statistical error in the stacked profile through the uncertainties in \eqref{eq:fitA_c} and \eqref{eq:fitS_c}.
} 
\label{fig:AS_compar}
\end{figure*}

The performance of the new fitting formulas given above is examined in Fig.~\ref{fig:AS_compar}, where the total density profile and their logarithmic slope are again plotted for the four subgroups. Similarly to the case of mass-selected samples in Fig.~\ref{fig:dpfitsum}, the sum of the profiles with the parameters given by the formulas in equations~\eqref{eq:fitA_c} and \eqref{eq:fitS_c} reproduces the total profile measured from the HR run well beyond the resolution limit of the LR run, which we actually use to calibrate the fitting functions.

\begin{figure*}
\centering\includegraphics[width=2\columnwidth]{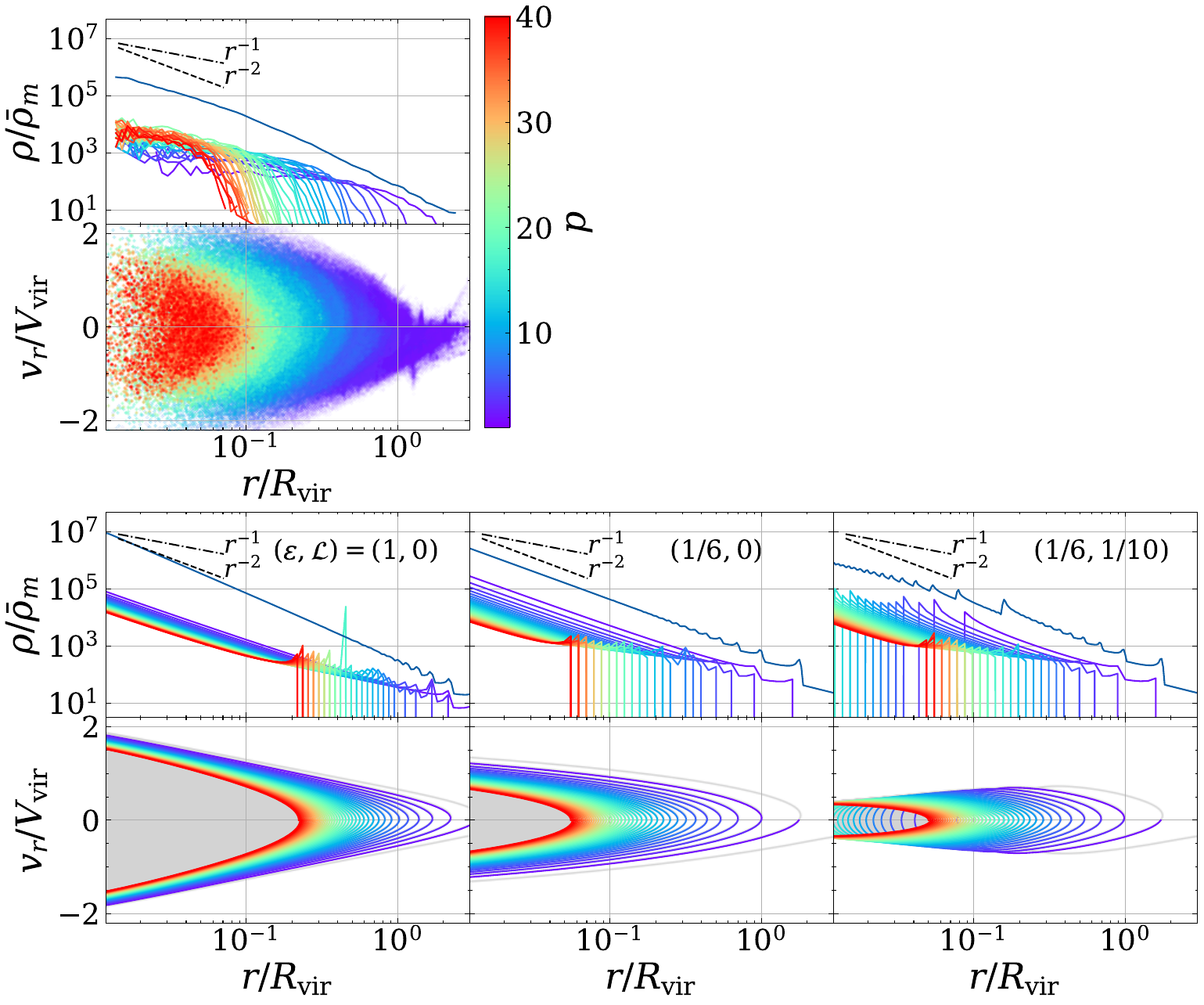}
    \caption{Density profiles and phase-space distributions of particles separated by $p$ (color-coded as the colour bar indicates), and total density profile (blue solid line).
    As a representative of $N$-body halo, we show the distributions of the same halo shown in Fig.~\ref{fig:radialprop18} in the top row.
    Note that the virial overdensity $\Delta_\mathrm{vir}$ is $18\pi^2$ in EdS universe (background metric of the self-similar solutions), and it is different from those in $\Lambda$CDM universe, $313$ at $z=0$.
    Here we set $\Delta_\mathrm{vir}=313$ and normalize the coordinates in the self-similar solution.
    This does not change the shape of density profiles.
    }
    \label{fig:sscomparison}
\end{figure*}

\begin{figure*}
\centering\includegraphics[width=2\columnwidth]{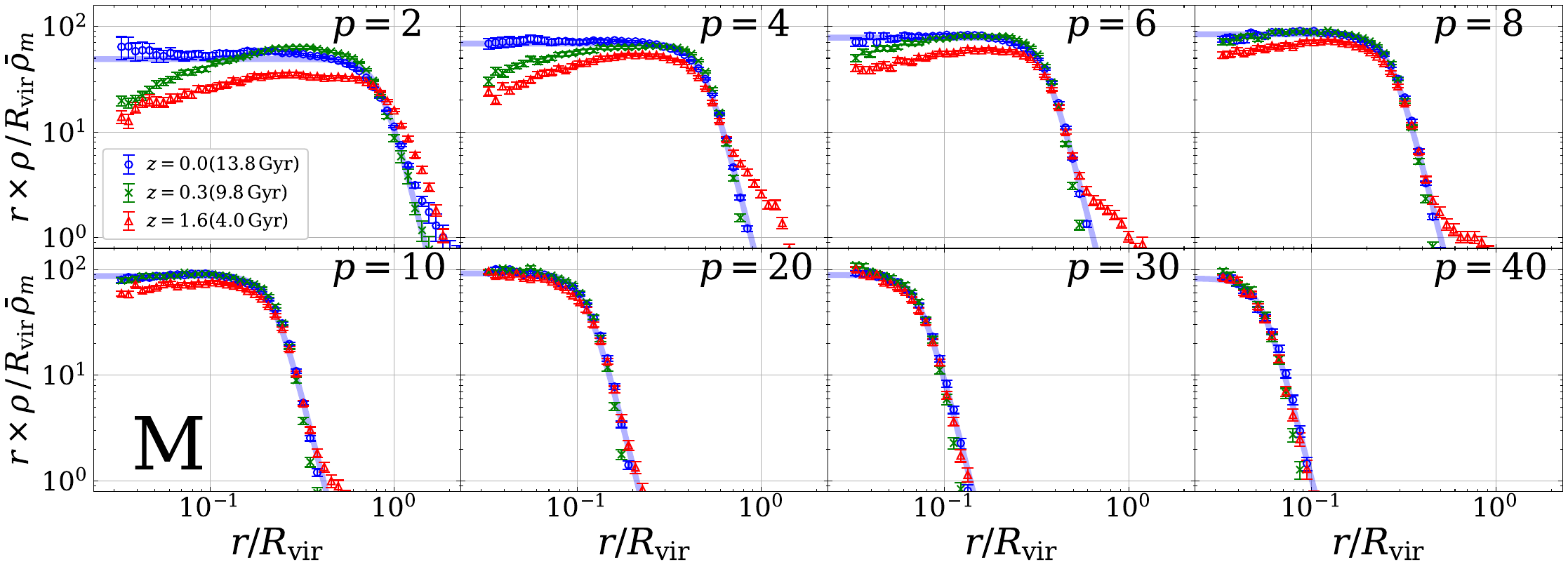}
    \caption{
    Stacked density profiles for the halo sample M measured at $z=0$ (blue), $0.3$ (green), and $1.6$ (red).
    In the same panel, each profile consists of identical particles divided by $p$ at $z=0$.
    We also show the best-fit curve of equation~\eqref{eq:stream} for the measured profiles at $z=0$ as thin blue lines.
    At every redshift, we use physical distance as units of radial coordinates and normalize them to $R_\mathrm{vir}$ at $z=0$.}
    \label{fig:timeevolution}
\end{figure*}

\subsection{Comparison with self-similar solutions}\label{subsec:SScompar}

The results in Section \ref{sec:results} and \ref{subsec:dep_on_MAR} show that the inner structure of halos exhibits common features, and each stream profile is characterised by a simple double-power law function  
in a rather universal manner, suggesting that haloes evolve in a self-similar fashion. Here, we compare the radial phase-space structure and density profile for each stream with those predicted by self-similar solutions. 
To be strict, self-similar solutions are valid only in the Einstein-de Sitter universe, and hence a face-to-face comparison with simulations in the $\Lambda$CDM model makes little mathematical sense. Nevertheless, the secondary infall model of \citet{1985ApJS...58...39B}, equivalent to the self-similar solution by \citet{1984ApJ...281....1F} with a specific parameter choice (see below), has been shown to reproduce the power-law slope of the pseudo-phase-space density, $Q(r)\propto r^{-1.875}$, found in simulations in the $\Lambda$CDM model. 

Along the lines of this, we focus on the qualitative trends in simulations discussed so far and compare them with predictions by self-similar solutions. Among various models considered in the literature, we adopt
spherically symmetric solutions by \citet{1984ApJ...281....1F} and \citet{1997PhRvD..56.1863S} (see also \cite{2001MNRAS.325.1397N}) as representative models of halo evolution under stationary matter accretion. Note that the latter model allows us to incorporate the non-zero angular momentum into the matter flow. The number of parameters characterising the halo structure is two. One is the slope of the initial linear overdensity, parameterised as $\delta_{\rm lin}(r)\propto r^{-3\epsilon}$ in the range $0\leq \epsilon\leq 1$, which is related to the accretion rate through $M_{\rm ta}(t)\propto \{a(t)\}^{1/\epsilon}$ with $M_{\rm ta}$ being the mass inside the turn-around radius. Another parameter is the dimensionless angular momentum $\mathcal{L}$, defined by $\mathcal{L}\equiv L/(GM_*r_*)^{1/2}$, with $r_*$ and $M_*$ being respectively the turn-around radius and the total mass inside the turn-around radius for each spherical shell accreting onto the halo.

The bottom panels of Fig.~\ref{fig:sscomparison} plot the radial density profile (upper) and phase-space distribution (lower) predicted by the self-similar solution for specific parameter choices, $(\epsilon,\,L)=(1,\,0)$ (left), $(1/6,\,0)$ (middle) and $(1/6,\,1/10)$ (right). Note that the second solution corresponds to the secondary infall model of \cite{1985ApJS...58...39B} as we mentioned above. These are obtained numerically based on the method described by \citet{2010PhRvD..82j4044Z}. For reference, we also show in the upper panel the results of a representative halo taken from Fig.~\ref{fig:radialprop18}.   

Because of the spherical symmetry, the radial phase space structure of the self-similar solution exhibits distinctive streams, and all of the plotted results show a qualitatively similar trend at the outer part. That is, for each stream, a spiky structure appears in density around the caustics, and this is shortly followed by a sharp drop.  On the other hand, the inner structure of self-similar solutions generically shows a power-law behavior, and in the cases with zero angular momentum, the slope of the total density profile as well as each stream profile is predicted to be nearly $-2$. To be strict, the asymptotic slope of the solution with $(\epsilon,\,\mathcal{L})=(1,\,0)$ is slightly different from $-2$: the total and each stream approach $\rho\propto r^{-9/4}$ and $r^{-15/8}$,  respectively\footnote{To derive the asymptotic slope of each stream profile, we first note that 
\citet{1984ApJ...281....1F} and \citet{2001MNRAS.325.1397N} analytically derived the asymptotic inner slope of total density profiles $\rho_{\rm tot}$ for self-similar solutions, summarised as:
\begin{align}
    \rho_{\rm tot}(r) \propto 
    \begin{cases}
    {r^{-9\epsilon/(1+3\epsilon)}\, (\frac{2}{3} \leq \epsilon \leq 1)}\\
    {r^{-2}\,(0 \leq \epsilon \leq \frac{2}{3})} \notag
    \end{cases}
\end{align}
for $L=0$, and 
\begin{align}
    \rho_{\rm tot}(r) \propto 
    r^{-9\epsilon/(1+3\epsilon)}\, (0 \leq \epsilon \leq 1)\notag
\end{align}
for $L \neq 0$.
Then, consider the interior mass of particles/shells having the number of apocentre passages $p$ inside the radius $r$, which we denote by $M_p(r)$. For a stationary halo composed of equal mass particles/shells, this is proportional to the time $\delta t(r)$ that a particle/shell with $p$ spends inside $r$ \citep{1984ApJ...281....1F,Dalal2010,2011ApJ...734..100L}. Except for the radii near the apoapsis, the time $\delta t(r)$ is proportional to $r/v_{\rm r}$ with $v_{\rm r}$ being the radial velocity. Since $v_{\rm r}\propto \sqrt{\Phi}$, we have $M_p(r)\propto r/\sqrt{\Phi(r)}$. For the total profile that has a slope less than or equal to $-2$, corresponding to the solutions with $\epsilon\leq2/3$, the potential becomes constant near the centre, and we obtain $M_p(r)\propto r$, which gives the stream profile of $r^{-2}$. On the other hand, if the inner slope of the total profile is steeper than $-2$, corresponding to the solutions with $\epsilon>2/3$, the potential diverges at the centre and we have $\Phi\propto r^{(2-3\epsilon)/(1+3\epsilon)}$. Taking this dependence into account, the interior mass with $p$ becomes $M_p(r)\propto r^{9\epsilon/2(1+3\epsilon)}$, which yields the stream profile of $\propto r^{-3(2+3\epsilon)/2(1+3\epsilon)}$. Hence, for $(\epsilon,\,{\mathcal L})=(1,\,0)$, we obtain the slope of the stream profile, $-15/8$.}. 
Setting aside a subtle difference, the predicted steep slopes for the total and stream profiles are rather contrasted with those found in the simulations. In the case of $(\epsilon,\,\mathcal{L})=(1/6,\,1)$, the non-zero angular momentum yields a potential barrier near the centre in each mass shell, making the pericentre radius finite. As a result, a sharp cutoff appears at the inner radii for each stream profile, and this results in the total profile being shallower than the slope of $-2$. Thus, apart from small bumpy structures, the resultant total profile resembles the one obtained from simulations. However, the density profile of each stream still possesses a steep slope close to $-2$ above the pericentre radius, which contradicts the asymptotic inner slope seen in our simulation results.

These results indicate that the self-similar solutions considered here miss something essential or some complexities inherent in the cosmological $N$-body simulations. Therefore, a more comprehensive study is necessary taking into account properly the missing ingredients in the self-similar solutions. This may involve relaxing the symmetry assumptions \citep{1993ApJ...418....4R,2011ApJ...734..100L} or exploring non-zero tidal torques \citep{2010PhRvD..82j4044Z}. Also, the angular momentum distribution may be the key. Although we have considered the self-similar solution with angular momentum, this allows for the angular momentum to be provided in a very specific manner. Introducing a broad angular momentum distribution yields a new radial dependence of the particle trajectories that potentially leads to a shallower stream profile consistent with simulations. This could be achieved perhaps by further breaking the self-similarity, as studied by \cite{Lu2006}. We leave these investigations to future work.

\subsection{On the emergence of double-power law nature}\label{subsec:dp_emerge}

As a final discussion toward a better understanding of the origin of the universal double-power law nature, we focus on the halo sample M in Table \ref{tab:halo_catalog}, and select the particles with $p=2$, $4$, $6$, $8$, $10$, $20$, $30$ and $40$ at $z=0$. Then we trace their trajectories to higher redshifts and measure the density profiles for each value of $p$ stacked over different halos. Here, the values of $p$ always refer to those counted until $z=0$ instead of the redshift at which the profiles are plotted. Namely, we investigate the time evolution of the profile for the same set of particles over time. Fig.~\ref{fig:timeevolution} overplots the results at $z=0.3$ (green) and $1.6$ (red), on top of those at $z=0$ already shown in Fig.~\ref{fig:rdpfit_rhor_all} (blue). We observe that the amplitude of the curve increases as the redshift decreases. 
Interestingly, however, the evolution of the inner profiles becomes significantly weaker as the value of $p$ increases, and at $p=40$, the profiles almost converge even at the outermost part. 
This suggests that the double-power law nature was established at an early stage of halo formation and remains stable against matter accretion, which can only affect the outer part of the density profile represented by particles with $p$ smaller than or equal to $6$. 
Apart from the origin of the universal profile, this picture is consistent with previous studies that show that the accreting matter mainly accumulates in the outer region \citep[e.g., ][]{2001ApJ...557..533F,2003MNRAS.339...12Z,Wang2011}, and partly explains why the characteristic scale $S(p)$ in equation (\ref{eq:fitS}) is a decreasing function of $p$; particles with larger $p$ have accreted earlier and their distribution tends to be relaxed in the inner part of haloes. 
In this respect, the dynamics at the early stage of halo formation would clarify the origin of the double-power law nature. 

\section{Conclusions}
\label{sec:conclusions}

Due to its cold nature, CDM haloes inherently possess multi-stream regions, where the velocity of DM at a given position becomes multi-valued at a macroscopic level in the phase-space distribution. Recently, the outer boundary of the multi-stream regions called the splashback radius has attracted much attention, and there are numerous theoretical and observational studies to clarify its nature as well as to test the CDM paradigm. Beyond the splashback radius, there is a limited number of works to characterise the phase-space structure even in $N$-body simulations. 

In this paper, focusing mainly on the inner structure of CDM haloes in cosmological $N$-body simulations, we have quantified the phase-space distribution of DM particles in the multi-stream regions. Based on the methodology developed by \citet{2020MNRAS.493.2765S}, which is considered an extension of the SPARTA algorithm by \citet{2017ApJS..231....5D}, we have classified DM particles inside haloes at $z=0$ by the number of apocentre passages, which we denote by $p$. Making use of $1,001$ snapshots, the analysis with the improved identification for halo centres allows us to keep a precise track of the DM trajectories, and we successfully count the number of their apocentre passages up to $p=40$ in a robust manner over the halo mass range of $3\times 10^{11} \leq M_{\rm vir}/(h^{-1}\,M_\odot) \leq 1\times 10^{14}$. Provided the particle distribution classified by $p$, the multi-stream structure of haloes inside the splashback radius becomes clearly visible (see Figs.~\ref{fig:radialprop1}, \ref{fig:radialprop18}, and \ref{fig:radialprop235}), and we are able to analyse the individual density profile for each stream. In addition, by stacking a number of haloes over each mass bin, we have quantified the statistical nature of each stream profile in a more quantitative manner. 

Our important findings are summarised as follows:

\begin{enumerate}
\item  
The density profiles for particles having the same value of $p$ generally exhibit a double-power law nature, consisting of a shallow cusp with an asymptotic slope around $-1$ in the inner part and a steep density drop with the slope $\lesssim -7$ in the outer part. These features commonly appear over a wide range of halo mass.  

\item 
The analysis with stacked halo profiles reveals that the profiles of each stream can be accurately characterised by the fitting function in equation~\eqref{eq:stream}, which has fixed inner and outer slopes, $-1$ and $-8$, respectively. The characteristic density $A$ and the scale $S$ of this fitting function are given as a simple function of the number of apocentre passages $p$, with a weak halo mass dependence (see equations~\ref{eq:fitA} and \ref{eq:fitS}). Interestingly, summing up the function in equation~\eqref{eq:stream} over $p$, we can reconstruct the total density profiles consistent with the total profiles measured from the HR run, even beyond the resolution limit of the LR run used to calibrate the density $A$ and scale $S$. As a result, our prediction of the total density profile based on equation~\eqref{eq:stream} closely matches the best-fitted Einasto profile with an inner cusp of the slope $-2\sim-1$, which is slightly steeper than that of the NFW profile. 

\item The double-power law nature of each stream profile appears persistent not only in mass-selected haloes but also in haloes selected based on different criteria. While the functional form of the profile is described by equation~\eqref{eq:stream} in a universal manner, the characteristic density $A$ and scale $S$ of the double-power law function, which are both given as a function of $p$, exhibit an explicit dependence on the selection criterion. As an illustrative example, we re-calibrated these parameters in the halo samples divided by the concentration parameter $c_{\rm vir}$, and summarise their fitting formulas in equations (\ref{eq:fitA_c}) and (\ref{eq:fitS_c}). 

\item A class of self-similar solutions that describe the stationary accretion of DM under a spherical symmetry is compared to our simulation results, but fails to reproduce their radial multi-stream structure. In particular, the asymptotic slope of the stream profiles is predicted to be steeper than what we measured in the simulations and hence contradicts the model described by equation \eqref{eq:stream}. This remains true even when we introduce a non-zero angular momentum, suggesting that taking into account the dynamical complexities associated with halo accretion/merger history or relaxing the symmetry assumptions would be important. Nevertheless, tracing the DM particles having the same number of apocentre passages $p$ determined at $z=0$ to higher redshifts, we find that their density profiles have already converged well at an earlier time especially for larger values of $p$. This is consistent with previous numerical studies and indicates that the double-power law nature appears to have been established during an early accretion phase and remains stable.
\end{enumerate}

The universal features of haloes found in this paper are a direct consequence of the cold nature of dark matter and serve as valuable insights into the physical properties of CDM haloes. 
While this study has utilised $N$-body simulations and investigated the inner multi-stream structure up to $p=40$, 
recent developments in simulating collisionless self-gravitating systems through Vlasov-Poisson equations offer a promising way to further probe the phase-space structure~\citep{Yoshikawa_Yoshida_Umemura2013, Hahn_Angulo2016, Sousbie_Colombi2016}. 
This would provide a deeper understanding of the physics behind the universal features. Of particular interest would be to clarify the phase-space nature of the so-called prompt cusp, the central cusp of proto-haloes having the density slope of $-1.5$ that has formed quasi-instantaneously, which is first found by \citep{2010Ishiyama} and is recently analyzed in detail by \citet{2023DelosWhite} (see also \citet{Anderhalden2013, Ishiyama2014, Ogiya2016, Angulo2017, Ishiyama2020, Colombi2021, Ondaro-Mallea2023}).
Since the cusp can survive until the present time and can be a dominant site for dark matter annihilation radiation \citep{2022arXiv220911237D}, a more quantitative theoretical study of the prompt cusp would give a huge impact on the indirect dark matter search through the observations of annihilation radiation such as gamma-ray excess \citep{2023arXiv230713023D}. 

In order to search for observational evidence of this universality, the impact of baryonic feedback would be crucial, especially for galactic haloes. The AGN or supernova feedback are known to change the inner density structure, and hence they would alter the multi-stream structure in phase space. To investigate their quantitative impact, an analysis using simulations involving galaxy formation processes would be useful \citep[e.g.,][]{2016MNRAS.457.1931S,2018MNRAS.475..676S} and beneficial to understand the stability of the universal properties found in the dark matter only simulations.
In this respect, it is notable that recent cosmological hydrodynamical simulation shows the phase-space structure of dark halos can be inferred from those of stellar halos \citep{Genina2023}.
Further research on the correlation between stellar and dark components will be the basis for observational verification of the universal features we have discovered.

Finally, it is worth stressing that the present method to reveal multi-stream structures is generally applied to other particle-based simulation data, meaning that one can also scrutinize the radial phase-space structures for alternative dark matter models, including warm dark matter and self-interacting dark matter \citep[e.g.,][as recent progress]{Stucker2020,Banerjee2020,Stucker2022,Correa2022}. Since the nature of dark matter 
alters small-scale structure formation~\citep[e.g.,][for a review]{Bullock_Boylan-Kolchin2017}, there would certainly be differences in the inner multi-stream structures of dark matter haloes, which can be valuable observational probes to clarify the nature of dark matter. Thus, a further investigation based on the present method would be very important, and this is left to our future work.

\section*{Acknowledgements}
We thank St\'ephane Colombi and Takashi Hiramatsu for insightful suggestions and discussions, Shogo Ishikawa and Satoshi Tanaka for useful comments and discussions. This work was
supported in part by MEXT/JSPS KAKENHI Grant
Number JP19H00677 (TN), JP20H05861, JP21H01081
(AT and TN), and JP22K03634 (TN). We also acknowledge financial support from Japan Science and Technology Agency (JST) AIP Acceleration Research Grant Number JP20317829
(AT and TN). 
YE is also supported by JST, the establishment of university fellowships towards the creation of science technology innovation, Grant Number JPMJFS2123.
Numerical computations were carried out 
at Yukawa Institute Computer Facility, and Cray XC50 at
centre for Computational Astrophysics, National Astronomical Observatory of Japan.
Finally, YE gratefully thanks TAP colleagues and neighbors in the Kumano Dormitory for their warm support and valuable discussions.
\section*{Data Availability}

The data on which this study is based will be provided on request as appropriate.



\bibliographystyle{mnras}
\bibliography{202207library} 

\begin{thebibliography}{}
\makeatletter
\relax
\def\mn@urlcharsother{\let\do\@makeother \do\$\do\&\do\#\do\^\do\_\do\%\do\~}
\def\mn@doi{\begingroup\mn@urlcharsother \@ifnextchar [ {\mn@doi@}
  {\mn@doi@[]}}
\def\mn@doi@[#1]#2{\def\@tempa{#1}\ifx\@tempa\@empty \href
  {http://dx.doi.org/#2} {doi:#2}\else \href {http://dx.doi.org/#2} {#1}\fi
  \endgroup}
\def\mn@eprint#1#2{\mn@eprint@#1:#2::\@nil}
\def\mn@eprint@arXiv#1{\href {http://arxiv.org/abs/#1} {{\tt arXiv:#1}}}
\def\mn@eprint@dblp#1{\href {http://dblp.uni-trier.de/rec/bibtex/#1.xml}
  {dblp:#1}}
\def\mn@eprint@#1:#2:#3:#4\@nil{\def\@tempa {#1}\def\@tempb {#2}\def\@tempc
  {#3}\ifx \@tempc \@empty \let \@tempc \@tempb \let \@tempb \@tempa \fi \ifx
  \@tempb \@empty \def\@tempb {arXiv}\fi \@ifundefined
  {mn@eprint@\@tempb}{\@tempb:\@tempc}{\expandafter \expandafter \csname
  mn@eprint@\@tempb\endcsname \expandafter{\@tempc}}}

\bibitem[\protect\citeauthoryear{{Adhikari}, {Dalal}  \&
  {Chamberlain}}{{Adhikari} et~al.}{2014}]{2014JCAP...11..019A}
{Adhikari} S.,  {Dalal} N.,   {Chamberlain} R.~T.,  2014, \mn@doi [\jcap]
  {10.1088/1475-7516/2014/11/019}, \href
  {https://ui.adsabs.harvard.edu/abs/2014JCAP...11..019A} {2014, 019}

\bibitem[\protect\citeauthoryear{{Anderhalden} \& {Diemand}}{{Anderhalden} \&
  {Diemand}}{2013}]{Anderhalden2013}
{Anderhalden} D.,  {Diemand} J.,  2013, \mn@doi [\jcap]
  {10.1088/1475-7516/2013/04/009}, \href
  {https://ui.adsabs.harvard.edu/abs/2013JCAP...04..009A} {2013, 009}

\bibitem[\protect\citeauthoryear{{Angulo}, {Hahn}, {Ludlow}  \&
  {Bonoli}}{{Angulo} et~al.}{2017}]{Angulo2017}
{Angulo} R.~E.,  {Hahn} O.,  {Ludlow} A.~D.,   {Bonoli} S.,  2017, \mn@doi
  [\mnras] {10.1093/mnras/stx1658}, \href
  {https://ui.adsabs.harvard.edu/abs/2017MNRAS.471.4687A} {471, 4687}

\bibitem[\protect\citeauthoryear{{Banerjee}, {Adhikari}, {Dalal}, {More}  \&
  {Kravtsov}}{{Banerjee} et~al.}{2020}]{Banerjee2020}
{Banerjee} A.,  {Adhikari} S.,  {Dalal} N.,  {More} S.,   {Kravtsov} A.,  2020,
  \mn@doi [\jcap] {10.1088/1475-7516/2020/02/024}, \href
  {https://ui.adsabs.harvard.edu/abs/2020JCAP...02..024B} {2020, 024}

\bibitem[\protect\citeauthoryear{{Baxter} et~al.,}{{Baxter}
  et~al.}{2017}]{2017ApJ...841...18B}
{Baxter} E.,  et~al., 2017, \mn@doi [\apj] {10.3847/1538-4357/aa6ff0}, \href
  {https://ui.adsabs.harvard.edu/abs/2017ApJ...841...18B} {841, 18}

\bibitem[\protect\citeauthoryear{{Behroozi}, {Wechsler}  \& {Wu}}{{Behroozi}
  et~al.}{2013}]{2013ApJ...762..109B}
{Behroozi} P.~S.,  {Wechsler} R.~H.,   {Wu} H.-Y.,  2013, \mn@doi [\apj]
  {10.1088/0004-637X/762/2/109}, \href
  {https://ui.adsabs.harvard.edu/abs/2013ApJ...762..109B} {762, 109}

\bibitem[\protect\citeauthoryear{{Bertschinger}}{{Bertschinger}}{1985}]{1985ApJS...58...39B}
{Bertschinger} E.,  1985, \mn@doi [\apjs] {10.1086/191028}, \href
  {https://ui.adsabs.harvard.edu/abs/1985ApJS...58...39B} {58, 39}

\bibitem[\protect\citeauthoryear{{Binney} \& {Tremaine}}{{Binney} \&
  {Tremaine}}{2008}]{2008gady.book.....B}
{Binney} J.,  {Tremaine} S.,  2008, {Galactic Dynamics: Second Edition}

\bibitem[\protect\citeauthoryear{{Blas}, {Lesgourgues}  \& {Tram}}{{Blas}
  et~al.}{2011}]{Blas2011}
{Blas} D.,  {Lesgourgues} J.,   {Tram} T.,  2011, \mn@doi [\jcap]
  {10.1088/1475-7516/2011/07/034}, \href
  {https://ui.adsabs.harvard.edu/abs/2011JCAP...07..034B} {2011, 034}

\bibitem[\protect\citeauthoryear{{Bryan} \& {Norman}}{{Bryan} \&
  {Norman}}{1998}]{1998ApJ...495...80B}
{Bryan} G.~L.,  {Norman} M.~L.,  1998, \mn@doi [\apj] {10.1086/305262}, \href
  {https://ui.adsabs.harvard.edu/abs/1998ApJ...495...80B} {495, 80}

\bibitem[\protect\citeauthoryear{{Bullock} \& {Boylan-Kolchin}}{{Bullock} \&
  {Boylan-Kolchin}}{2017}]{Bullock_Boylan-Kolchin2017}
{Bullock} J.~S.,  {Boylan-Kolchin} M.,  2017, \mn@doi [\araa]
  {10.1146/annurev-astro-091916-055313}, \href
  {https://ui.adsabs.harvard.edu/abs/2017ARA&A..55..343B} {55, 343}

\bibitem[\protect\citeauthoryear{{Colombi}}{{Colombi}}{2021}]{Colombi2021}
{Colombi} S.,  2021, \mn@doi [\aap] {10.1051/0004-6361/202039719}, \href
  {https://ui.adsabs.harvard.edu/abs/2021A&A...647A..66C} {647, A66}

\bibitem[\protect\citeauthoryear{{Contigiani}, {Hoekstra}  \&
  {Bah{\'e}}}{{Contigiani} et~al.}{2019}]{2019MNRAS.485..408C}
{Contigiani} O.,  {Hoekstra} H.,   {Bah{\'e}} Y.~M.,  2019, \mn@doi [\mnras]
  {10.1093/mnras/stz404}, \href
  {https://ui.adsabs.harvard.edu/abs/2019MNRAS.485..408C} {485, 408}

\bibitem[\protect\citeauthoryear{{Correa}, {Schaller}, {Ploeckinger}, {Anau
  Montel}, {Weniger}  \& {Ando}}{{Correa} et~al.}{2022}]{Correa2022}
{Correa} C.~A.,  {Schaller} M.,  {Ploeckinger} S.,  {Anau Montel} N.,
  {Weniger} C.,   {Ando} S.,  2022, \mn@doi [\mnras] {10.1093/mnras/stac2830},
  \href {https://ui.adsabs.harvard.edu/abs/2022MNRAS.517.3045C} {517, 3045}

\bibitem[\protect\citeauthoryear{{Crocce}, {Pueblas}  \&
  {Scoccimarro}}{{Crocce} et~al.}{2006}]{Crocce2006}
{Crocce} M.,  {Pueblas} S.,   {Scoccimarro} R.,  2006, \mn@doi [\mnras]
  {10.1111/j.1365-2966.2006.11040.x}, \href
  {https://ui.adsabs.harvard.edu/abs/2006MNRAS.373..369C} {373, 369}

\bibitem[\protect\citeauthoryear{{Dalal}, {Lithwick}  \& {Kuhlen}}{{Dalal}
  et~al.}{2010}]{Dalal2010}
{Dalal} N.,  {Lithwick} Y.,   {Kuhlen} M.,  2010, arXiv e-prints, \href
  {https://ui.adsabs.harvard.edu/abs/2010arXiv1010.2539D} {p. arXiv:1010.2539}

\bibitem[\protect\citeauthoryear{{Delos} \& {White}}{{Delos} \&
  {White}}{2022}]{2022arXiv220911237D}
{Delos} M.~S.,  {White} S. D.~M.,  2022, \mn@doi [arXiv e-prints]
  {10.48550/arXiv.2209.11237}, \href
  {https://ui.adsabs.harvard.edu/abs/2022arXiv220911237D} {p. arXiv:2209.11237}

\bibitem[\protect\citeauthoryear{{Delos} \& {White}}{{Delos} \&
  {White}}{2023}]{2023DelosWhite}
{Delos} M.~S.,  {White} S. D.~M.,  2023, \mn@doi [\mnras]
  {10.1093/mnras/stac3373}, \href
  {https://ui.adsabs.harvard.edu/abs/2023MNRAS.518.3509D} {518, 3509}

\bibitem[\protect\citeauthoryear{{Delos}, {Korsmeier}, {Widmark}, {Blanco},
  {Linden}  \& {White}}{{Delos} et~al.}{2023}]{2023arXiv230713023D}
{Delos} M.~S.,  {Korsmeier} M.,  {Widmark} A.,  {Blanco} C.,  {Linden} T.,
  {White} S. D.~M.,  2023, \mn@doi [arXiv e-prints]
  {10.48550/arXiv.2307.13023}, \href
  {https://ui.adsabs.harvard.edu/abs/2023arXiv230713023D} {p. arXiv:2307.13023}

\bibitem[\protect\citeauthoryear{{Diemer}}{{Diemer}}{2017}]{2017ApJS..231....5D}
{Diemer} B.,  2017, \mn@doi [\apjs] {10.3847/1538-4365/aa799c}, \href
  {https://ui.adsabs.harvard.edu/abs/2017ApJS..231....5D} {231, 5}

\bibitem[\protect\citeauthoryear{{Diemer}}{{Diemer}}{2022}]{Diemer2022}
{Diemer} B.,  2022, \mn@doi [\mnras] {10.1093/mnras/stac878}, \href
  {https://ui.adsabs.harvard.edu/abs/2022MNRAS.513..573D} {513, 573}

\bibitem[\protect\citeauthoryear{{Diemer} \& {Kravtsov}}{{Diemer} \&
  {Kravtsov}}{2014}]{2014ApJ...789....1D}
{Diemer} B.,  {Kravtsov} A.~V.,  2014, \mn@doi [\apj]
  {10.1088/0004-637X/789/1/1}, \href
  {https://ui.adsabs.harvard.edu/abs/2014ApJ...789....1D} {789, 1}

\bibitem[\protect\citeauthoryear{{Dutton} \& {Macci{\`o}}}{{Dutton} \&
  {Macci{\`o}}}{2014}]{2014MNRAS.441.3359D}
{Dutton} A.~A.,  {Macci{\`o}} A.~V.,  2014, \mn@doi [\mnras]
  {10.1093/mnras/stu742}, \href
  {https://ui.adsabs.harvard.edu/abs/2014MNRAS.441.3359D} {441, 3359}

\bibitem[\protect\citeauthoryear{Einasto}{Einasto}{1965}]{einasto1965construction}
Einasto J.,  1965, Trudy Astrofizicheskogo Instituta Alma-Ata, 5, 87

\bibitem[\protect\citeauthoryear{{Enomoto}, {Nishimichi}  \&
  {Taruya}}{{Enomoto} et~al.}{2023}]{2023ApJ...950L..13E}
{Enomoto} Y.,  {Nishimichi} T.,   {Taruya} A.,  2023, \mn@doi [\apjl]
  {10.3847/2041-8213/acd7ee}, \href
  {https://ui.adsabs.harvard.edu/abs/2023ApJ...950L..13E} {950, L13}

\bibitem[\protect\citeauthoryear{{Fillmore} \& {Goldreich}}{{Fillmore} \&
  {Goldreich}}{1984}]{1984ApJ...281....1F}
{Fillmore} J.~A.,  {Goldreich} P.,  1984, \mn@doi [\apj] {10.1086/162070},
  \href {https://ui.adsabs.harvard.edu/abs/1984ApJ...281....1F} {281, 1}

\bibitem[\protect\citeauthoryear{{Fukushige} \& {Makino}}{{Fukushige} \&
  {Makino}}{2001}]{2001ApJ...557..533F}
{Fukushige} T.,  {Makino} J.,  2001, \mn@doi [\apj] {10.1086/321666}, \href
  {https://ui.adsabs.harvard.edu/abs/2001ApJ...557..533F} {557, 533}

\bibitem[\protect\citeauthoryear{{Gao}, {Navarro}, {Cole}, {Frenk}, {White},
  {Springel}, {Jenkins}  \& {Neto}}{{Gao} et~al.}{2008}]{2008MNRAS.387..536G}
{Gao} L.,  {Navarro} J.~F.,  {Cole} S.,  {Frenk} C.~S.,  {White} S. D.~M.,
  {Springel} V.,  {Jenkins} A.,   {Neto} A.~F.,  2008, \mn@doi [\mnras]
  {10.1111/j.1365-2966.2008.13277.x}, \href
  {https://ui.adsabs.harvard.edu/abs/2008MNRAS.387..536G} {387, 536}

\bibitem[\protect\citeauthoryear{{Genina}, {Deason}  \& {Frenk}}{{Genina}
  et~al.}{2023}]{Genina2023}
{Genina} A.,  {Deason} A.~J.,   {Frenk} C.~S.,  2023, \mn@doi [\mnras]
  {10.1093/mnras/stad397}, \href
  {https://ui.adsabs.harvard.edu/abs/2023MNRAS.520.3767G} {520, 3767}

\bibitem[\protect\citeauthoryear{{Hahn} \& {Angulo}}{{Hahn} \&
  {Angulo}}{2016}]{Hahn_Angulo2016}
{Hahn} O.,  {Angulo} R.~E.,  2016, \mn@doi [\mnras] {10.1093/mnras/stv2304},
  \href {https://ui.adsabs.harvard.edu/abs/2016MNRAS.455.1115H} {455, 1115}

\bibitem[\protect\citeauthoryear{{Ishiyama}}{{Ishiyama}}{2014}]{Ishiyama2014}
{Ishiyama} T.,  2014, \mn@doi [\apj] {10.1088/0004-637X/788/1/27}, \href
  {https://ui.adsabs.harvard.edu/abs/2014ApJ...788...27I} {788, 27}

\bibitem[\protect\citeauthoryear{{Ishiyama} \& {Ando}}{{Ishiyama} \&
  {Ando}}{2020}]{Ishiyama2020}
{Ishiyama} T.,  {Ando} S.,  2020, \mn@doi [\mnras] {10.1093/mnras/staa069},
  \href {https://ui.adsabs.harvard.edu/abs/2020MNRAS.492.3662I} {492, 3662}

\bibitem[\protect\citeauthoryear{{Ishiyama}, {Fukushige}  \&
  {Makino}}{{Ishiyama} et~al.}{2009}]{2009PASJ...61.1319I}
{Ishiyama} T.,  {Fukushige} T.,   {Makino} J.,  2009, \mn@doi [\pasj]
  {10.1093/pasj/61.6.1319}, \href
  {https://ui.adsabs.harvard.edu/abs/2009PASJ...61.1319I} {61, 1319}

\bibitem[\protect\citeauthoryear{{Ishiyama}, {Makino}  \&
  {Ebisuzaki}}{{Ishiyama} et~al.}{2010}]{2010Ishiyama}
{Ishiyama} T.,  {Makino} J.,   {Ebisuzaki} T.,  2010, \mn@doi [\apjl]
  {10.1088/2041-8205/723/2/L195}, \href
  {https://ui.adsabs.harvard.edu/abs/2010ApJ...723L.195I} {723, L195}

\bibitem[\protect\citeauthoryear{{Ishiyama}, {Nitadori}  \&
  {Makino}}{{Ishiyama} et~al.}{2012}]{2012arXiv1211.4406I}
{Ishiyama} T.,  {Nitadori} K.,   {Makino} J.,  2012, arXiv e-prints, \href
  {https://ui.adsabs.harvard.edu/abs/2012arXiv1211.4406I} {p. arXiv:1211.4406}

\bibitem[\protect\citeauthoryear{{Iwasawa}, {Tanikawa}, {Hosono}, {Nitadori},
  {Muranushi}  \& {Makino}}{{Iwasawa} et~al.}{2016}]{2016PASJ...68...54I}
{Iwasawa} M.,  {Tanikawa} A.,  {Hosono} N.,  {Nitadori} K.,  {Muranushi} T.,
  {Makino} J.,  2016, \mn@doi [\pasj] {10.1093/pasj/psw053}, \href
  {https://ui.adsabs.harvard.edu/abs/2016PASJ...68...54I} {68, 54}

\bibitem[\protect\citeauthoryear{{Klypin}, {Trujillo-Gomez}  \&
  {Primack}}{{Klypin} et~al.}{2011}]{Klypin_etal2011}
{Klypin} A.~A.,  {Trujillo-Gomez} S.,   {Primack} J.,  2011, \mn@doi [\apj]
  {10.1088/0004-637X/740/2/102}, \href
  {https://ui.adsabs.harvard.edu/abs/2011ApJ...740..102K} {740, 102}

\bibitem[\protect\citeauthoryear{{Klypin}, {Yepes}, {Gottl{\"o}ber}, {Prada}
  \& {He{\ss}}}{{Klypin} et~al.}{2016}]{2016MNRAS.457.4340K}
{Klypin} A.,  {Yepes} G.,  {Gottl{\"o}ber} S.,  {Prada} F.,   {He{\ss}} S.,
  2016, \mn@doi [\mnras] {10.1093/mnras/stw248}, \href
  {https://ui.adsabs.harvard.edu/abs/2016MNRAS.457.4340K} {457, 4340}

\bibitem[\protect\citeauthoryear{{Lithwick} \& {Dalal}}{{Lithwick} \&
  {Dalal}}{2011}]{2011ApJ...734..100L}
{Lithwick} Y.,  {Dalal} N.,  2011, \mn@doi [\apj]
  {10.1088/0004-637X/734/2/100}, \href
  {https://ui.adsabs.harvard.edu/abs/2011ApJ...734..100L} {734, 100}

\bibitem[\protect\citeauthoryear{{Lu}, {Mo}, {Katz}  \& {Weinberg}}{{Lu}
  et~al.}{2006}]{Lu2006}
{Lu} Y.,  {Mo} H.~J.,  {Katz} N.,   {Weinberg} M.~D.,  2006, \mn@doi [\mnras]
  {10.1111/j.1365-2966.2006.10270.x}, \href
  {https://ui.adsabs.harvard.edu/abs/2006MNRAS.368.1931L} {368, 1931}

\bibitem[\protect\citeauthoryear{{Ludlow}, {Navarro}, {Angulo},
  {Boylan-Kolchin}, {Springel}, {Frenk}  \& {White}}{{Ludlow}
  et~al.}{2014}]{2014MNRAS.441..378L}
{Ludlow} A.~D.,  {Navarro} J.~F.,  {Angulo} R.~E.,  {Boylan-Kolchin} M.,
  {Springel} V.,  {Frenk} C.,   {White} S. D.~M.,  2014, \mn@doi [\mnras]
  {10.1093/mnras/stu483}, \href
  {https://ui.adsabs.harvard.edu/abs/2014MNRAS.441..378L} {441, 378}

\bibitem[\protect\citeauthoryear{{More} et~al.,}{{More}
  et~al.}{2016}]{2016ApJ...825...39M}
{More} S.,  et~al., 2016, \mn@doi [\apj] {10.3847/0004-637X/825/1/39}, \href
  {https://ui.adsabs.harvard.edu/abs/2016ApJ...825...39M} {825, 39}

\bibitem[\protect\citeauthoryear{{Namekata} et~al.,}{{Namekata}
  et~al.}{2018}]{2018PASJ...70...70N}
{Namekata} D.,  et~al., 2018, \mn@doi [\pasj] {10.1093/pasj/psy062}, \href
  {https://ui.adsabs.harvard.edu/abs/2018PASJ...70...70N} {70, 70}

\bibitem[\protect\citeauthoryear{{Navarro}, {Frenk}  \& {White}}{{Navarro}
  et~al.}{1997}]{1997ApJ...490..493N}
{Navarro} J.~F.,  {Frenk} C.~S.,   {White} S. D.~M.,  1997, \mn@doi [\apj]
  {10.1086/304888}, \href
  {https://ui.adsabs.harvard.edu/abs/1997ApJ...490..493N} {490, 493}

\bibitem[\protect\citeauthoryear{{Navarro} et~al.,}{{Navarro}
  et~al.}{2004}]{2004MNRAS.349.1039N}
{Navarro} J.~F.,  et~al., 2004, \mn@doi [\mnras]
  {10.1111/j.1365-2966.2004.07586.x}, \href
  {https://ui.adsabs.harvard.edu/abs/2004MNRAS.349.1039N} {349, 1039}

\bibitem[\protect\citeauthoryear{Nitadori, Makino  \& Hut}{Nitadori
  et~al.}{2006}]{Nitadori2006-ek}
Nitadori K.,  Makino J.,   Hut P.,  2006, \mn@doi [New Astron.]
  {10.1016/j.newast.2006.07.007}, 12, 169

\bibitem[\protect\citeauthoryear{{Nusser}}{{Nusser}}{2001}]{2001MNRAS.325.1397N}
{Nusser} A.,  2001, \mn@doi [\mnras] {10.1046/j.1365-8711.2001.04527.x}, \href
  {https://ui.adsabs.harvard.edu/abs/2001MNRAS.325.1397N} {325, 1397}

\bibitem[\protect\citeauthoryear{{Ogiya}, {Nagai}  \& {Ishiyama}}{{Ogiya}
  et~al.}{2016}]{Ogiya2016}
{Ogiya} G.,  {Nagai} D.,   {Ishiyama} T.,  2016, \mn@doi [\mnras]
  {10.1093/mnras/stw1551}, \href
  {https://ui.adsabs.harvard.edu/abs/2016MNRAS.461.3385O} {461, 3385}

\bibitem[\protect\citeauthoryear{{Ondaro-Mallea}, {Angulo}, {St{\"u}cker},
  {Hahn}  \& {White}}{{Ondaro-Mallea} et~al.}{2023}]{Ondaro-Mallea2023}
{Ondaro-Mallea} L.,  {Angulo} R.~E.,  {St{\"u}cker} J.,  {Hahn} O.,   {White}
  S. D.~M.,  2023, \mn@doi [arXiv e-prints] {10.48550/arXiv.2309.05707}, \href
  {https://ui.adsabs.harvard.edu/abs/2023arXiv230905707O} {p. arXiv:2309.05707}

\bibitem[\protect\citeauthoryear{{Planck Collaboration} et~al.,}{{Planck
  Collaboration} et~al.}{2016}]{2016A&A...594A..13P}
{Planck Collaboration} et~al., 2016, \mn@doi [\aap]
  {10.1051/0004-6361/201525830}, \href
  {https://ui.adsabs.harvard.edu/abs/2016A&A...594A..13P} {594, A13}

\bibitem[\protect\citeauthoryear{{Rees} \& {Ostriker}}{{Rees} \&
  {Ostriker}}{1977}]{1977MNRAS.179..541R}
{Rees} M.~J.,  {Ostriker} J.~P.,  1977, \mn@doi [\mnras]
  {10.1093/mnras/179.4.541}, \href
  {https://ui.adsabs.harvard.edu/abs/1977MNRAS.179..541R} {179, 541}

\bibitem[\protect\citeauthoryear{{Ryden}}{{Ryden}}{1993}]{1993ApJ...418....4R}
{Ryden} B.~S.,  1993, \mn@doi [\apj] {10.1086/173365}, \href
  {https://ui.adsabs.harvard.edu/abs/1993ApJ...418....4R} {418, 4}

\bibitem[\protect\citeauthoryear{{Sawala} et~al.,}{{Sawala}
  et~al.}{2016}]{2016MNRAS.457.1931S}
{Sawala} T.,  et~al., 2016, \mn@doi [\mnras] {10.1093/mnras/stw145}, \href
  {https://ui.adsabs.harvard.edu/abs/2016MNRAS.457.1931S} {457, 1931}

\bibitem[\protect\citeauthoryear{{Scoccimarro}}{{Scoccimarro}}{1998}]{Scoccimarro1998}
{Scoccimarro} R.,  1998, \mn@doi [\mnras] {10.1046/j.1365-8711.1998.01845.x},
  \href {https://ui.adsabs.harvard.edu/abs/1998MNRAS.299.1097S} {299, 1097}

\bibitem[\protect\citeauthoryear{{Shin} et~al.,}{{Shin}
  et~al.}{2021}]{Shin2021}
{Shin} T.,  et~al., 2021, \mn@doi [\mnras] {10.1093/mnras/stab2505}, \href
  {https://ui.adsabs.harvard.edu/abs/2021MNRAS.507.5758S} {507, 5758}

\bibitem[\protect\citeauthoryear{{Sikivie}, {Tkachev}  \& {Wang}}{{Sikivie}
  et~al.}{1997}]{1997PhRvD..56.1863S}
{Sikivie} P.,  {Tkachev} I.~I.,   {Wang} Y.,  1997, \mn@doi [\prd]
  {10.1103/PhysRevD.56.1863}, \href
  {https://ui.adsabs.harvard.edu/abs/1997PhRvD..56.1863S} {56, 1863}

\bibitem[\protect\citeauthoryear{{Sousbie} \& {Colombi}}{{Sousbie} \&
  {Colombi}}{2016}]{Sousbie_Colombi2016}
{Sousbie} T.,  {Colombi} S.,  2016, \mn@doi [Journal of Computational Physics]
  {10.1016/j.jcp.2016.05.048}, \href
  {https://ui.adsabs.harvard.edu/abs/2016JCoPh.321..644S} {321, 644}

\bibitem[\protect\citeauthoryear{{Springel} et~al.,}{{Springel}
  et~al.}{2018}]{2018MNRAS.475..676S}
{Springel} V.,  et~al., 2018, \mn@doi [\mnras] {10.1093/mnras/stx3304}, \href
  {https://ui.adsabs.harvard.edu/abs/2018MNRAS.475..676S} {475, 676}

\bibitem[\protect\citeauthoryear{{St{\"u}cker}, {Hahn}, {Angulo}  \&
  {White}}{{St{\"u}cker} et~al.}{2020}]{Stucker2020}
{St{\"u}cker} J.,  {Hahn} O.,  {Angulo} R.~E.,   {White} S. D.~M.,  2020,
  \mn@doi [\mnras] {10.1093/mnras/staa1468}, \href
  {https://ui.adsabs.harvard.edu/abs/2020MNRAS.495.4943S} {495, 4943}

\bibitem[\protect\citeauthoryear{{St{\"u}cker}, {Angulo}, {Hahn}  \&
  {White}}{{St{\"u}cker} et~al.}{2022}]{Stucker2022}
{St{\"u}cker} J.,  {Angulo} R.~E.,  {Hahn} O.,   {White} S. D.~M.,  2022,
  \mn@doi [\mnras] {10.1093/mnras/stab3078}, \href
  {https://ui.adsabs.harvard.edu/abs/2022MNRAS.509.1703S} {509, 1703}

\bibitem[\protect\citeauthoryear{{Sugiura}, {Nishimichi}, {Rasera}  \&
  {Taruya}}{{Sugiura} et~al.}{2020}]{2020MNRAS.493.2765S}
{Sugiura} H.,  {Nishimichi} T.,  {Rasera} Y.,   {Taruya} A.,  2020, \mn@doi
  [\mnras] {10.1093/mnras/staa413}, \href
  {https://ui.adsabs.harvard.edu/abs/2020MNRAS.493.2765S} {493, 2765}

\bibitem[\protect\citeauthoryear{{Tanikawa}, {Yoshikawa}, {Okamoto}  \&
  {Nitadori}}{{Tanikawa} et~al.}{2012}]{2012NewA...17...82T}
{Tanikawa} A.,  {Yoshikawa} K.,  {Okamoto} T.,   {Nitadori} K.,  2012, \mn@doi
  [\na] {10.1016/j.newast.2011.07.001}, \href
  {https://ui.adsabs.harvard.edu/abs/2012NewA...17...82T} {17, 82}

\bibitem[\protect\citeauthoryear{{Tanikawa}, {Yoshikawa}, {Nitadori}  \&
  {Okamoto}}{{Tanikawa} et~al.}{2013}]{2013NewA...19...74T}
{Tanikawa} A.,  {Yoshikawa} K.,  {Nitadori} K.,   {Okamoto} T.,  2013, \mn@doi
  [\na] {10.1016/j.newast.2012.08.009}, \href
  {https://ui.adsabs.harvard.edu/abs/2013NewA...19...74T} {19, 74}

\bibitem[\protect\citeauthoryear{{Vogelsberger} \& {White}}{{Vogelsberger} \&
  {White}}{2011}]{Vogelsberger2011}
{Vogelsberger} M.,  {White} S. D.~M.,  2011, \mn@doi [\mnras]
  {10.1111/j.1365-2966.2011.18224.x}, \href
  {https://ui.adsabs.harvard.edu/abs/2011MNRAS.413.1419V} {413, 1419}

\bibitem[\protect\citeauthoryear{{Vogelsberger}, {White}, {Mohayaee}  \&
  {Springel}}{{Vogelsberger} et~al.}{2009}]{CausticVogelsberger2009}
{Vogelsberger} M.,  {White} S. D.~M.,  {Mohayaee} R.,   {Springel} V.,  2009,
  \mn@doi [\mnras] {10.1111/j.1365-2966.2009.15615.x}, \href
  {https://ui.adsabs.harvard.edu/abs/2009MNRAS.400.2174V} {400, 2174}

\bibitem[\protect\citeauthoryear{{Wang} et~al.,}{{Wang}
  et~al.}{2011}]{Wang2011}
{Wang} J.,  et~al., 2011, \mn@doi [\mnras] {10.1111/j.1365-2966.2011.18220.x},
  \href {https://ui.adsabs.harvard.edu/abs/2011MNRAS.413.1373W} {413, 1373}

\bibitem[\protect\citeauthoryear{{Wang}, {Bose}, {Frenk}, {Gao}, {Jenkins},
  {Springel}  \& {White}}{{Wang} et~al.}{2020}]{2020Natur.585...39W}
{Wang} J.,  {Bose} S.,  {Frenk} C.~S.,  {Gao} L.,  {Jenkins} A.,  {Springel}
  V.,   {White} S.~D.~M.,  2020, \mn@doi [\nat] {10.1038/s41586-020-2642-9},
  \href {https://ui.adsabs.harvard.edu/abs/2020Natur.585...39W} {585, 39}

\bibitem[\protect\citeauthoryear{{White} \& {Rees}}{{White} \&
  {Rees}}{1978}]{1978MNRAS.183..341W}
{White} S.~D.~M.,  {Rees} M.~J.,  1978, \mn@doi [\mnras]
  {10.1093/mnras/183.3.341}, \href
  {https://ui.adsabs.harvard.edu/abs/1978MNRAS.183..341W} {183, 341}

\bibitem[\protect\citeauthoryear{{White} \& {Vogelsberger}}{{White} \&
  {Vogelsberger}}{2009}]{CausticsWhite2009}
{White} S. D.~M.,  {Vogelsberger} M.,  2009, \mn@doi [\mnras]
  {10.1111/j.1365-2966.2008.14038.x}, \href
  {https://ui.adsabs.harvard.edu/abs/2009MNRAS.392..281W} {392, 281}

\bibitem[\protect\citeauthoryear{{Xhakaj}, {Diemer}, {Leauthaud}, {Wasserman},
  {Huang}, {Luo}, {Adhikari}  \& {Singh}}{{Xhakaj}
  et~al.}{2020}]{2020MNRAS.499.3534X}
{Xhakaj} E.,  {Diemer} B.,  {Leauthaud} A.,  {Wasserman} A.,  {Huang} S.,
  {Luo} Y.,  {Adhikari} S.,   {Singh} S.,  2020, \mn@doi [\mnras]
  {10.1093/mnras/staa3046}, \href
  {https://ui.adsabs.harvard.edu/abs/2020MNRAS.499.3534X} {499, 3534}

\bibitem[\protect\citeauthoryear{Yoshikawa \& Fukushige}{Yoshikawa \&
  Fukushige}{2005}]{Yoshikawa_2005}
Yoshikawa K.,  Fukushige T.,  2005, \mn@doi [Publications of the Astronomical
  Society of Japan] {10.1093/pasj/57.6.849}, 57, 849–860

\bibitem[\protect\citeauthoryear{{Yoshikawa}, {Yoshida}  \&
  {Umemura}}{{Yoshikawa} et~al.}{2013}]{Yoshikawa_Yoshida_Umemura2013}
{Yoshikawa} K.,  {Yoshida} N.,   {Umemura} M.,  2013, \mn@doi [\apj]
  {10.1088/0004-637X/762/2/116}, \href
  {https://ui.adsabs.harvard.edu/abs/2013ApJ...762..116Y} {762, 116}

\bibitem[\protect\citeauthoryear{{Zhao}, {Mo}, {Jing}  \& {B{\"o}rner}}{{Zhao}
  et~al.}{2003}]{2003MNRAS.339...12Z}
{Zhao} D.~H.,  {Mo} H.~J.,  {Jing} Y.~P.,   {B{\"o}rner} G.,  2003, \mn@doi
  [\mnras] {10.1046/j.1365-8711.2003.06135.x}, \href
  {https://ui.adsabs.harvard.edu/abs/2003MNRAS.339...12Z} {339, 12}

\bibitem[\protect\citeauthoryear{{Zukin} \& {Bertschinger}}{{Zukin} \&
  {Bertschinger}}{2010}]{2010PhRvD..82j4044Z}
{Zukin} P.,  {Bertschinger} E.,  2010, \mn@doi [\prd]
  {10.1103/PhysRevD.82.104044}, \href
  {https://ui.adsabs.harvard.edu/abs/2010PhRvD..82j4044Z} {82, 104044}

\makeatother
\end{thebibliography}




\appendix

\section{Convergence tests I: the number of particles determining the centres and the number of snapshots}\label{app:1}
In this Appendix, we demonstrate that the number of particles used to determine the centres of halo progenitors,$\boldsymbol{x}_\mathrm{h}$ and $\boldsymbol{v}_\mathrm{h}$, as well as the number of snapshots employed, do not affect our results. 
As introduced in Section~\ref{subsec:halo_centre}, we define $\boldsymbol{x}_\mathrm{h}$ and $\boldsymbol{v}_\mathrm{h}$ as the average position and velocity of $1000$ particles in the progenitor as a fiducial choice.
Here, the number of particles used to determine $\boldsymbol{x}_\mathrm{h}$ and $\boldsymbol{v}_\mathrm{h}$ (hereafter denoted as $N_\mathrm{det}$) can be arbitrary. Therefore, we investigate the effects of varying $N_\mathrm{det}$ by employing values of $500$ and $2000$ in addition to the fiducial value of $1000$.

\begin{figure*}
    \begin{tabular}{ccc}
        \begin{minipage}{0.66\columnwidth}
        \centering            \includegraphics[scale=0.29]{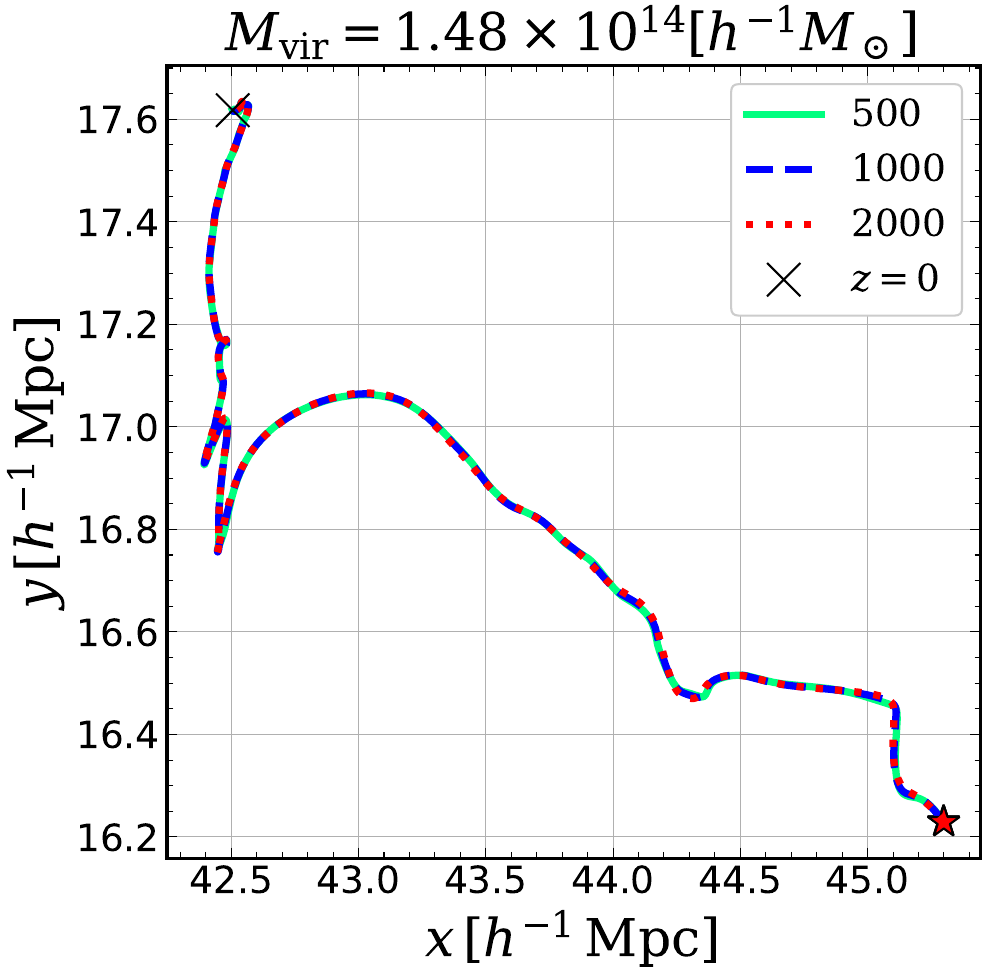}
        \end{minipage} &
        \begin{minipage}{0.66\columnwidth}
        \centering
        \includegraphics[scale=0.29]{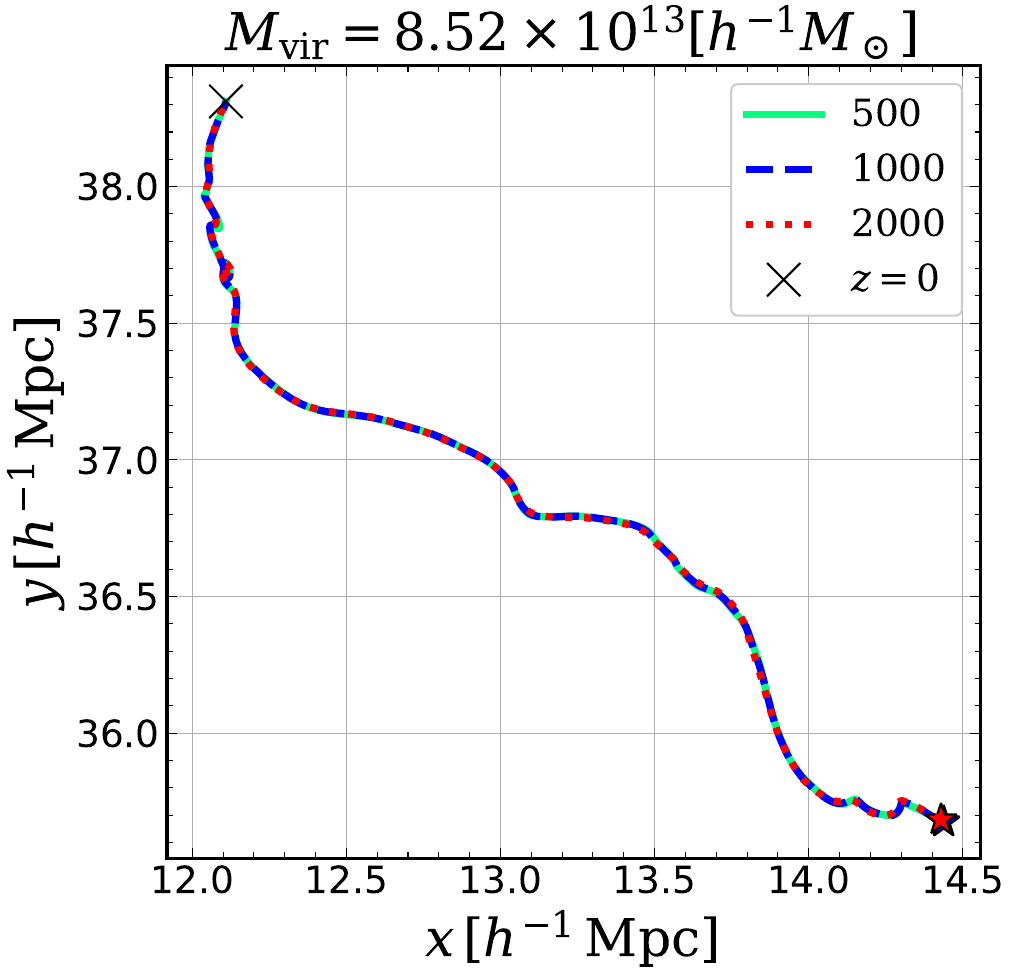}
        \end{minipage} &
        \begin{minipage}{0.66\columnwidth}
        \centering            \includegraphics[scale=0.29]{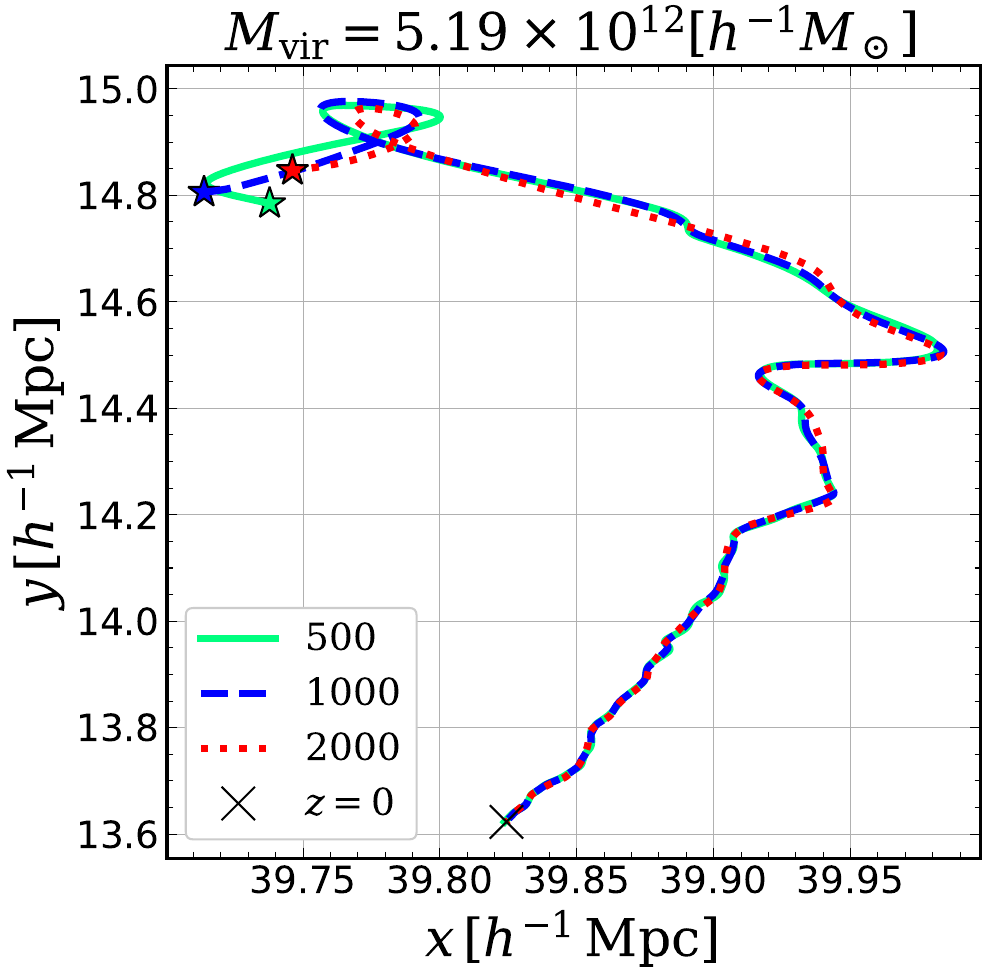}
        \end{minipage} \\
        \begin{minipage}{0.66\columnwidth}
        \centering            \includegraphics[scale=0.29]{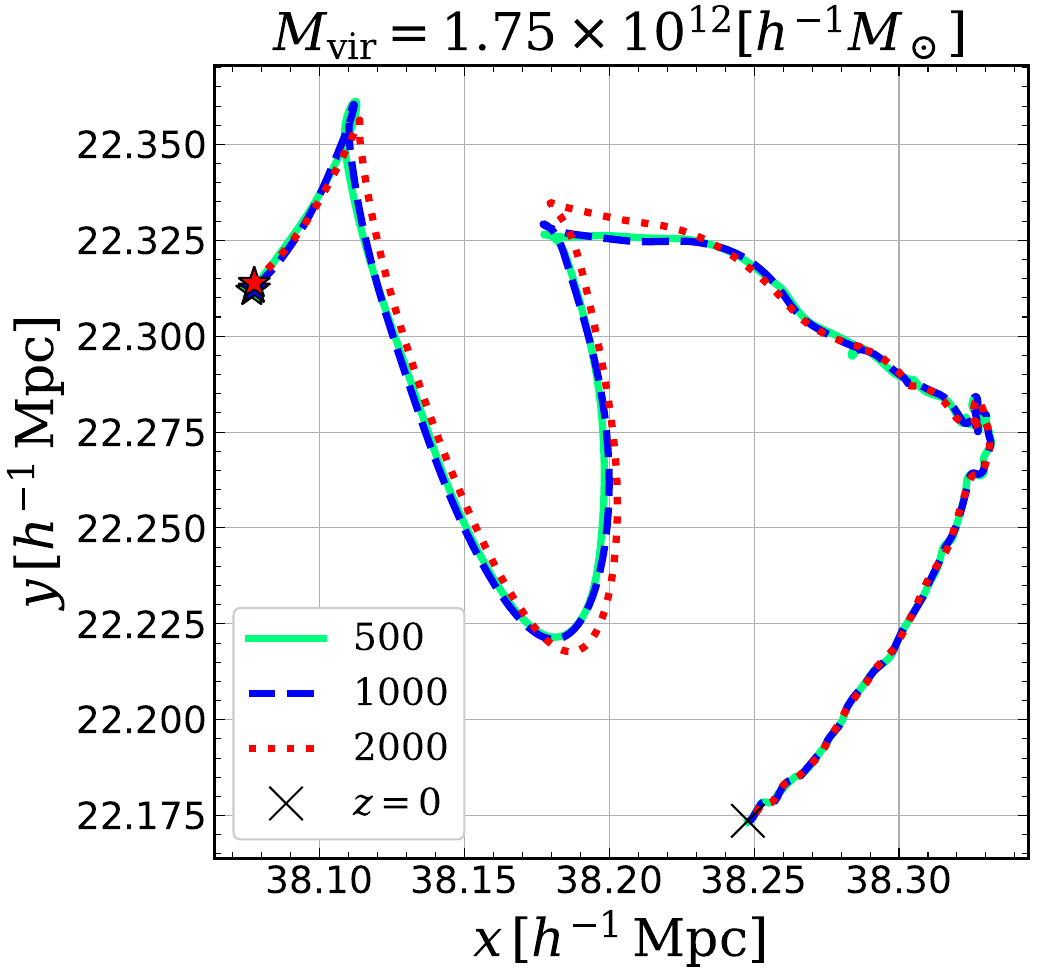}
        \end{minipage} &
        \begin{minipage}{0.66\columnwidth}
        \centering            \includegraphics[scale=0.29]{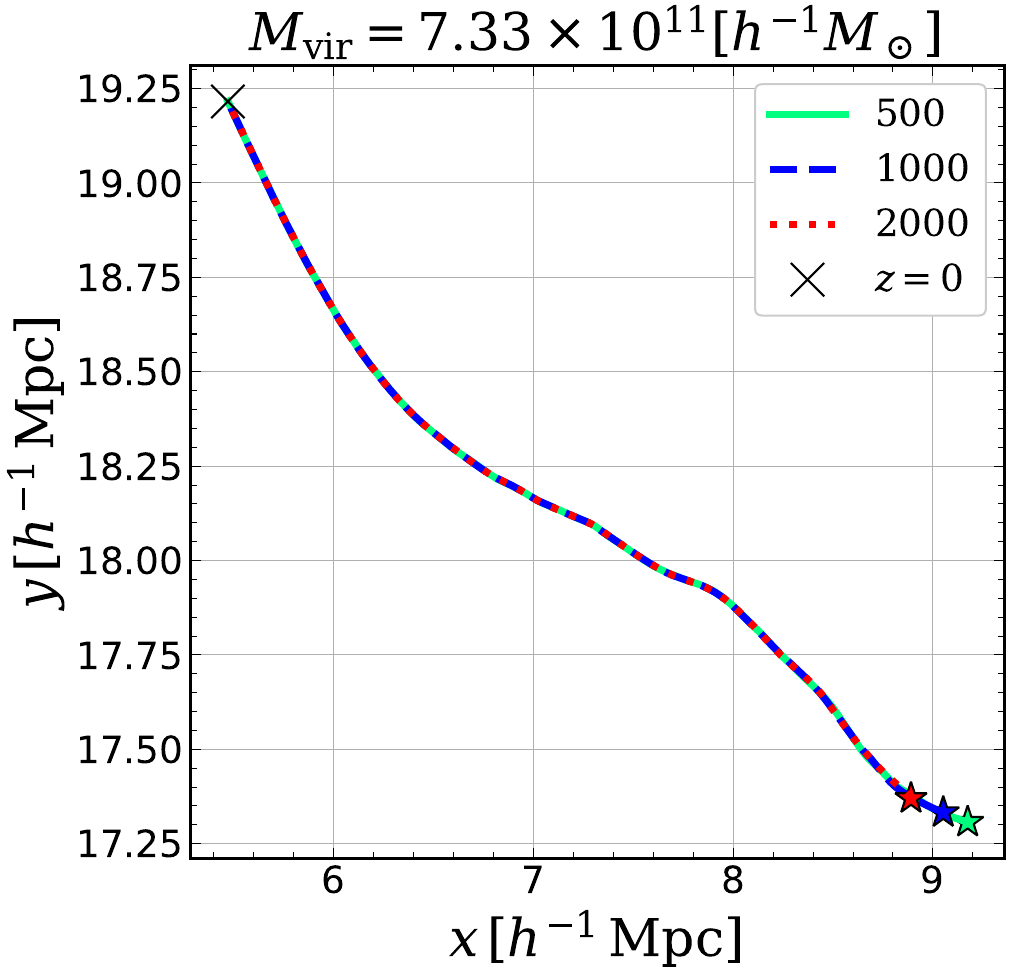}
        \end{minipage} &
        \begin{minipage}{0.66\columnwidth}
        \centering            \includegraphics[scale=0.29]{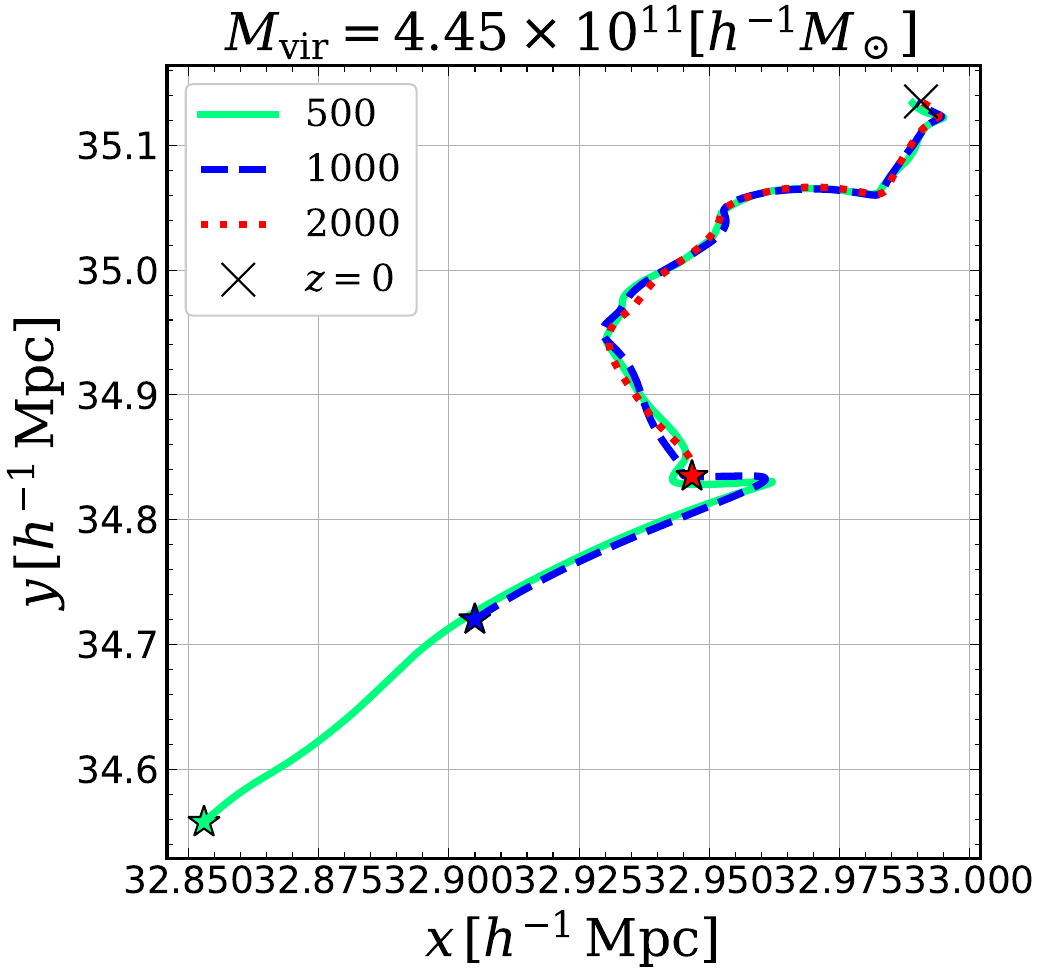}
        \end{minipage}
    \end{tabular}
\caption{The representative trajectories of $\boldsymbol{x}_\mathrm{h}$ from $t=t_s$ to $z=0$.
The colours and the line types correspond to each $N_\mathrm{det}$ shown in the legends.
The stars denote $\boldsymbol{x}_\mathrm{h}$ at $t=t_s$, and the black crosses denote $\boldsymbol{x}_\mathrm{h}$ at $z=0$.
In the case of $t_s$ equals the cosmic time at $z=5$, we select $N_\mathrm{det}$ particles according to the method using the phase-space metric Eq.~\ref{eq:dps} introduced in Section~\ref{subsec:halo_centre}.}
\label{fig:trajectory}
\end{figure*}

In Fig.~\ref{fig:trajectory}, we present representative trajectories of $\boldsymbol{x}_\mathrm{h}$ for each $N_\mathrm{det}$ and, to the best of our visual assessment, observe that the trajectories for different $N_\mathrm{det}$ closely match each other.
However, its impact on the number of apocentre passages is not trivial, especially for particles with large values of $p$ given their small orbital size. Also, the different choice of $N_\mathrm{det}$ can indirectly affect the value of $p$ through its effects on $t_s$, the starting time of the counting.
Remind that we define $t_s$ as the point when the number of particles in the progenitors decreases below $N_\mathrm{det}$; hence, with lower values of $N_\mathrm{det}$, $t_s$ decreases (i.e., we can track the progenitor to higher redshifts), and we start counting $p$ earlier.
The earlier initiation of $p$ counting may potentially result in higher $p$ values for individual particles.

The two massive haloes shown in the upper left and upper middle panels in Fig.~\ref{fig:trajectory} have substantial progenitors, and we observe that the starting position of the counting, $\boldsymbol{x}_\mathrm{h}$ at $t=t_s$ (indicated by star symbols), remains nearly the same for every $N_\mathrm{det}$.
In contrast, some of the less massive haloes, particularly the one in the lower right panel, exhibit substantial variations in $\boldsymbol{x}_\mathrm{h}(t_s)$. This indicates that we can track the progenitors to different times depending on the value of $N_\mathrm{det}$.

\begin{figure*}
    \begin{tabular}{ccc}
        \begin{minipage}{0.66\columnwidth}
        \centering            \includegraphics[scale=0.33]{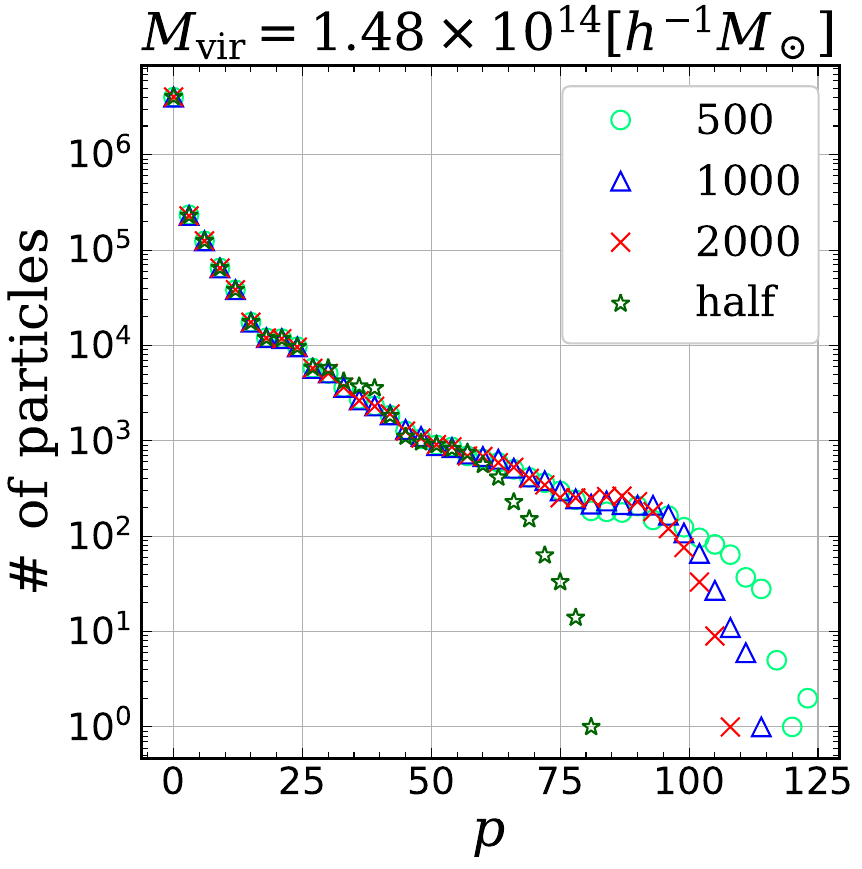}
        \end{minipage} &
        \begin{minipage}{0.66\columnwidth}
        \centering
        \includegraphics[scale=0.33]{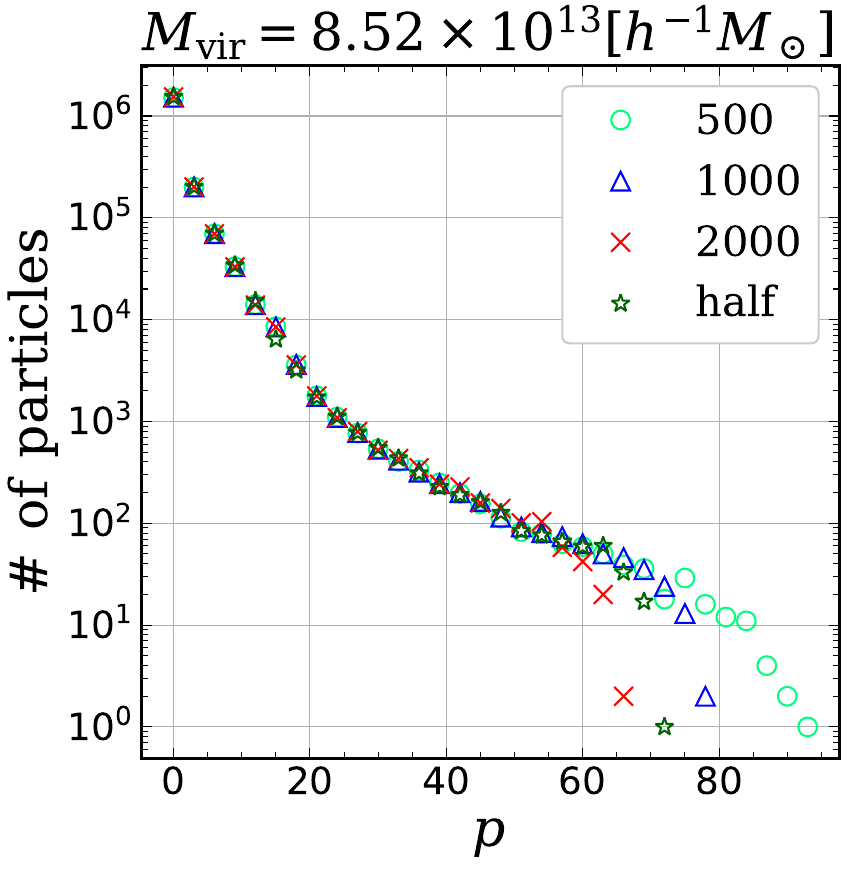}
        \end{minipage} &
        \begin{minipage}{0.66\columnwidth}
        \centering            \includegraphics[scale=0.33]{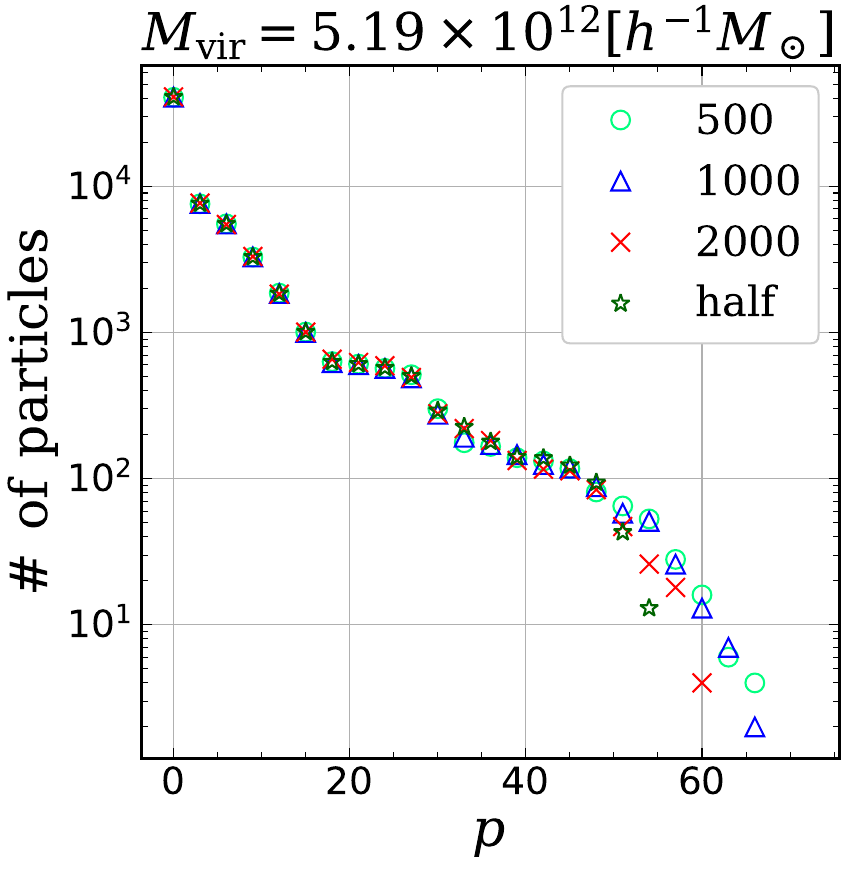}
        \end{minipage} \\
        \begin{minipage}{0.66\columnwidth}
        \centering            \includegraphics[scale=0.33]{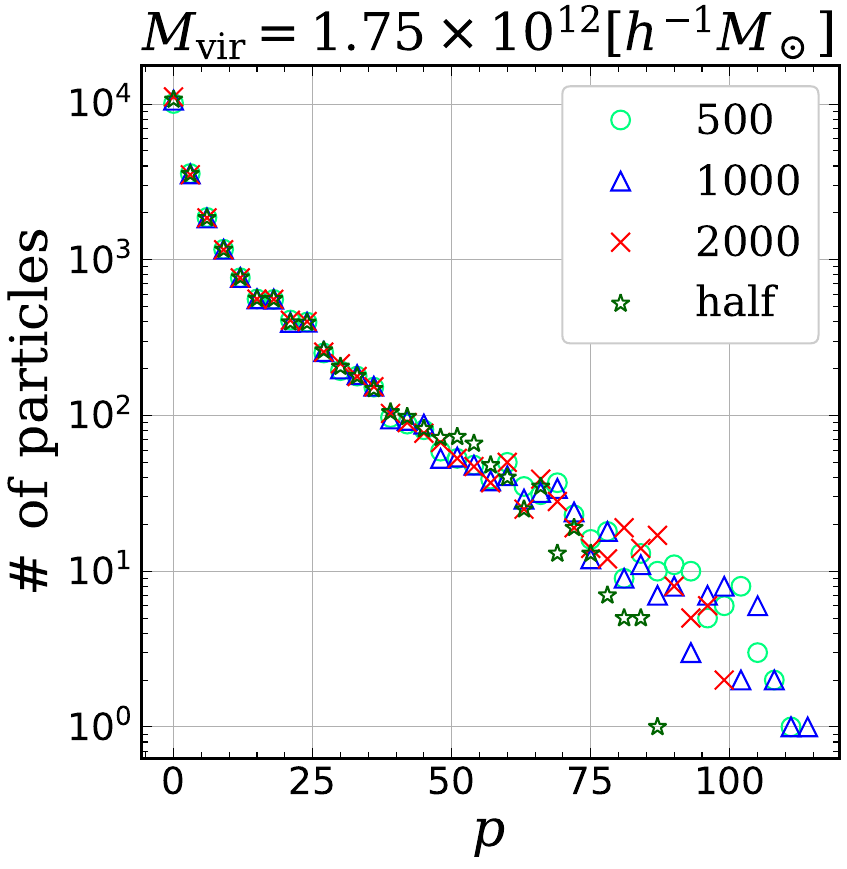}
        \end{minipage} &
        \begin{minipage}{0.66\columnwidth}
        \centering            \includegraphics[scale=0.33]{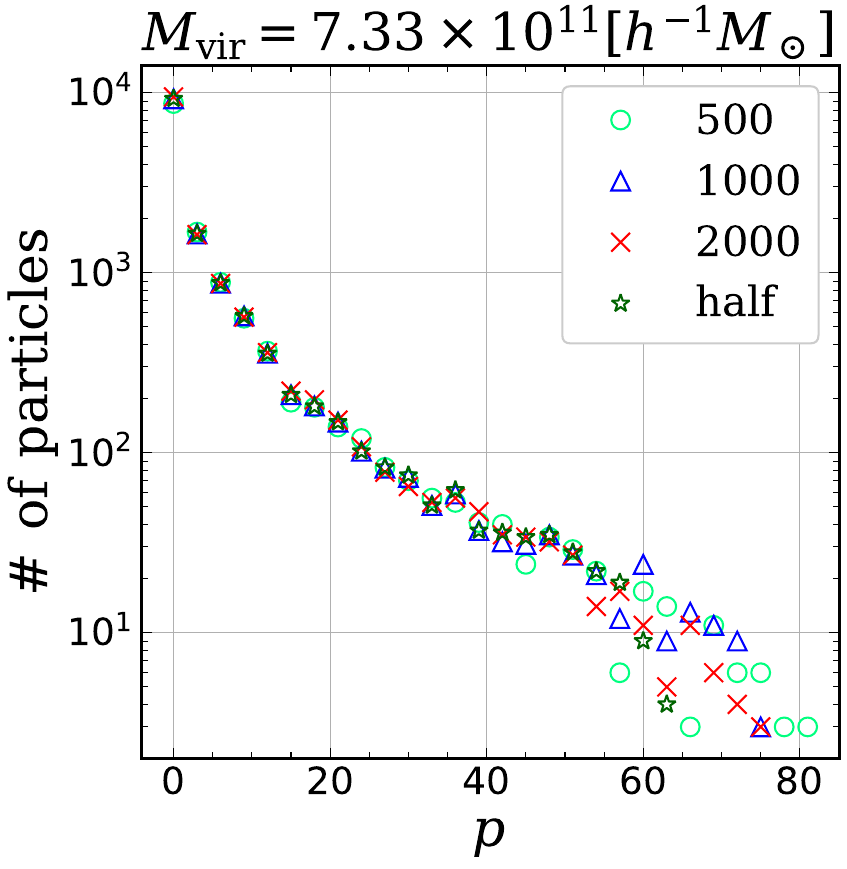}
        \end{minipage} &
        \begin{minipage}{0.66\columnwidth}
        \centering            \includegraphics[scale=0.33]{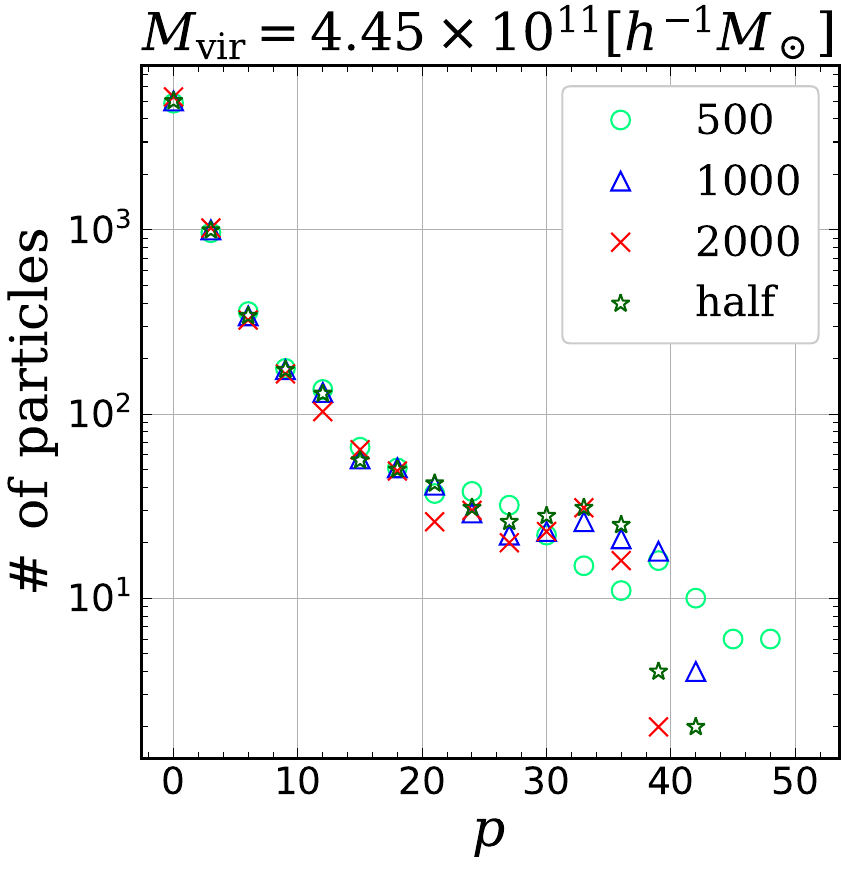}
        \end{minipage}
    \end{tabular}
\caption{Distribution of particles for each $p$ of the halos shown in Fig.~\ref{fig:trajectory}.
The circle, triangle, and cross markers correspond to each $N_\mathrm{det}$ fiducial number 1000.
shown in the legends, and the colours correspond to Fig.~\ref{fig:trajectory}.
The green star denotes the case we track trajectories using every other one of $1001$ snapshots.}
\label{fig:pdistri_compare}
\end{figure*}

We then plot the distribution of $p$ for particles around the same six haloes in Fig.~\ref{fig:pdistri_compare}. Despite the differences discussed in Fig.~\ref{fig:trajectory}, the overall trend remains consistent. For low values of $p$, roughly below $50$, different symbols show good agreement within $\sim 10\%$. However, the dependence on $N_\mathrm{det}$ becomes more noticeable for larger values of $p$. Therefore, for particles with $p\lesssim 50$, both the value of $t_s$ and the precise trajectories of progenitors do not affect much the counts of $p$. In the case of the halo shown in the upper left panel, we can always track the progenitor up to $z=5$ for all the three values of $N_\mathrm{det}$. Therefore, the dependence of particle counts on $N_\mathrm{det}$ for $p\gtrsim 50$ solely arises from the slighly different trajectories of the centre. Note that for less massive haloes, even for $p\lesssim 50$, the counts exhibit fluctuations due to Poisson noise, particularly when the number of particles for each $p$ is less than $\sim30$.

Finally, we investigate the impact of the number of snapshots used in the analysis. 
If our snapshots lack sufficient time resolution for tracking particles, some particles may complete multiple orbits between the snapshots, potentially leading to miscounts of their $p$ values.
Therefore, it is crucial to ensure that our set of 1001 snapshots equally sampled in redshift retains adequate time resolution to mitigate this effect.
To assess this, we track the trajectories using every other one of the 1001 snapshots, and the results are depicted as green star symbols in Fig.~\ref{fig:pdistri_compare}.
Note that $N_\mathrm{det}$ here is set to 1000, which serves as our fiducial number.
We observe that, for most of the halos, the number of particles with large $p$ values is smaller than in the cases where we employ 1001 snapshots (as indicated by the triangles).
However, we can confirm that the distribution for $p\lesssim 50$ even when we use half of the snapshots.

Based on the results presented in this Appendix, we choose to primarily focus on particles with $p\leq 40$ in the main text to ensure a conservative approach.

\section{Convergence tests II: mass resolution}\label{app:2}

In this Appendix, we further examine a convergence study and check if our simulation setup has a sufficient mass resolution to faithfully track particle trajectories without miscounting the number of apocenter passages. 

For this purpose, in addition to the LR and the HR simulations in Table~\ref{tab:simulation_parameters}, we ran two other simulations that have the same initial condition, but with different mass resolutions, halving the box size of simulations (i.e., setting the side length to $10.25\,h^{-1}$\,Mpc).  One is N125, which has the DM particles of $N=125^3$, giving the same mass resolution as in the LR simulation. Another is N250, having the DM particles of $250^3$. With the same box as in N125, the latter provides eight times higher resolution than N125 and LR. 

\begin{figure*}
    \begin{tabular}{ccc}
        \begin{minipage}{0.66\columnwidth}
        \centering            
        \includegraphics[scale=0.7]{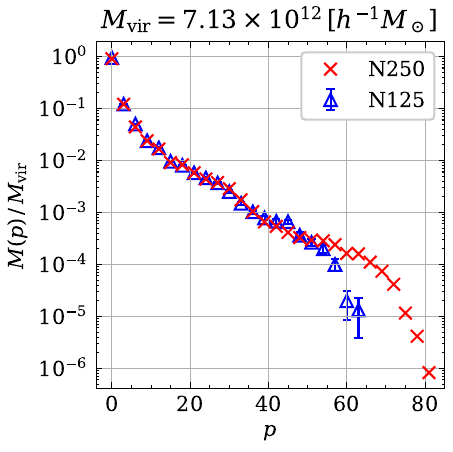}
        \end{minipage} &
        \begin{minipage}{0.66\columnwidth}
        \centering
        \includegraphics[scale=0.7]{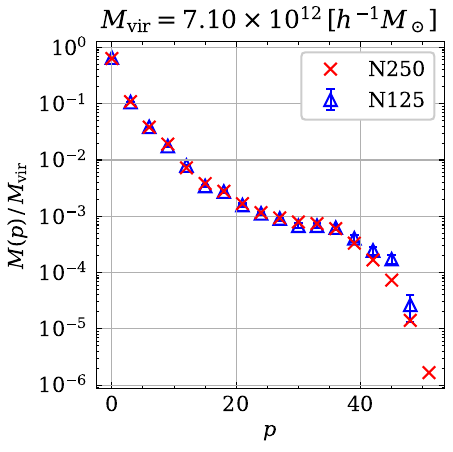}
        \end{minipage} &
        \begin{minipage}{0.66\columnwidth}
        \centering
        \includegraphics[scale=0.7]{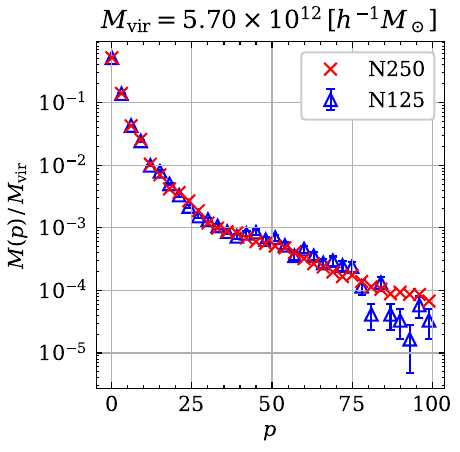}
        \end{minipage} \\
        \begin{minipage}{0.66\columnwidth}
        \centering
        \includegraphics[scale=0.7]{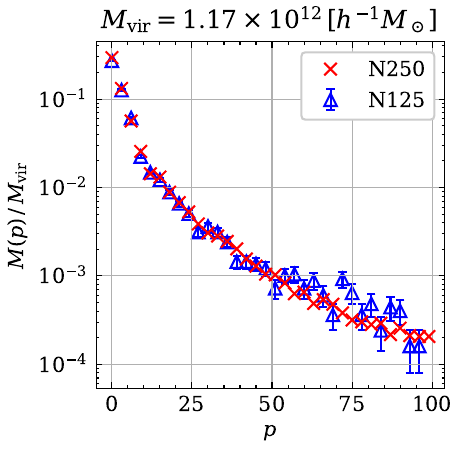}
        \end{minipage} &
        \begin{minipage}{0.66\columnwidth}
        \centering
        \includegraphics[scale=0.7]{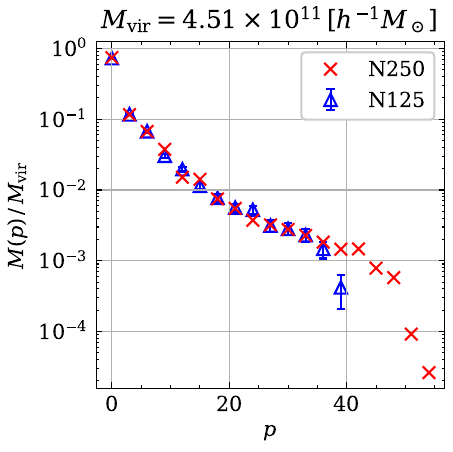}
        \end{minipage} &
        \begin{minipage}{0.66\columnwidth}
        \centering
        \includegraphics[scale=0.7]{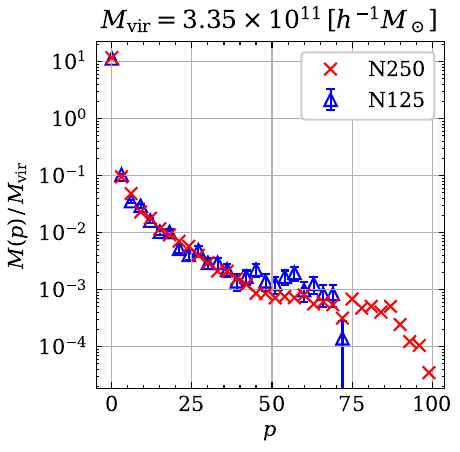}
        \end{minipage}
    \end{tabular}
\caption{
Mass distribution of the DM particles having the same number of apocenter passages $p$, plotted against $p$. Here, we compare the results between two additional simulations, N125 (open triangles) and N250 (crosses). While the former has the same mass resolution as in LR simulation, the latter simulation has a better mass resolution, eight times higher than that of the LR simulation. The errorbars indicate the Poisson noise estimated from the number of particles in each $p$ bin. Note that we basically follow the methods to identify the halo centres and trajectories as well as to count the number of apocenter passages in Sections~\ref{subsec:halo_centre} and \ref{subsec:counting}, but the number of particles used to determine the halo centre in N250 is changed from $1000$ to $8000$ in order to be consistent with its mass resolution. 
}
\label{fig:pdistri_compar_mass}
\end{figure*}

Repeating the same analysis as described in Section~\ref{sec:methods}, we pick up $6$ representative haloes and compare their multi-stream properties between N250 and N125. 
Fig.~\ref{fig:pdistri_compar_mass} shows the particle mass distribution as a function of $p$ for each halo. We see that the results almost coincide with each other at $p\leq 40$. The exception may be the middle-lower panel, where the rightmost bin of N125 shows a large scatter even at $p\leq40$ and deviates significantly from the result of N250. However, it turns out that this bin contains only $4$ particles, and hence the large discrepancy is attributed to the Poisson noise. 

\begin{figure*}
    \begin{tabular}{ccc}
        \begin{minipage}{0.66\columnwidth}
        \centering            
        \includegraphics[scale=0.47]{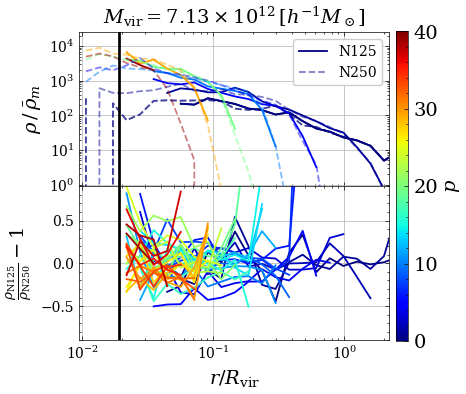}
        \end{minipage} &
        \begin{minipage}{0.66\columnwidth}
        \centering
        \includegraphics[scale=0.47]{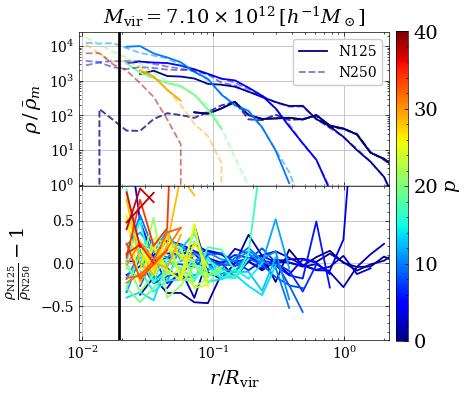}
        \end{minipage} &
        \begin{minipage}{0.66\columnwidth}
        \centering
        \includegraphics[scale=0.47]{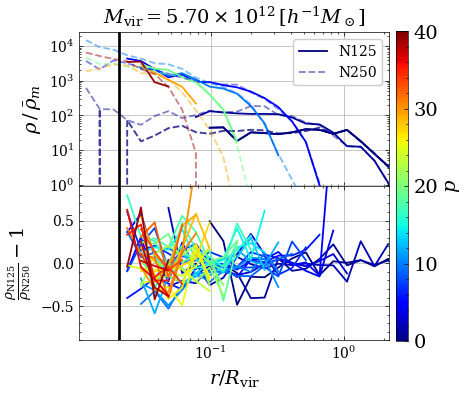}
        \end{minipage} \\
        \begin{minipage}{0.66\columnwidth}
        \centering
        \includegraphics[scale=0.47]{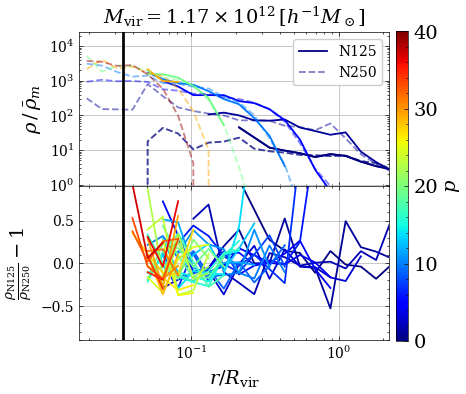}
        \end{minipage} &
        \begin{minipage}{0.66\columnwidth}
        \centering
        \includegraphics[scale=0.47]{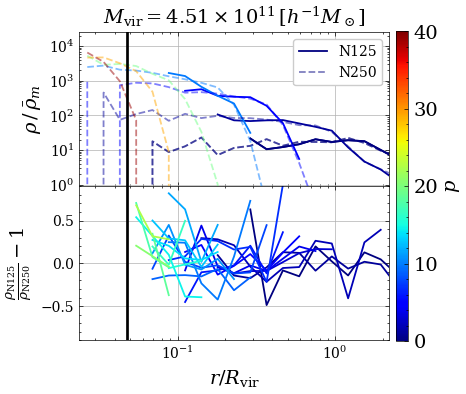}
        \end{minipage} &
        \begin{minipage}{0.66\columnwidth}
        \centering
        \includegraphics[scale=0.47]{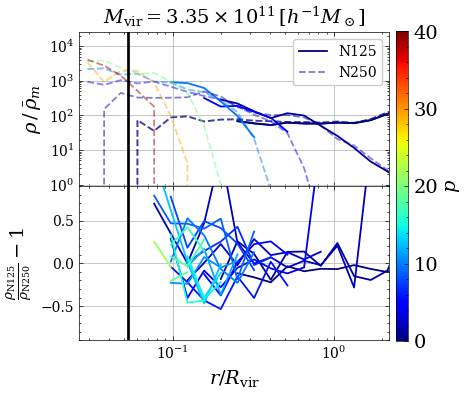}
        \end{minipage}
    \end{tabular}
\caption{
Density profiles of DM particles classified by the number of apocenter passages, $p$. Similar to Fig.~\ref{fig:pdistri_compar_mass}, we compare the results from two additional simulation data between N125 (solid) and N250 (dashed) for six representative haloes.  
The upper panels show the profiles for $p=0,1,5,10,20,30,40$, but excluding bins containing below $10$ particles for N125. The vertical black lines indicate three times the softening length of N125. Lower panels plot the fractional differences of the results between N125 and N250.}
\label{fig:pdens_compar_mass}
\end{figure*}

Next look at Fig.~\ref{fig:pdens_compar_mass}, in which the stream profiles for various values of $p$ are plotted for $6$ representative haloes, with $p$ indicated by color scales. In each panel, fractional differences between measured profiles from N125 and N250 are estimated and plotted in the lower sub-panel. The result shows no systematic trend, and most of the profiles converge well within $50\%$. We have also checked other haloes in N125 and confirmed that for haloes of the mass larger than $3.2\times 10^{11}\,h^{-1}\,M_{\rm vir}$, corresponding to the lowest mass of the mass range S, the behaviors remain the same as shown in Fig.~\ref{fig:pdens_compar_mass}. We therefore conclude that the mass resolution in our LR simulation is sufficient to track particle trajectories and give a converged result for counting the number of apocenter passages at least in a statistical sense.

\section{The Exceptional halo I}\label{excep1}
In Fig.~\ref{fig:residual_mass}, we observe an exceptional halo (represented by the black dot near the top left corner) that remains in the catalogue even after applying the criteria defined by equations~\eqref{eq:equilibrium1} and \eqref{eq:subhaloes}. However, these criteria are designated to filter out subhaloes and haloes undergoing major mergers. This particular halo exhibits a significant discrepancy between the two estimates of its centre, $\boldsymbol{x}_\mathrm{h,ss}$ and $\boldsymbol{x}_\mathrm{h,ROCK}$.

\begin{figure}
\begin{tabular}{c}
    \includegraphics[width=0.95\columnwidth]{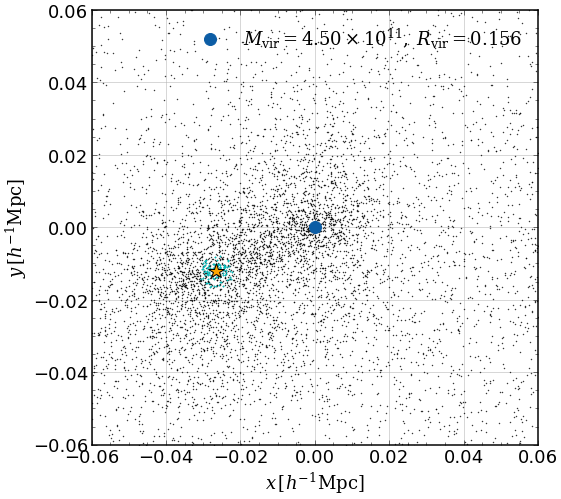} 
    \\
    \includegraphics[width=0.95\columnwidth]{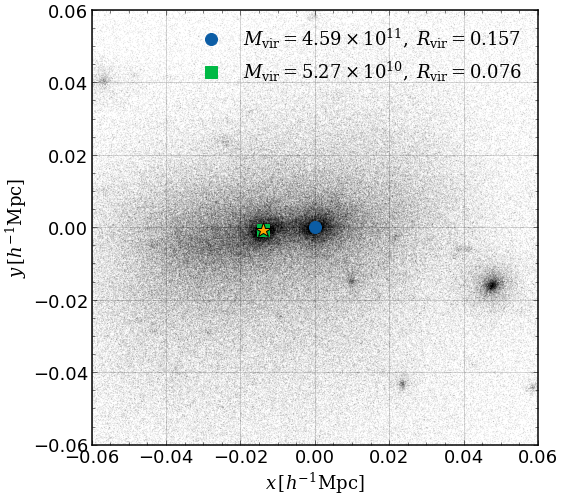} 
\end{tabular}
    \caption{Projected snapshots of the exceptional halo in found near the top left corner of Fig.~\ref{fig:residual_mass}. 
    The upper (lower) panel displays the halo in the LR (HR) run. 
    In both panels, the orange star symbol denotes the density peak where the shrinking-sphere method converges. The simulation particles depicted by cyan dots correspond to the 100 particles inside the sphere in the final iteration of the shrinking-sphere method. The blue-filled circle and green square denote the haloes identified by the \textsc{Rockstar} finder. The mass ($M_\mathrm{vir}$) and radius ($R_\mathrm{vir}$) of these haloes are shown in the figure legend in units of $[h^{-1}M_\odot]$ and $[h^{-1}\mathrm{Mpc}]$, respectively. Note that the cyan dots are completely hidden in the lower panel behind the green square. In the LR run, \textsc{Rockstar} fails to identify any halo corresponding to the density peak marked by the orange star, whereas it is successfully identified in the HR run. The overdensities relative to the background density $\bar{\rho}_\mathrm{m}$ for the nearest 100 particles from the blue points and orange stars are as follows: LR -- $1.0\times10^{5}$ and $1.2\times10^{5}$, and HR -- $2.2\times10^{6}$ and $1.6\times10^{6}$, respectively. Notablly, all the three haloes in the two runs identified by \textsc{Rockstar} satisfy the criteria defined by equations~\eqref{eq:equilibrium1} and ~\eqref{eq:subhaloes}. 
    However, it is evident that the distance $|\boldsymbol{x}_\mathrm{h,ss}-\boldsymbol{x}_\mathrm{h,ROCK}|$ still reaches up to $8\%$ of $R_\mathrm{vir}$ in HR.}
    \label{fig:ex_residual_N500}
\end{figure}

To delve deeper into this halo, we investigate the corresponding halo in the HR run. Here, we confirm the presence of another nearby halo identified by \textsc{Rockstar} (the green square symbol in the lower panel of Fig.~\ref{fig:ex_residual_N500}), which has no counterpart in the \textsc{Rockstar} catalog from the LR run. When we apply the shrinking-sphere method to determine the density peak in this region in the HR run, the centre converges to this nearby halo, despite its virial mass being one order of magnitude smaller than the halo of interest. This outcome is not surprising, as the peak density, determined by the 100 particles near the centre of the less massive nearby halo in the last step of the shrinking-sphere method, is comparable to that around the more massive halo. Therefore, we consider that, even though the nearby halo is not identified by the \textsc{Rockstar} finder in the LR run, the shrinking-sphere method selects this nearby halo instead of the more massive one, leading to a significant discrepancy in the two estimated centres. 

In this sense, although this halo is not excluded by equations~\eqref{eq:equilibrium1} or \eqref{eq:subhaloes}, we consider it likely to undergo a major merger. To address potential technical issues due to the uncertainty of the estimation of the centre, we apply the third criterion, i.e., equation~\eqref{eq:residual_mass}, to identify and exclude such cases from our catalogue.

\section{The Exceptional halo II}\label{excep2}
In the main text, we have identified another problematic halo that remains in the catalog even after applying the first three conditions. This halo exhibits an undesired property when it comes to accurately tracking its trajectory to define the apocentre passages (the halo depicted by the black dot near the top left corner of Fig.~\ref{fig:fract_cendif}). In this Appendix, we conduct a detailed investigation of this particular halo.

\begin{figure}
\begin{tabular}{c}
     \includegraphics[width=0.9\columnwidth]{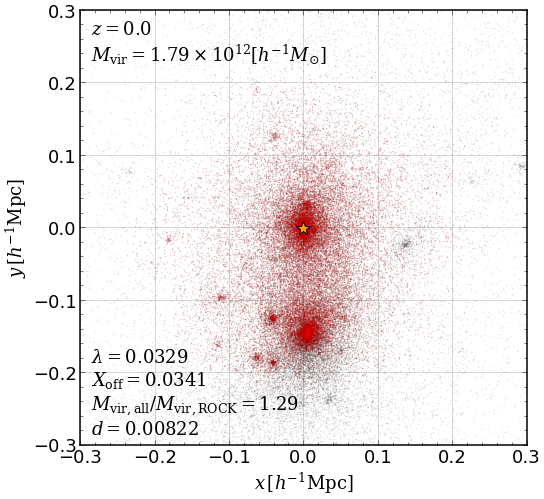} \\
     \includegraphics[width=0.9\columnwidth]{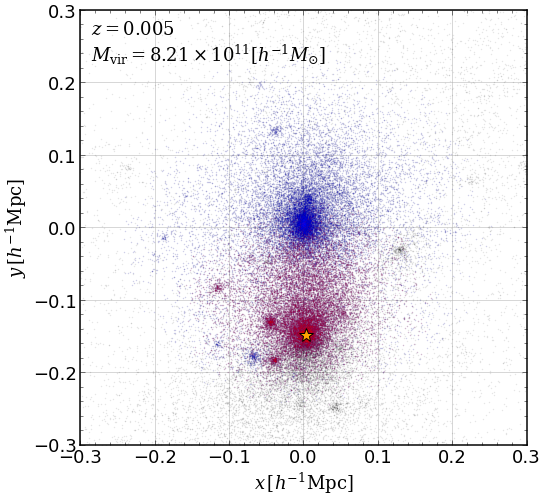}
\end{tabular}
    \caption{Projected snapshots of the simulation particles around an exceptional halo in Fig.~\ref{fig:fract_cendif} at $z=0$ and $0.005$, centreed at $\boldsymbol{x}_\mathrm{h,ROCK}$ at $z=0$. The particles referred to as members of the halo at each redshift (i.e., those within the sphere of virial overdensity) are coloured in red. The blue dots in the lower panel are the members at $z=0$ but non-members at $z=0.005$. The orange stars denote the centre of halo, $\boldsymbol{x}_\mathrm{h,ss}$, determined by the shrinking-sphere method using red particles at each redshift. 
    This halo has $R_\mathrm{vir,ROCK} = 0.206\,h^{-1}\mathrm{Mpc}$ at $z=0$, while the position of the star symbol moves by $0.239[h^{-1}\mathrm{Mpc}]$ between the two snapshots. 
    In the top panel, $\lambda$ denotes the spin parameter, $X_\mathrm{off}$ stands for the offset parameter, $d$ indicates the distance $|\boldsymbol{x}_\mathrm{h,ss}-\boldsymbol{x}_\mathrm{h,ROCK}|/R_\mathrm{vir}$; these parameters satisfy the criteria given by equations~\eqref{eq:equilibrium1}, \eqref{eq:subhaloes} and \eqref{eq:residual_mass}. 
    In both panels, $M_\mathrm{vir}$ denotes the mass corresponding to the red coloured particles, not $M_\mathrm{vir,ROCK}$.
    At $z=0.005$, the secondary peak observed at $z=0$ is regarded as the primary peak by the shrinking-sphere method, leading to a significant displacement of $\boldsymbol{x}_\mathrm{h,ss}$ between the two snapshots. 
    The secondary peak is also found by \textsc{Rockstar}, with a virial mass of $1.05\times10^{12}[h^{-1}M_\odot]$, but it has $\lambda =0.0879$ and is excluded by the criterion in equation~\eqref{eq:equilibrium1}.}
    \label{fig:exceptionalhalo1}
\end{figure}

We present particle snapshots of this exceptional halo at two redshifts, $z=0$ and $z=0.005$, in Fig.~\ref{fig:exceptionalhalo1}. 
These snapshots reveal the merger of two prominent structures. At $z=0$, these structures possess comparable masses, with the primary one plotted near the origin having a mass of $1.79\times10^{12}\,h^{-1}M_\odot$ and the secondary one located below with $1.05\times10^{12}\,h^{-1}M_\odot$. The primary structure is retained after the criterion given by equation~\eqref{eq:equilibrium1}, but the secondary structure does not. 
The bottom panel of Fig.~\ref{fig:exceptionalhalo1} illustrates the inconsistent tracking of the centre of the secondary structure (indicated by the star symbol) at $z=0.005$ through the shrinking-sphere method.
In this case, a major merger event hinders us from consistently tracking the centre over time.
Therefore, we introduce the last criteria, i.e., equation~\eqref{eq:fract_cendif}, to identify and exclude such halos from our catalog.

\end{document}